\documentclass[aps,prd,twocolumn,tightenlines,superscriptaddress,showpacs,preprintnumbers,amsmath,amssymb]{revtex4}

\usepackage[dvips]{color}

\usepackage{graphicx} 
\usepackage{dcolumn}  
\usepackage{bm}
\usepackage{pstricks}
\usepackage{pst-node}

\graphicspath{{ps}}

\usepackage{mathrsfs}
\usepackage{amssymb}
\usepackage{amsmath}
\usepackage{tabularx}
\usepackage{rotating}
\usepackage{multirow}

\newcommand{\bz}{{B^0}}
\newcommand{\bzb}{{\overline{B}{}^0}}

\newcommand{\dE}{{\Delta E}}
\newcommand{\mb}{{M_{\rm bc}}}
\newcommand{\mbc}{{M_{\rm bc}}}
\newcommand{\Dt}{\Delta t}

\newcommand{\fCP}{f_{CP}}

\newcommand{\ftag}{f_{\rm tag}}

\newcommand{\dm}{\Delta m_d}
\newcommand{\dmd}{\dm}
\newcommand{\taubz}{{\tau_\bz}}

\newcommand*{\dwl}{\ensuremath{{\Delta w_l}}}

\newcommand{\pip}{{\pi^+}}
\newcommand{\pim}{{\pi^-}}
\newcommand{\piz}{{\pi^0}}

\newcommand{\bbar}{{\overline{B}}}

\newcommand{\lsig}{{\cal L}_{\rm sig}}
\newcommand{\lbkg}{{\cal L}_{\rm bkg}}
\newcommand{\rsigbkg}{{\cal R}}

\newcommand{\pbstar}{p_B^{\rm cms}}

\newcommand{\nbb}{449}
\newcommand{\nbbsvdone}{152}
\newcommand{\nbbsvdtwo}{297}
\newcommand{\lint}{414}

\newcommand{\kakkoOverlineBeta}{\raise1ex\hbox{\scriptsize ${}^($}
\overline{\beta} \raise1ex\hbox{\scriptsize ${}^)$} \! \!}
\newcommand{\kakkoOverlinef}{\raise1ex\hbox{\scriptsize ${}^($}
\overline{f} \raise1ex\hbox{\scriptsize ${}^{)}$} \! \!}
\newcommand{\kakkoOverlineF}{\raise1ex\hbox{\scriptsize ${}^($}
\overline{F} \raise1ex\hbox{\scriptsize ${}^{)}$} \! \!}
\newcommand{\kakkoOverlineA}{\raise1ex\hbox{\scriptsize ${}^($}
\overline{A} \raise1ex\hbox{\scriptsize ${}^{)}$} \! \!}
\newcommand{\kakkoOverlineGamma}{{\raise0.5ex\hbox{\scriptsize ${}^($}
\overline{\gamma} \raise0.5ex\hbox{\scriptsize ${}^)$} \! \!}}

\newcommand{\pipipi}{\pi^+\pi^-\pi^0}
\newcommand{\BB}{B\overline{B}}
\newcommand{\qq}{q\overline{q}}

\newcommand{\RE}{\text{Re}}
\newcommand{\IM}{\text{Im}}

\newcommand{\BFall}{25.8 \pm 1.2 \pm 3.6}
\newcommand{\BFpm}{22.6 \pm 1.1 \pm 4.4}
\newcommand{\BFz}{3.0 \pm 0.5 \pm 0.7}

\newcommand{\BFpmss}{22.6 \pm 1.1 [\text{stat.}] \pm 4.4 [\text{syst.}]}
\newcommand{\BFzss}{3.0 \pm 0.5 [\text{stat.}] \pm 0.7 [\text{syst.}]}

\begin{document}

\preprint{\vbox{ \hbox{   }
\hbox{Belle Preprint 2007-43}
\hbox{KEK Preprint 2007-47}
}}

\title{ \quad\\[0.5cm]
\boldmath Measurement of $CP$ Asymmetries and Branching Fractions
in a Time-Dependent Dalitz Analysis
of $B^0 \to (\rho \pi)^0$ and a Constraint on the Quark Mixing Angle $\phi_2$}

\affiliation{Budker Institute of Nuclear Physics, Novosibirsk}
\affiliation{Chiba University, Chiba}
\affiliation{University of Cincinnati, Cincinnati, Ohio 45221}
\affiliation{The Graduate University for Advanced Studies, Hayama}
\affiliation{Gyeongsang National University, Chinju}
\affiliation{Hanyang University, Seoul}
\affiliation{University of Hawaii, Honolulu, Hawaii 96822}
\affiliation{High Energy Accelerator Research Organization (KEK), Tsukuba}
\affiliation{Institute of High Energy Physics, Chinese Academy of Sciences, Beijing}
\affiliation{Institute of High Energy Physics, Vienna}
\affiliation{Institute of High Energy Physics, Protvino}
\affiliation{Institute for Theoretical and Experimental Physics, Moscow}
\affiliation{J. Stefan Institute, Ljubljana}
\affiliation{Kanagawa University, Yokohama}
\affiliation{Korea University, Seoul}
\affiliation{Kyungpook National University, Taegu}
\affiliation{\'Ecole Polytechnique F\'ed\'erale de Lausanne (EPFL), Lausanne}
\affiliation{University of Ljubljana, Ljubljana}
\affiliation{University of Maribor, Maribor}
\affiliation{University of Melbourne, School of Physics, Victoria 3010}
\affiliation{Nagoya University, Nagoya}
\affiliation{Nara Women's University, Nara}
\affiliation{National Central University, Chung-li}
\affiliation{National United University, Miao Li}
\affiliation{Department of Physics, National Taiwan University, Taipei}
\affiliation{H. Niewodniczanski Institute of Nuclear Physics, Krakow}
\affiliation{Nippon Dental University, Niigata}
\affiliation{Niigata University, Niigata}
\affiliation{University of Nova Gorica, Nova Gorica}
\affiliation{Osaka City University, Osaka}
\affiliation{Osaka University, Osaka}
\affiliation{Panjab University, Chandigarh}
\affiliation{Saga University, Saga}
\affiliation{University of Science and Technology of China, Hefei}
\affiliation{Seoul National University, Seoul}
\affiliation{Sungkyunkwan University, Suwon}
\affiliation{University of Sydney, Sydney, New South Wales}
\affiliation{Toho University, Funabashi}
\affiliation{Tohoku Gakuin University, Tagajo}
\affiliation{Department of Physics, University of Tokyo, Tokyo}
\affiliation{Tokyo Institute of Technology, Tokyo}
\affiliation{Tokyo Metropolitan University, Tokyo}
\affiliation{Tokyo University of Agriculture and Technology, Tokyo}
\affiliation{Virginia Polytechnic Institute and State University, Blacksburg, Virginia 24061}
\affiliation{Yonsei University, Seoul}
\author{A.~Kusaka}\affiliation{Department of Physics, University of Tokyo, Tokyo} 
\author{C.~C.~Wang}\affiliation{Department of Physics, National Taiwan University, Taipei} 
\author{I.~Adachi}\affiliation{High Energy Accelerator Research Organization (KEK), Tsukuba} 
\author{H.~Aihara}\affiliation{Department of Physics, University of Tokyo, Tokyo} 
\author{K.~Arinstein}\affiliation{Budker Institute of Nuclear Physics, Novosibirsk} 
\author{V.~Aulchenko}\affiliation{Budker Institute of Nuclear Physics, Novosibirsk} 
\author{T.~Aushev}\affiliation{\'Ecole Polytechnique F\'ed\'erale de Lausanne (EPFL), Lausanne}\affiliation{Institute for Theoretical and Experimental Physics, Moscow} 
\author{A.~M.~Bakich}\affiliation{University of Sydney, Sydney, New South Wales} 
\author{V.~Balagura}\affiliation{Institute for Theoretical and Experimental Physics, Moscow} 
\author{E.~Barberio}\affiliation{University of Melbourne, School of Physics, Victoria 3010} 
% ...
\collaboration{The Belle Collaboration}


\begin{abstract}
We present the results of a time-dependent Dalitz plot analysis of
$B^0 \to \pi^+ \pi^- \pi^0$ decays based on a $\lint {\rm fb}^{-1}$ data
sample
that contains $\nbb\times 10^6 B\bbar$ pairs. The data were collected
on the $\Upsilon(4S)$ resonance with the Belle detector at the KEKB
asymmetric energy $e^+ e^-$ collider.
Combining our analysis with information on charged $B$ decay modes,
we perform a full Dalitz and isospin analysis
and obtain a constraint on the quark mixing angle $\phi_2$,
$68^\circ < \phi_2 < 95^\circ$ at the 68.3\% confidence level for the
$\phi_2$ solution consistent with the standard model (SM).
A large SM-disfavored region also remains.
The branching fractions for the decay processes
$\bz \to \rho^\pm(770) \pi^\mp$ and $\bz \to \rho^0(770) \pi^0$ are
measured
to be
$\left( \BFpmss \right) \times 10^{-6}$ and
$\left( \BFzss \right) \times 10^{-6}$, respectively.
These are the first branching fraction measurements of the process
$\bz \to \rho(770)\pi$
with the lowest resonance $\rho(770)$ explicitly separated from the
radial excitations.
\end{abstract}

\pacs{11.30.Er, 12.15.Hh, 13.25.Hw}

\maketitle

{\renewcommand{\thefootnote}{\fnsymbol{footnote}}}
\setcounter{footnote}{0}

\section{Introduction
\label{sec:introduction}
}
In the standard model (SM),
$CP$ violation arises from an irreducible complex phase in the
Cabibbo-Kobayashi-Maskawa (CKM) matrix~\cite{Kobayashi:1973fv}.
The SM predicts
that measurement of a time-dependent $CP$ asymmetry
between the
decay rates of $\bz$ and $\bzb$
gives access to the $CP$ violating phase
in the CKM matrix~\cite{Carter:1980hr,Carter:1980tk,Bigi:1981qs}.
The angle $\phi_2$ of the CKM unitarity triangle
can be measured via
the tree diagram contribution
in $b\rightarrow u \overline{u} d$ decay processes,
such as
$B^0 \rightarrow \pi^+ \pi^-$,
$B^0 \rightarrow \rho^\pm\pi^\mp$, or
$B^0 \rightarrow \rho^+ \rho^-$~\cite{ChargeConjugate}.
In these decay processes,
however,
contributions from so-called $b\rightarrow d$ penguin diagrams
could contaminate
the measurement of $\phi_2$.
Snyder and Quinn pointed out
that a Dalitz plot analysis
of $B^0 \rightarrow \rho \pi$,
which includes $B^0 \rightarrow \rho^+\pi^-$,
$B^0 \rightarrow \rho^-\pi^+$, and
$B^0 \rightarrow \rho^0\pi^0$,
offers a unique way to determine $\phi_2$
without ambiguity.
The Dalitz plot analysis takes into account
a possible
contamination from the penguin
contribution~\cite{Snyder:1993mx}.
In addition, an isospin analysis~\cite{Lipkin:1991st,Gronau:1991dq}
involving the charged decay modes,
$B^+\rightarrow \rho^+\pi^0$ and $B^+\rightarrow \rho^0\pi^+$,
provides further improvement
of the $\phi_2$ determination.

The Belle~\cite{kusaka:221602} and BaBar~\cite{Aubert:2007jn}
Collaborations recently reported the first measurements employing a
time-dependent Dalitz plot analysis technique. In this paper we describe
the
details of the time-dependent Dalitz plot analysis with the Belle
detector at the KEKB asymmetric energy $e^+ e^-$ collider reported in
Ref.~\cite{kusaka:221602}. We also
present the first measurements of the branching fractions of
$\bz \to \rho^\pm(770) \pi^\mp$ and $\bz \to \rho^0(770) \pi^0$ decay
processes obtained from the Dalitz plot analysis, where the $\rho(770)$
is separated from radial excitations. These results can be compared
with the branching fraction of the process
$B^\pm \to \rho^0(770)\pi^\pm$~\cite{Aubert:2005sk}.

\subsection{KEKB and Belle Detector
\label{subsec:intro_detector}
}
KEKB~\cite{KEKB} operates at the $\Upsilon(4S)$ resonance
($\sqrt{s}=10.58$~GeV) with a peak luminosity that exceeds
$1.6\times 10^{34}~{\rm cm}^{-2}{\rm s}^{-1}$.
At KEKB, the $\Upsilon(4S)$ is produced
with a Lorentz boost of $\beta\gamma=0.425$ nearly along
the electron beamline ($z$).
Since the $B^0$ and $\bzb$ mesons are approximately at
rest in the $\Upsilon(4S)$ center-of-mass system (cms),
$\Delta t$ can be determined from the displacement in the $z$ direction,
$\Delta z$, between the vertices of the two $B$ mesons:
$\Delta t \simeq  \Delta z/\beta\gamma c$.

The Belle detector is a large-solid-angle magnetic
spectrometer that
consists of a silicon vertex detector (SVD),
a 50-layer central drift chamber (CDC), an array of
aerogel threshold \v{C}erenkov counters (ACC),
a barrel-like arrangement of time-of-flight
scintillation counters (TOF), and an electromagnetic calorimeter
comprised of CsI(Tl) crystals (ECL) located inside
a super-conducting solenoid coil that provides a 1.5~T
magnetic field.  An iron flux-return located outside of
the coil is instrumented to detect $K_L^0$ mesons and to identify
muons (KLM).  The detector
is described in detail elsewhere~\cite{Belle}.
Two inner detector configurations were used. A 2.0 cm beampipe
and a 3-layer silicon vertex detector were used for the first data sample
of $\nbbsvdone\times 10^6 B\bbar$ pairs (DS-I),
while a 1.5 cm beampipe, a 4-layer
silicon detector and a small-cell inner drift chamber were used to record
the remaining $\nbbsvdtwo\times 10^6 B\bbar$ pairs
(DS-II)~\cite{SVD2}.

\subsection{Outline of the analysis}
The analysis proceeds in the following steps.
First, we extract the signal
fraction (Sec.~\ref{sec:selection_and_reconstruction}).
We then determine the sizes
and phases of the contributions from radial excitations
(Sec.~\ref{sec:lineshape_determine}).
Using the parameters determined in the steps above,
we perform a time-dependent Dalitz plot analysis
(Secs.~\ref{sec:time_depend_analysis_pdf}--\ref{sec:systematic_errors}).
The fit results are interpreted as
quasi-two-body $CP$ violation parameters
(Sec.~\ref{sec:quasi_two_body}) and as branching fractions of
$\bz \to \rho^\pm(770) \pi^\mp$ and $\bz \to \rho^0(770) \pi^0$ decays
(Sec.~\ref{sec:branch}).
We subsequently use these results to constrain the CKM angle $\phi_2$
(Sec.~\ref{sec:constraint_on_ph2}).

\subsection{Differential decay width of time-dependent Dalitz plot}
We measure the decay process $B^0 \rightarrow \pi^+ \pi^- \pi^0$,
where we denote the four-momenta of the $\pi^+$, $\pi^-$, and
$\pi^0$ by $p_+$, $p_-$, and $p_0$, respectively.
The invariant-mass squared of their combinations
\begin{equation}
\begin{split}
s_+ = (p_+ + p_0)^2, & \quad \quad
s_- = (p_- + p_0)^2, \\
s_0 = (p_+ & + p_-)^2
\end{split}
\end{equation}
satisfies the following equation
\begin{equation}
s_+ + s_- + s_0 = m_{B^0}^2 + 2 m_{\pi^+}^2 + m_{\pi^0}^2
\label{equ:dalitz_var_add_relation}
\end{equation}
by energy and momentum conservation.
The differential (time-integrated) decay width with respect to the
variables above (Dalitz plot) is
\begin{equation}
d\Gamma = \frac{1}{(2\pi)^3}
\frac{|\kakkoOverlineA \, {}_{3\pi}|^2}{8 m_{B^0}^2} ds_+ ds_- \;,
\label{equ:amplitude_dalitz_width}
\end{equation}
where $\kakkoOverlineA \, {}_{3\pi}$ is the Lorentz-invariant
amplitude of the $\bz (\bzb) \to \pipipi$ decay.

In the decay chain
$\Upsilon(4S) \rightarrow \bz \bzb \rightarrow f_1 f_2$,
where one of the $B$'s decays into final state $f_1$ at time $t_1$
and the other decays into another final state $f_2$ at time $t_2$,
the time-dependent amplitude is
\begin{eqnarray}
\begin{split}
A(t_1, & t_2) \sim e^{-(\Gamma/2 + iM)(t_1 + t_2)} \\
\times & \Biggl\{
\cos[\dmd (t_1-t_2)/2]
\left(A_1 \overline{A}_2 - \overline{A}_1 A_2\right) \\
& \; \; - i \sin[\dmd (t_1-t_2)/2]
\left(\frac{p}{q} A_1 A_2 - \frac{q}{p} \overline{A}_1 \overline{A}_2
\right)
\Biggr\} \;. \\
\end{split}
\end{eqnarray}
Here, $p$ and $q$ define the mass eigenstates of neutral $B$ mesons
as $p \bz \pm q \bzb$, with average mass $M$ and width $\Gamma$,
and mass difference $\Delta m_d$.
The width difference is assumed to be zero.

The decay amplitudes are defined as follows,
\begin{eqnarray}
A_1 & \equiv & A(B^0 \rightarrow f_1) \;, \\
\overline{A}_1 & \equiv & A(\bzb \rightarrow f_1) \;, \\
A_2 & \equiv & A(B^0 \rightarrow f_2) \;, \\
\overline{A}_2 & \equiv & A(\bzb \rightarrow f_2) \;.
\end{eqnarray}
In this analysis,
we take $A_{3\pi}$ as $A_1$
and choose $f_2$ to be
a flavor eigenstate,
i.e., $A_2 = 0$ or $\overline{A}_2 = 0$.
Here we call the $B$ decaying into $f_1 = \fCP = \pi^+ \pi^- \pi^0$
the $CP$ side $B$ while the other $B$ is the tag side $B$,
$f_2 = f_\mathrm{tag}\:(\overline{f}_\mathrm{tag})$.
The differential decay width dependence on time difference
$\Delta t \equiv t_{CP} - t_\mathrm{tag}$ is then,
\begin{equation}
\begin{split}
d \Gamma  \sim
& e^{-\Gamma |\Delta t|}  \Bigl\{
\left(|A_{3\pi}|^2 + |\overline{A}_{3\pi}|^2\right)
\\
& - q_\mathrm{tag} \cdot
\left(|A_{3\pi}|^2 - |\overline{A}_{3\pi}|^2\right)
\cos (\dmd \Delta t)
\\
& +
q_\mathrm{tag}\cdot
2 \mathrm{Im}\left(
\frac{q}{p}
A_{3\pi}^*
\overline{A}_{3\pi}
\right)
\sin (\dmd \Delta t)
\Bigr\}
\, d \Dt \;,
\\
\end{split}
\label{equ:amplitude_dt_width}
\end{equation}
where we assume $|q/p| = 1$ ($CP$ and $CPT$ conservation in mixing)
and
$|A(B^0\rightarrow f_\mathrm{tag})| = |A(\bzb\rightarrow \overline{f}_\mathrm{tag})|$,
and integrate
over $t_\mathrm{sum} = t_{CP} + t_\mathrm{tag}$.
Here $q_\mathrm{tag}$ is the $b$-flavor charge
and $q_\mathrm{tag} = +1\: (-1)$  when the tag-side
$B$ decays as a $\bz$ ($\bzb$) flavor eigenstate.

Combining the Dalitz plot decay width (\ref{equ:amplitude_dalitz_width})
and the time dependent decay width
(\ref{equ:amplitude_dt_width}),
we obtain the time dependent Dalitz plot decay width
\begin{equation}
d\Gamma \sim |A(\Delta t; s_+, s_-)|^2 \: d\Delta t \: ds_+ \: ds_- \;,
\end{equation}
where
\begin{equation}
\begin{split}
|A(\Delta t; & s_+, s_-)|^2
=
e^{-\Gamma |\Delta t|}
\Bigl\{
\left(|A_{3\pi}|^2 + |\overline{A}_{3\pi}|^2\right)
\\
&  - q_\mathrm{tag} \cdot
\left(|A_{3\pi}|^2 - |\overline{A}_{3\pi}|^2\right)
\cos (\dmd \Delta t)
\\
&  + q_\mathrm{tag} \cdot
2 \mathrm{Im}\left(
\frac{q}{p}
A_{3\pi}^*
\overline{A}_{3\pi}
\right)
\sin (\dmd \Delta t)
\Bigr\} \;,
\label{equ:amplitude_time_dep_three_pi}
\end{split}
\end{equation}
\begin{equation}
A_{3\pi} = A_{3\pi}(s_+, s_-) \;, \quad
\overline{A}_{3\pi} = \overline{A}_{3\pi}(s_+, s_-) \;.
\end{equation}

We assume that the $B^0 \to \pi^+\pi^-\pi^0$ decay is dominated by
the $B^0 \rightarrow (\rho \pi)^0$ amplitudes:
$B^0 \rightarrow \rho^+ \pi^-$, $B^0 \rightarrow \rho^- \pi^+$,
and $B^0 \rightarrow \rho^0 \pi^0$,
where $\rho$ can be $\rho(770)$, $\rho(1450)$,
or $\rho(1700)$.
Although there could exist
contributions from $B^0$ decays into
non-$\rho\pi$ $\pi^+\pi^-\pi^0$ final states, such as
$f_0(980)\pi^0$, $f_0(600) \pi^0$, $\omega \pi^0$, and
non-resonant $\pi^+\pi^-\pi^0$,
we confirm that these contributions are small; their effects are taken
into account as systematic
uncertainties (Sec.~6-5).
The Dalitz plot amplitude $A_{3\pi}(s_+, s_-)$
can then be written as
\begin{eqnarray}
A_{3\pi}(s_+, s_-) & = &
\sum_{\kappa = (+, -, 0)}
f_\kappa(s_+, s_-) A^\kappa \;,
\label{equ:amplitude_q2b_b}
\\
\frac{q}{p}
\overline{A}_{3\pi}(s_+, s_-)
& = &
\sum_{\kappa = (+, -, 0)}
\overline{f}{}_\kappa(s_+, s_-) \overline{A}{}^\kappa \;,
\label{equ:amplitude_q2b_bbar}
\end{eqnarray}
where $A^\kappa (\overline{A}{}^\kappa)$
are complex amplitudes corresponding to
$\bz(\bzb) \to \rho^+ \pi^-, \rho^- \pi^+, \rho^0 \pi^0$
for $\kappa = +, -, 0$
and the functions $f_\kappa(s_+, s_-)$
incorporate the kinematic and dynamical properties
of the $B^0$ decay into a vector $\rho$ and a pseudoscalar
$\pi$.
The goal of this analysis is to measure
the complex amplitudes $A^+$, $A^-$, $A^0$,
$\overline{A}{}^+$, $\overline{A}{}^-$, and $\overline{A}{}^0$;
we then constrain the CKM angle $\phi_2$ using these amplitudes.

In contrast to a quasi-two-body $CP$ violation analysis,
the time-dependent Dalitz analysis
includes measurements of the sizes of the interferences
among the final states $\rho^+\pi^-$, $\rho^-\pi^+$
and $\rho^0\pi^0$,
and $CP$-violating asymmetries in the mixed final states.
In principle,
these measurements allow us to determine all the relative
sizes and phases of the amplitudes $A^\kappa$ and $\overline{A}{}^\kappa$,
which are related to $\phi_2$ through an isospin
relation~\cite{Lipkin:1991st,Gronau:1991dq} by
\begin{equation}
e^{+2i\phi_2} =
\frac{\overline{A}{}^+ + \overline{A}{}^- + 2 \overline{A}{}^0}
{A^+ + A^- + 2 A^0} \;.
\end{equation}
Consequently, in the limit of high statistics,
we can constrain $\phi_2$ without discrete ambiguities.

\subsection{\boldmath Kinematics of
$B^0 \rightarrow (\rho \pi)^0$
\label{sec:square_dalitz_kinema_angular}
}
The function $\kakkoOverlinef {}_\kappa(s_+, s_-)$
can be factorized into two parts as
\begin{equation}
\kakkoOverlinef {}_\kappa(s_+, s_-) = T_J^\kappa F_{\pi}(s_\kappa)
\quad (\kappa = +, -, 0) \;,
\label{equ:f_kappa_definition_with_unique_lineshape_assumption}
\end{equation}
where $F_{\pi}(s_\kappa)$
and $T_J^\kappa$
correspond to the lineshape of the $\rho$
and the helicity distribution of the $\rho$,
respectively.
Here we assume that
a single unique functional form for
the lineshape $F_{\pi}(s)$ can be used
for all six $\kakkoOverlinef {}_\kappa$~\cite{Fpi_strictly_speakiing}.
Since this assumption has no good theoretical or experimental
foundation,
we check the validity of the assumption with data
and assign systematic errors.
\par
The lineshape is parameterized with Breit-Wigner functions
corresponding to the $\rho(770)$, $\rho(1450)$, and $\rho(1700)$
resonances:
\begin{equation}
F_{\pi}(s) =
BW_{\rho(770)}^{\mathrm{GS}}
+ \beta \cdot BW_{\rho(1450)}^{\mathrm{GS}}
+ \gamma \cdot BW_{\rho(1700)}^{\mathrm{GS}} \;,
\label{equ:fpi_lineshape_beta_gamma}
\end{equation}
where the amplitudes $\beta$ and $\gamma$ (denoting the relative size of two
resonances)
are complex numbers.
We use the Gounaris-Sakurai (GS)
model~\cite{Gounaris:1968mw} for the Breit-Wigner shape
of each resonance~\cite{ResonanceModel}.
\par
\begin{figure}[htbp]
\begin{center}
\includegraphics[width=0.15\textwidth]{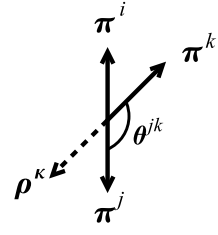}
\caption{
\label{fig:amplitude_rho_pi_kinema_in_rho_rest}
The relation between three pions in the rest frame of $\rho^\kappa$.
}
\end{center}
\end{figure}

In the case of a pseudoscalar-vector ($J=1$) decay,
$T_J^\kappa$ is given by
\begin{equation}
T_1^\kappa = - 4 |\vec{p}_j| |\vec{p}_k| \cos \theta^{jk} \;,
\end{equation}
\begin{equation}
\left(
\begin{split}
T_1^+ & = - 4 |\vec{p}_+| |\vec{p}_-| \cos \theta^{+-} \\
T_1^- & = - 4 |\vec{p}_0| |\vec{p}_+| \cos \theta^{0+} \\
T_1^0 & = - 4 |\vec{p}_-| |\vec{p}_0| \cos \theta^{-0} \\
\end{split}
\right) ,
\end{equation}
where $\vec{p}_j, \vec{p}_k$ are the three momenta
of the $\pi^j$ and $\pi^k$ in the rest frame of $\rho^\kappa$
(or the $\pi^i\pi^j$ system),
and $\theta^{jk} (\equiv \theta_\kappa)$
is the angle between $\vec{p}_j$ and $\vec{p}_k$
(see Fig.~\ref{fig:amplitude_rho_pi_kinema_in_rho_rest}).

\subsection{Fitting parameters
\label{sec:square_dlz_fitting_parameters}}
After
(\ref{equ:f_kappa_definition_with_unique_lineshape_assumption})
is inserted
into expressions
(\ref{equ:amplitude_q2b_b}) and (\ref{equ:amplitude_q2b_bbar}),
the coefficients of Eq.~(\ref{equ:amplitude_time_dep_three_pi})
become
\begin{equation}
\begin{split}
& |A_{3\pi}|^2 \pm |\overline{A}_{3\pi}|^2
= \sum_{\kappa \in \{+,-,0\}} |f_\kappa|^2 U^\pm_\kappa
\\
& \quad
+ 2 \hspace{-5mm} \sum_{\kappa<\sigma \in \{+,-,0\}} \hspace{-3mm}
\left(
\mathrm{Re}[f_\kappa f^*_\sigma] U^{\pm,\mathrm{Re}}_{\kappa\sigma}
- \mathrm{Im}[f_\kappa f^*_\sigma] U^{\pm,\mathrm{Im}}_{\kappa\sigma}
\right)
\;,
\end{split}
\label{equ:dalitz_parameterization_1}
\end{equation}
\begin{equation}
\begin{split}
& \mathrm{Im}\left(\frac{q}{p} A_{3\pi}^* \overline{A}_{3\pi} \right)
= \sum_{\kappa \in \{+,-,0\}} |f_\kappa|^2 I_\kappa
\\
& \hspace{12mm}
+ \hspace{-4mm} \sum_{\kappa<\sigma \in \{+,-,0\}} \hspace{-3mm}
\left(
\mathrm{Re}[f_\kappa f^*_\sigma] I^\mathrm{Im}_{\kappa\sigma}
+ \mathrm{Im}[f_\kappa f^*_\sigma] I^\mathrm{Re}_{\kappa\sigma}
\right)
\;,
\end{split}
\label{equ:dalitz_parameterization_2}
\end{equation}
with
\begin{eqnarray}
U^\pm_{\kappa}
& = & |A^\kappa|^2 \pm |\overline{A}{}^\kappa|^2 \;,
\label{equ:fit_params_first}
\\
I_\kappa & = & \mathrm{Im}
\left[
\overline{A}{}^{\kappa} A^{\kappa *}
\right] \;,
\label{equ:fit_params_2}
\\
U^{\pm,\mathrm{Re}\:(\mathrm{Im})}_{\kappa\sigma}
& = & \mathrm{Re}\:(\mathrm{Im})
\left[ A^\kappa A^{\sigma *} \pm \overline{A}{}^\kappa
\overline{A}{}^{\sigma *}
\right] \;,
\label{equ:fit_params_3}
\\
I^\mathrm{Re\:(Im)}_{\kappa \sigma} & = & \mathrm{Re\:(Im)}
\left[
\overline{A}{}^\kappa A^{\sigma *}
\! - \! (+) \, \overline{A}{}^\sigma A^{\kappa *}
\right] \;.
\label{equ:fit_params_last}
\end{eqnarray}
The 27 coefficients
(\ref{equ:fit_params_first})--(\ref{equ:fit_params_last})
are the parameters determined by the fit~\cite{Quinn:2000by}.
The parameters (\ref{equ:fit_params_first})--(\ref{equ:fit_params_2})
and (\ref{equ:fit_params_3})--(\ref{equ:fit_params_last}) are called
noninterfering and interfering parameters, respectively.
This parameterization allows us
to describe the differential decay width as a linear combination
of independent functions,
whose coefficients are the fit parameters
in a well-behaved fit.
We fix the overall normalization by requiring $U^+_{+} = 1$.
Thus, 26 of the 27 coefficients are free parameters in the fit.

\subsection{Square Dalitz plot (SDP)}
\begin{figure}[tbp]
\begin{center}
\includegraphics[width=0.38\textwidth]{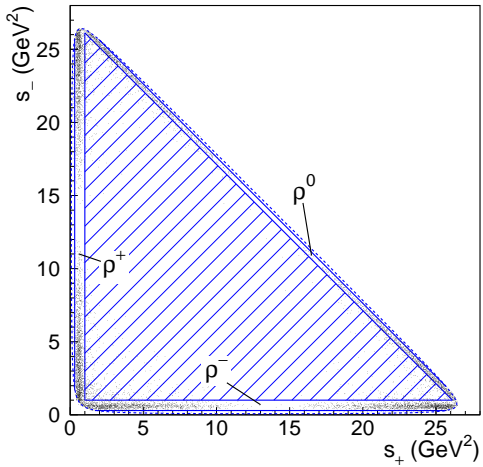}
\caption{
\label{fig:sdp_sig_distr_usual_dalitz}
Distribution of signal Monte Carlo (MC) (without detector efficiency and
smearing)
in the usual Dalitz plot.
The dashed line is the kinematic boundary
while the hatched region corresponds to the region
rejected by the mass cut described in
Sec.~\ref{sec:selection_and_reconstruction}.
}
\end{center}
\end{figure}

\begin{figure}[bp]
\begin{center}
\includegraphics[width=0.23\textwidth]{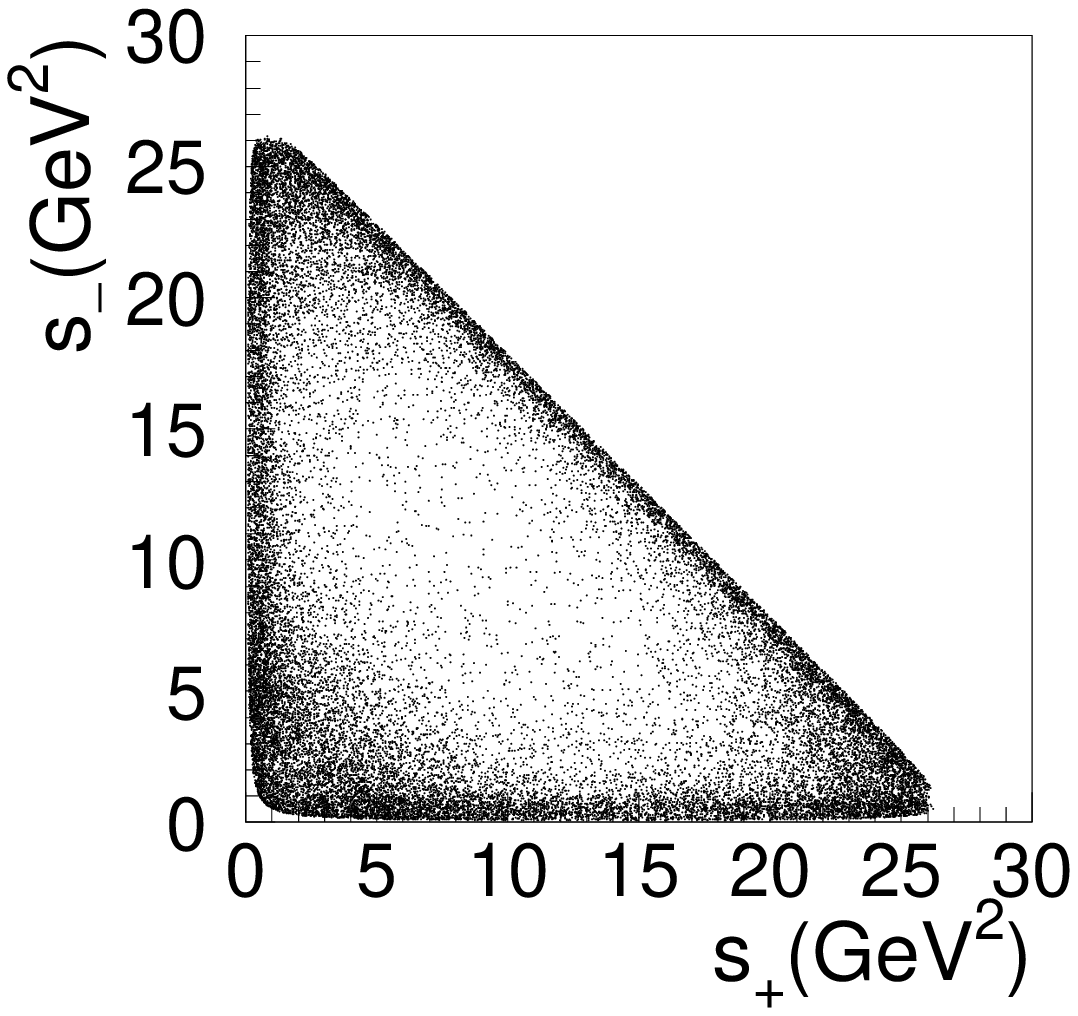}
\includegraphics[width=0.23\textwidth]{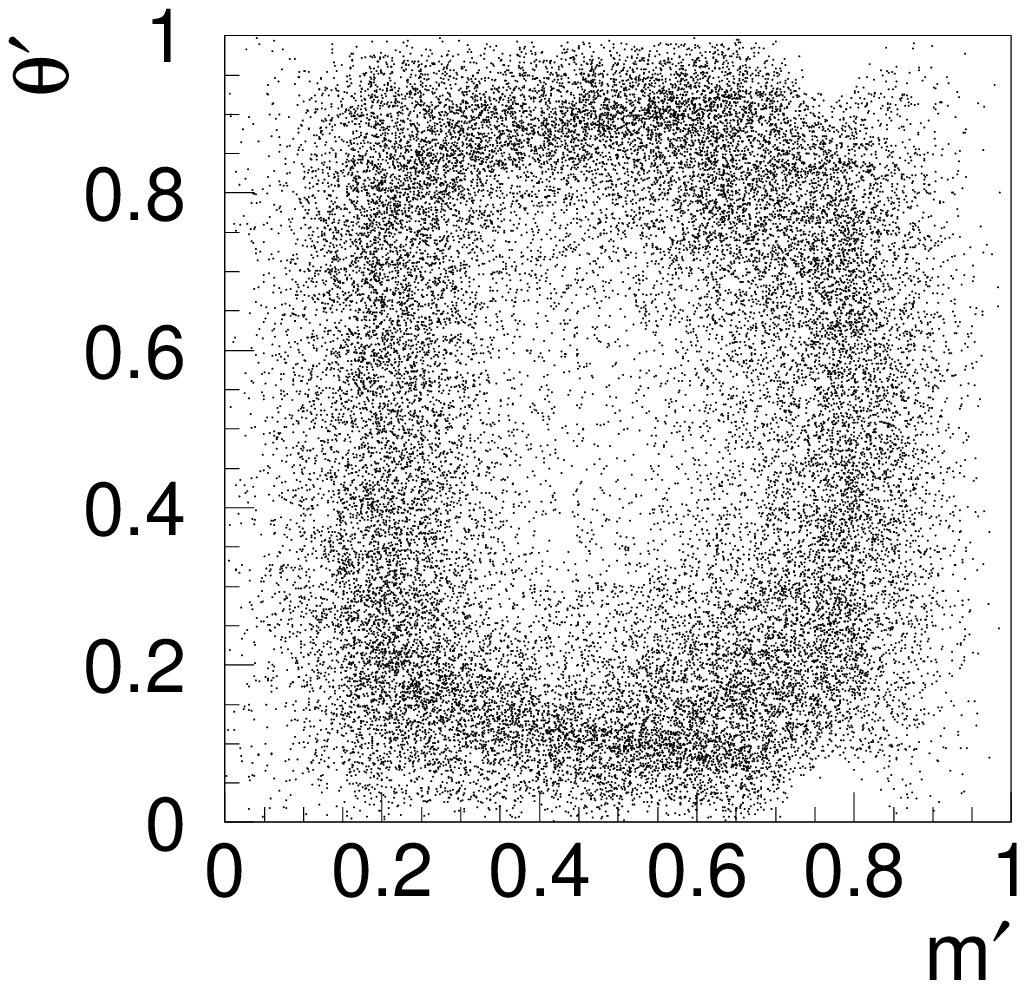}
\caption{
\label{fig:sdp_qq_distr}
Distribution of
$q\overline{q}$ background (from the data $\mbc$ sideband)
in the
usual Dalitz plot (left)
and square Dalitz plot (right).
}
\end{center}
\end{figure}

\begin{figure}[tp]
\begin{center}
\includegraphics[width=0.38\textwidth]{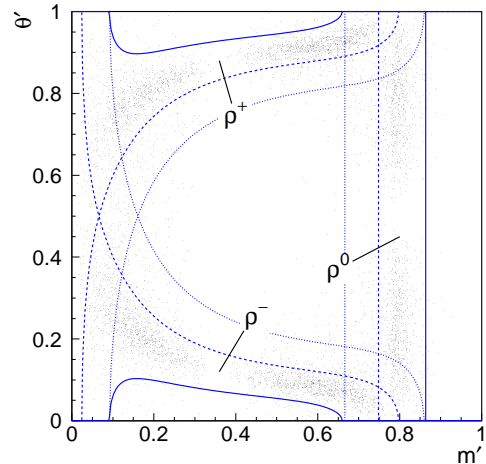}
\caption{
\label{fig:sdp_sig_distr_square_dalitz}
Distribution of signal MC (without detector efficiency and
smearing)
in the square Dalitz plot.
The solid, dashed, and dotted lines correspond to the iso-contours
of $\sqrt{s_\kappa}=0.5\,\mathrm{GeV}$, $1.0\,\mathrm{GeV}$, and
$1.5\,\mathrm{GeV}$, respectively,
for each $\rho^\kappa$ resonance.
}
\end{center}
\end{figure}

The signal and the continuum background
$e^+e^- \rightarrow q\overline{q} \: (q=u, d, s, c)$,
which is the dominant background in this analysis,
populate the kinematic boundaries of the usual Dalitz plot
as shown in Figs.~\ref{fig:sdp_sig_distr_usual_dalitz} and
\ref{fig:sdp_qq_distr} (left).
Since we model part of the Dalitz plot probability density function (PDF)
with a binned histogram,
the part of the distribution that is concentrated in a narrow region
near the edge of the usual Dalitz plot
is not easy to treat.
We therefore apply the transformation
\begin{equation}
ds_+ ds_- \rightarrow |\det \boldsymbol{J}| dm' d\theta' \;,
\end{equation}
which defines the square Dalitz plot (SDP)~\cite{Aubert:2005sk}.
The new coordinates are
\begin{equation}
\label{equ:square_dalitz_mprim_definition}
m' \equiv \frac{1}{\pi} \arccos
\left(
2 \frac{m_0 - m_0^\mathrm{min}}{m_0^\mathrm{max} - m_0^\mathrm{min}}
- 1
\right) \;,
\end{equation}
\begin{equation}
\theta' \equiv \frac{1}{\pi} \theta_0
\quad
\left( = \frac{1}{\pi} \theta^{-0} \right) \;,
\end{equation}
where $m_0 = \sqrt{s_0}$,
$m_0^\mathrm{max} = m_{B^0} - m_{\pi^0}$
and  $m_0^\mathrm{min} = 2m_{\pi^+}$
are the kinematic limits of $m_0$,
and $\boldsymbol{J}$ is the Jacobian of the transformation.
The determinant of the Jacobian is given by
\begin{equation}
\begin{split}
& |\det \boldsymbol{J}| \\
& \; = 4|\vec{p}_+| |\vec{p}_0| m_0
\cdot \frac{m_0^\mathrm{max} - m_0^\mathrm{min}}{2} \pi \sin (\pi m')
\cdot \pi \sin(\pi \theta') \;,
\end{split}
\end{equation}
where $\vec{p}_+$ and $\vec{p}_0$
are the three momenta of $\pi^+$ and $\pi^0$
in the $\pi^+\pi^-$ rest frame.
Figures~\ref{fig:sdp_sig_distr_square_dalitz} and
\ref{fig:sdp_qq_distr} (right)
show the distributions of the signal and
continuum events, respectively, in the square Dalitz plot.


\section{Event Selection and Reconstruction
\label{sec:selection_and_reconstruction}
}
To reconstruct candidate $\bz\to\pip\pim\piz$ decays,
charged tracks reconstructed with the CDC and SVD are required to
originate from the interaction point (IP) and to have
transverse momenta greater than 0.1 GeV/$c$.
Using kaon identification (KID) information,
we distinguish charged kaons from pions based on
a kaon (pion) likelihood $\mathcal{L}_{K(\pi)}$
derived from the TOF, ACC and $dE/dx$ measurements in the CDC.
Tracks that are positively identified as electrons are rejected.

Photons are identified as isolated ECL clusters
that are not matched to any charged track.
We reconstruct $\piz$ candidates from pairs of photons
detected in the barrel (end-cap) ECL with $E_\gamma > 0.05$ (0.1) GeV,
where $E_\gamma$ is the photon energy measured with the ECL.
Photon pairs with momenta greater than 0.1 GeV/$c$ in the
laboratory frame and
with an invariant mass between 0.1178 GeV/$c^2$ and 0.1502 GeV/$c^2$,
roughly corresponding to $\pm 3\sigma$ in the mass resolution,
are used as $\piz$ candidates.

We identify $B$ meson decays
using the energy difference $\dE\equiv E_B^{\rm cms}-E_{\rm beam}^{\rm cms}$ and
the beam-energy-constrained mass $\mb\equiv\sqrt{(E_{\rm beam}^{\rm cms})^2-
(\pbstar)^2}$, where $E_{\rm beam}^{\rm cms}$ is
the beam energy in the cms, and
$E_B^{\rm cms}$ and $\pbstar$ are the cms energy and momentum,
respectively, of the
reconstructed $B$ candidate.

We select candidates in a fit region defined as
$-0.2\,\mathrm{GeV} < \dE < 0.2\,\mathrm{GeV}$
and $5.2\,\mathrm{GeV}/c^2 < \mb < 5.3\,\mathrm{GeV}/c^2$.
The fit region consists of a signal region
defined as $-0.1\,\mathrm{GeV} < \dE < 0.08\,\mathrm{GeV}$
and $\mb >5.27\,\mathrm{GeV}/c^2$,
and its complement, called the sideband region, which is
dominated by background events.

The vertex position for the $\bz\to\pip\pim\piz$ decay
is reconstructed using charged tracks that have
enough SVD hits~\cite{Tajima:2003bu}.
The $\ftag$ vertex is obtained with well-reconstructed
tracks that are not assigned to $\fCP$.
A constraint on the interaction-region profile
in the plane perpendicular to the beam axis
is also used with the selected tracks.

The $b$-flavor of the accompanying $B$ meson is identified
from inclusive properties of particles
that are not associated with the reconstructed $\bz \to \fCP$
decay.
We use two parameters, the $b$-flavor charge $q_\mathrm{tag}$ and $r$,
to represent the tagging information~\cite{Kakuno:2004cf}.
The parameter $r$ is an event-by-event,
Monte Carlo (MC) determined flavor-tagging dilution factor
that ranges from $r=0$ for no flavor
discrimination to $r=1$ for unambiguous flavor assignment.
It is used only to sort data into six $r$ intervals.
The wrong tag fractions for the six $r$ intervals,
$w_l~(l=1, 2, \cdots, 6)$, and the differences
between $\bz$ and $\bzb$ decays, $\dwl$,
are determined
using a high-statistics control sample of semileptonic and
hadronic $b\to c$ decays~\cite{Kakuno:2004cf,Abe:2004mz,Chen:2005dr}.

The dominant background for the $\bz \to \pi^+\pi^-\pi^0$ signal is from
continuum.
To distinguish these jet-like events from the spherical $B$ signal
events, we combine a set of variables that characterize the event topology
into a signal (background) likelihood variable $\cal L_{\rm sig\:(bkg)}$,
and impose requirements on the likelihood ratio
$\rsigbkg \equiv \lsig/(\lsig+\lbkg)$.
Due to a correlation between $\mathcal{R}$ and $r$,
these requirements depend on the quality of flavor tagging.

When more than one candidate in the same event is found
in the fit region, we select the best candidate
based on the reconstructed $\piz$ mass and $\rsigbkg$.
About 30\% of the signal events have multiple candidates.

After the best candidate selection, we reconstruct
the Dalitz variables $s_+$, $s_0$ and $s_-$
from 1) the four momenta of the $\pi^+$ and $\pi^-$,
2) the helicity angle of the $\rho^0$
(i.e., the helicity angle of the $\pi^+ \pi^-$ system),
and 3) the relation of Eq.~(\ref{equ:dalitz_var_add_relation}).
Note that the energy of the $\pi^0$ is not explicitly used here,
which improves the resolution of the Dalitz plot variables.
We reject candidates that are located in
one of the following regions in the Dalitz plot:
$\sqrt{s_0} > 0.95$ GeV/$c^2$ and $\sqrt{s_+}>1.0$ GeV$/c^2$ and $\sqrt{s_-} > 1.0$ GeV/$c^2$;
$\sqrt{s_0} < 0.55$ GeV/$c^2$ or $\sqrt{s_+}<0.55$ GeV$/c^2$ or
$\sqrt{s_-} < 0.55$ GeV/$c^2$ (see Fig.~\ref{fig:sdp_sig_distr_usual_dalitz}).
In these regions, the fraction of $\bz\to\rho\pi$ signal is small.
Moreover,
radial excitations
(the $\rho(1450)$ and $\rho(1700)$)
are the dominant contributions
to $\bz \rightarrow \pi^+\pi^-\pi^0$
in the region with $\sqrt{s} > 1.0$ GeV$/c^2$,
where $s$ is either $s_+$, $s_-$, or $s_0$.
Since the amplitudes
of the radial excitations
are in general independent of
the amplitude of the $\rho(770)$,
they are considered to be background in our analysis;
vetoing the high mass region considerably reduces
the systematic uncertainties due to their contributions.

\begin{figure}[hb]
\includegraphics[width=0.47\textwidth]{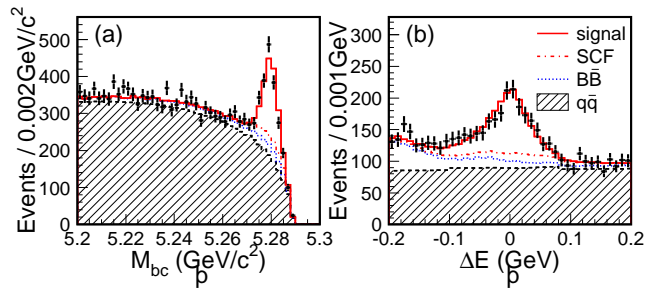}
\caption{The $\mbc$ (a) and $\dE$ (b) distributions
within the $\dE$ and $\mbc$ signal regions.
The histograms are cumulative. Solid, dot-dashed, dotted and dashed
hatched histograms correspond to correctly reconstructed signal,
SCF, $B\bbar$, and continuum PDFs, respectively.
}
\label{fig:mbc_and_de_plots}
\end{figure}

Figure \ref{fig:mbc_and_de_plots} shows the $\mbc$
and $\dE$ distributions for the reconstructed
$B^0 \rightarrow \pi^+\pi^-\pi^0$ candidates
within the $\dE$ and $\mbc$ signal regions, respectively.
The signal yield is determined from an unbinned four-dimensional
extended-maximum-likelihood fit to the $\dE$-$\mbc$
and Dalitz plot distribution in the fit region
defined above;
the Dalitz plot distribution is only used for the events
inside the $\dE$-$\mbc$ signal region.
The $\dE$-$\mbc$ distribution of signal is modeled with binned histograms
obtained from MC,
where the correlation between $\dE$ and $\mbc$,
the dependence on $p_{\pi^0}$,
and the difference between data and MC are taken into account.
We also take into account incorrectly reconstructed signal events,
which we call self-cross-feed (SCF)
and constitutes $\sim 20\%$ of the signal.
In a SCF event, either one of the three pions in $\fCP$ is swapped
with a pion in $f_\mathrm{tag}$, or else the $\pi^0$ in $\fCP$ is
misreconstructed.
We give the details of the $\dE$-$\mb$ and Dalitz plot PDF's of the SCF
component in appendix \ref{sec:appendix_pdf_definitions}.
For continuum,
we use the ARGUS parameterization~\cite{Albrecht:1990am}
for $\mbc$
and a linear function for $\dE$.
The $\dE$-$\mbc$ distribution of $B\overline{B}$ background
is modeled by binned histograms based on MC.
The Dalitz plot distributions for all components
are modeled in the same way as the time-dependent fit
described later,
but integrated over the proper time difference,
$\Delta t$, and summed over the flavor of the tag side $B$, $q_\mathrm{tag}$.
The fit yields $971 \pm 42$
$B^0 \rightarrow \pi^+\pi^-\pi^0$ events in the signal region,
where the error is statistical only.


\section{Determination of the contributions from radial excitations
\label{sec:lineshape_determine}
}
\begin{figure*}[t]
\includegraphics[width=0.288\textwidth]{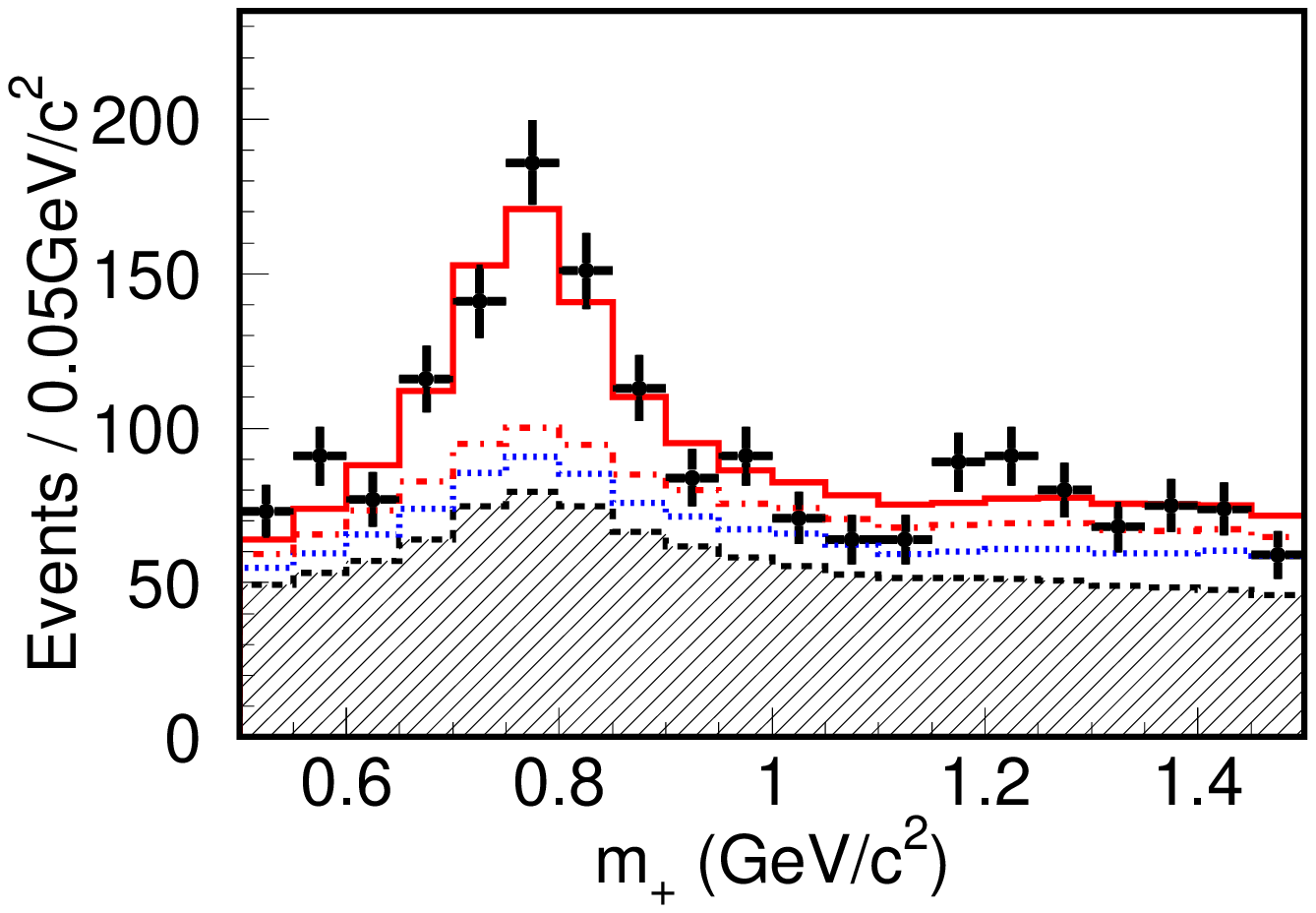}
\includegraphics[width=0.288\textwidth]{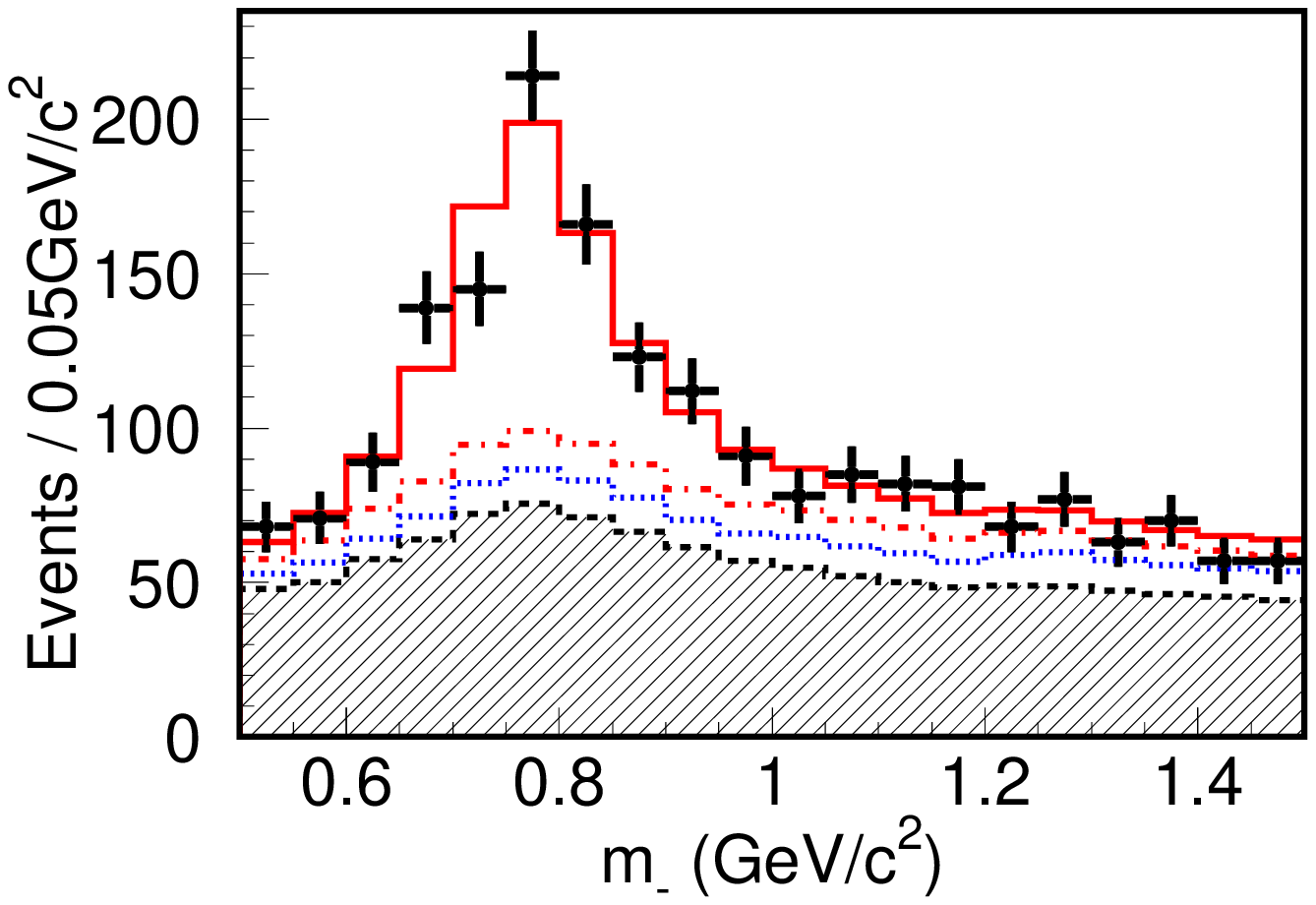}
\includegraphics[width=0.288\textwidth]{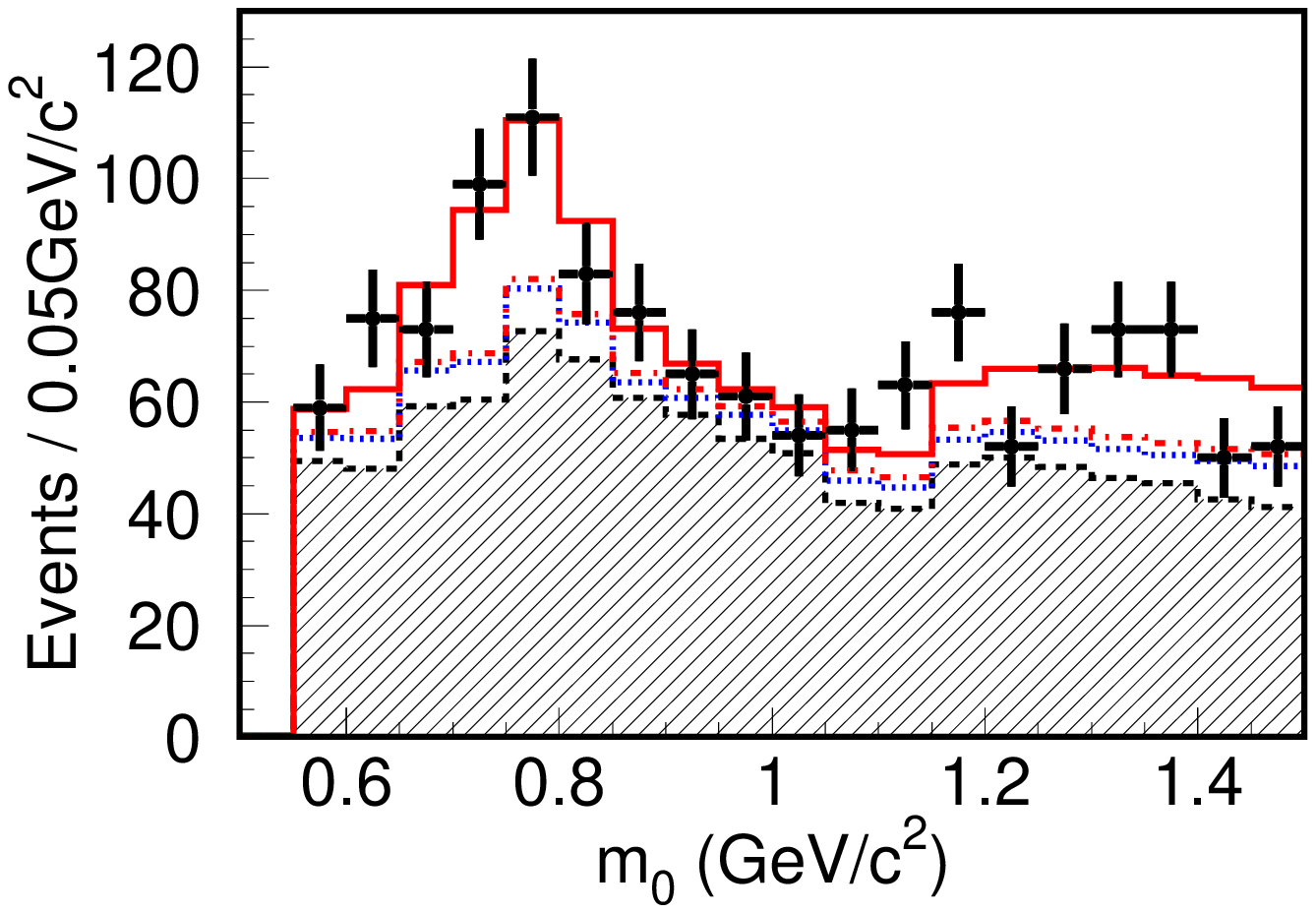}
\caption{
Mass distributions and fitted lineshapes
in $\rho^+\pi^-$ (left), $\rho^-\pi^+$ (middle),
and $\rho^0\pi^0$ (right) enhanced regions.
The histograms are cumulative. Solid, dot-dashed, dotted and dashed
hatched histograms correspond to correctly reconstructed signal,
SCF, $B\bbar$, and continuum PDFs, respectively.
Note
that there are feed-downs from
other quasi-two-body components than
those of interest,
especially in the high mass regions.
For example,
the
high mass region ($m_0 \gtrsim 1.0\, \mathrm{GeV}/c^2$)
of the $\rho^0\pi^0$ enhanced region (right)
includes large contributions from
$\rho^\pm \pi^\mp$.
}
\label{fig:mass_plot_upto_high_mass}
\end{figure*}

Although the contributions from radial excitations are suppressed by the
selections in the Dalitz plot described in the previous section,
there are still significant contributions from the long tails of the
radial
excitations and their interferences. We thus need to determine the
sizes of the radial excitations and their uncertainties to properly
model the signal PDF's and
systematic uncertainties associated with their degrees of freedom.

\begin{figure}[b]
\includegraphics[width=0.5\textwidth]{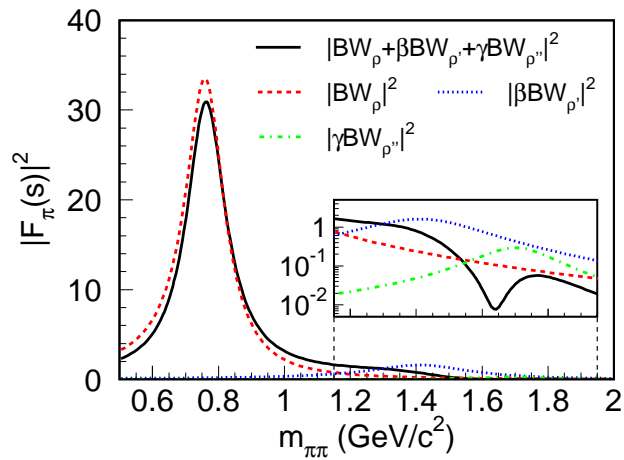}
\caption{
A schematic figure of the fit result of the lineshape
and the contributions from radial excitations.
Note that
our definition of
$F_{\pi}(s)$ does not include the factor $1/(1+\beta+\gamma)$
as in Eq.~(\ref{equ:fpi_lineshape_beta_gamma}).
The inset shows the high mass region,
$m_{\pi\pi}>1.15\,\mathrm{GeV}/c^2$,
on a semi-log scale where
the interference between the $\rho(770)$ and radial
excitations is visible.
One can see that the $\rho(770)$ and $\rho(1450)$
destructively interfere with each other
near $\sqrt{s} \equiv m_{\pi\pi} = 1.4\,\mathrm{GeV}/c^2$,
which means that
the $\rho(1450)$ has a large impact on
the phase of $F_{\pi}(s)$
although the absolute value of $|F_{\pi}(s)|$ is not much affected.
}
\label{fig:lineshape_fitted_schematic}
\end{figure}

Using the same data sample as described above
but performing a time-integrated Dalitz plot fit
with a wider Dalitz plot acceptance,
$0.55\,\mathrm{GeV}/c^2 < \sqrt{s_0} < 1.5\,\mathrm{GeV}/c^2$
or $\sqrt{s_+}<1.5\,\mathrm{GeV}/c^2$
or $\sqrt{s_-} < 1.5\,\mathrm{GeV}/c^2$,
we determine the $\rho$ lineshape,
i.e., the phases and amplitudes of the coefficients $\beta$
and $\gamma$ in Eq.~(\ref{equ:fpi_lineshape_beta_gamma}).
We use these for all of the decay amplitudes.
In this fit, we use the PDG values~\cite{Eidelman:2004wy}
for the masses and widths of the $\rho(1450)$ and $\rho(1700)$.
The fit yields
\begin{equation}
\label{equ:nominal_beta_gamma}
\begin{split}
& |\beta| = 0.31^{+0.07}_{-0.06}\;, \quad
\arg \beta = \left(219^{+16}_{-18}\right)^\circ\;, \\
& |\gamma| = 0.08^{+0.04}_{-0.03} \;, \quad
\arg \gamma = \left(102^{+26}_{-32}\right)^\circ \; .
\end{split}
\end{equation}
The mass distributions and fit results are shown in
Fig.~\ref{fig:mass_plot_upto_high_mass}.
Figure~\ref{fig:lineshape_fitted_schematic} schematically shows
how the radial excitations
contribute to our fit result.
Note that the above values are quantities used for time-dependent Dalitz
fit and we do not regard them as our measurements of $\beta$ and $\gamma$.
This is because these parameters are determined from
the region where
$\rho^+\pi^-$ and $\rho^-\pi^+$ modes, etc. interfere,
and they depend on the unfounded common lineshape assumption
of
Eq.~(\ref{equ:f_kappa_definition_with_unique_lineshape_assumption});
hence we do not give their systematic errors.
Because statistics are low,
we cannot determine $\beta$ and $\gamma$ for each decay mode without
imposing the common
lineshape assumption.
However, we include the effect of possible decay-mode dependent
differences in the values of $\beta$ and $\gamma$ in the systematic
errors, which are described in Sec.~6-1.

Thus, it is important to
determine the common or {\it average} lineshape as well as
to obtain an upper limit on the deviation
from the average lineshape for each of
the six decay amplitudes.
For this purpose,
we put constraints on
additional amplitudes that describe
1) the excess in the high mass region, $\sqrt{s} >
0.9\,\mathrm{GeV}/c^2$,
where $s$ is either $s_+$, $s_-$, or $s_0$;
and 2) interferences between radial excitations
and the lowest resonance $\rho(770)$
(e.g., interferences between $\rho(770)^+\pi^-$
and $\rho(1450)^-\pi^+$, etc.).
The nominal fit is performed
with the average lineshape determined above,
fixing all of the additional amplitudes to zero.
When floating the additional amplitudes for the other resonances,
we obtain results consistent with zero for all of the additional amplitudes
but with large uncertainties compared to the
errors for the average lineshape parameters above.
We use the fit result with the additional lineshape parameters floated
including their uncertainties in the systematic error study.


\section{Time-dependent Dalitz plot analysis
\label{sec:time_depend_analysis_pdf}
}
To determine the 26 Dalitz plot parameters,
we define the following event-by-event PDF:
\begin{equation}
\label{equ:dembc_dalitz_simultaneous-total_pdf}
P(\vec{x})
\equiv f_\mathrm{sig} \mathcal{P}_\mathrm{sig}(\vec{x})
+ f_{\BB} \mathcal{P}_{\BB}(\vec{x})
+ f_{\qq} \mathcal{P}_{\qq}(\vec{x})
\;,
\end{equation}
where
$\mathcal{P}_\mathrm{sig}$, $\mathcal{P}_{\BB}$ and $\mathcal{P}_{\qq}$
are PDF's
for signal, $\BB$ background and continuum background, respectively,
and
$f_\mathrm{sig}$, $f_{\BB}$ and $f_{\qq}$ are
the corresponding fractions that satisfy
\begin{equation}
f_\mathrm{sig} + f_{\BB} + f_{\qq} = 1 \;.
\end{equation}
The vector $\vec{x}$,
the arguments of the PDF's,
corresponds to a set of event-by-event variables:
\begin{equation}
\vec{x} \equiv
(\Delta E, \mbc; m', \theta';\Delta t, q_\mathrm{tag}, l; p_{\pi^0}) \;.
\end{equation}
A detailed description of the PDF can be found
in Appendix \ref{sec:appendix_pdf_definitions}.

With the PDF defined above,
we form the likelihood function
\begin{equation}
\mathcal{L} \equiv \prod_i P(\vec{x}_i) \;,
\end{equation}
where $i$ is an index over events.
We maximize $\mathcal{L}$ to
determine the 26 Dalitz plot parameters
using the likelihood function
with the signal fraction and the lineshape parameters
obtained in
Sec.~\ref{sec:selection_and_reconstruction}
and Sec.~\ref{sec:lineshape_determine}, respectively.


\section{Fit result}
An unbinned maximum likelihood fit to the 2824 events in the signal
region yields
the result listed in Table \ref{tbl:dt_all_data}.
The correlation matrix for the 26 parameters,
after combining statistical and systematic errors,
is shown in appendix~\ref{sec:appendix_correlation}.
Figure \ref{fig:dalitz_plot}
shows the projections of the square Dalitz plot in data
with the fit result superimposed.
We also show the mass and helicity distribution
for each $\rho \pi$ enhanced region
along with projections of the fit (Fig.~\ref{fig:mass_helicity_plot}).
We find that $U^+_0$ is $4.8\,\sigma$ above zero, corresponding to clear
evidence for the presence of the decay $\bz \to \rho^0\pi^0$ in
agreement with our previous measurement~\cite{Dragic:2006yv}
(see Sec.~\ref{sec:branch}).
Figure \ref{fig:dt_plot_result}
shows the $\Delta t$ distributions and background-subtracted
asymmetries.
We define the asymmetry in each $\Delta t$ bin
by
$(N_+ - N_-) / (N_+ + N_-)$,
where $N_{+ \, (-)}$ corresponds to the
background-subtracted number of events with $q_\mathrm{tag}=+1\,(-1)$.
The $\rho^- \pi^+$ enhanced region shows a significant cosine-like
asymmetry, arising from a nonzero value of $U^-_-$. Note that this is
not a $CP$-violating effect, since $\rho^- \pi^+$ is not a $CP$
eigenstate. No sinelike asymmetry is observed in any of the
$\rho^+\pi^-$, $\rho^-\pi^+$ or $\rho^0\pi^0$ enhanced regions.

As a check of our fit, we perform the time-dependent Dalitz
plot fit with the $\bz$ lifetime floated as a free parameter. We obtain
$1.41 \pm 0.07\,\mathrm{ps}$ for the lifetime, where the error is
statistical only, while the changes of the other parameters
are very small compared to their statistical errors.
The lifetime we obtain is consistent with world
average~\cite{Eidelman:2004wy} and thus validates our understanding of
the PDF's and the background fraction.

\begin{table}[tb]
\caption{ Results of the time-dependent Dalitz fit.}
\label{tbl:dt_all_data}
\begin{tabular}
{@{\hspace{0.5cm}}l@{\hspace{0.5cm}}|@{\hspace{0.5cm}}c@{\hspace{0.5cm}}}
\hline \hline
& Fit Result                       \\
\hline
$U^+_+$                   &  $+1$ (fixed)             \\
$U^+_-$                   & $+1.27 \pm 0.13\,(\text{stat.}) \pm 0.09\,(\text{syst.})$ \\
$U^+_0$                   & $+0.29 \pm 0.05\,(\text{stat.}) \pm 0.04\,(\text{syst.})$ \\
$U^{+, \mathrm{Re}}_{+-}$ & $+0.49 \pm 0.86\,(\text{stat.}) \pm 0.52\,(\text{syst.})$ \\
$U^{+, \mathrm{Re}}_{+0}$ & $+0.29 \pm 0.50\,(\text{stat.}) \pm 0.35\,(\text{syst.})$ \\
$U^{+, \mathrm{Re}}_{-0}$ & $+0.25 \pm 0.60\,(\text{stat.}) \pm 0.33\,(\text{syst.})$ \\
$U^{+, \mathrm{Im}}_{+-}$ & $+1.18 \pm 0.86\,(\text{stat.}) \pm 0.34\,(\text{syst.})$ \\
$U^{+, \mathrm{Im}}_{+0}$ & $-0.57 \pm 0.35\,(\text{stat.}) \pm 0.51\,(\text{syst.})$ \\
$U^{+, \mathrm{Im}}_{-0}$ & $-1.34 \pm 0.60\,(\text{stat.}) \pm 0.47\,(\text{syst.})$ \\
\hline
$U^-_+$                   & $+0.23 \pm 0.15\,(\text{stat.}) \pm 0.07\,(\text{syst.})$ \\
$U^-_-$                   & $-0.62 \pm 0.16\,(\text{stat.}) \pm 0.08\,(\text{syst.})$ \\
$U^-_0$                   & $+0.15 \pm 0.11\,(\text{stat.}) \pm 0.08\,(\text{syst.})$ \\
$U^{-, \mathrm{Re}}_{+-}$ & $-1.18 \pm 1.61\,(\text{stat.}) \pm 0.72\,(\text{syst.})$ \\
$U^{-, \mathrm{Re}}_{+0}$ & $-2.37 \pm 1.36\,(\text{stat.}) \pm 0.60\,(\text{syst.})$ \\
$U^{-, \mathrm{Re}}_{-0}$ & $-0.53 \pm 1.44\,(\text{stat.}) \pm 0.65\,(\text{syst.})$ \\
$U^{-, \mathrm{Im}}_{+-}$ & $-2.32 \pm 1.74\,(\text{stat.}) \pm 0.91\,(\text{syst.})$ \\
$U^{-, \mathrm{Im}}_{+0}$ & $-0.41 \pm 1.00\,(\text{stat.}) \pm 0.47\,(\text{syst.})$ \\
$U^{-, \mathrm{Im}}_{-0}$ & $-0.02 \pm 1.31\,(\text{stat.}) \pm 0.83\,(\text{syst.})$ \\
\hline
$I_+$                     & $-0.01 \pm 0.11\,(\text{stat.}) \pm 0.04\,(\text{syst.})$ \\
$I_-$                     & $+0.09 \pm 0.10\,(\text{stat.}) \pm 0.04\,(\text{syst.})$ \\
$I_0$                     & $+0.02 \pm 0.09\,(\text{stat.}) \pm 0.05\,(\text{syst.})$ \\
$I^{\mathrm{Re}}_{+-}$    & $+1.21 \pm 2.59\,(\text{stat.}) \pm 0.98\,(\text{syst.})$ \\
$I^{\mathrm{Re}}_{+0}$    & $+1.15 \pm 2.26\,(\text{stat.}) \pm 0.92\,(\text{syst.})$ \\
$I^{\mathrm{Re}}_{-0}$    & $-0.92 \pm 1.34\,(\text{stat.}) \pm 0.80\,(\text{syst.})$ \\
$I^{\mathrm{Im}}_{+-}$    & $-1.93 \pm 2.39\,(\text{stat.}) \pm 0.89\,(\text{syst.})$ \\
$I^{\mathrm{Im}}_{+0}$    & $-0.40 \pm 1.86\,(\text{stat.}) \pm 0.85\,(\text{syst.})$ \\
$I^{\mathrm{Im}}_{-0}$    & $-2.03 \pm 1.62\,(\text{stat.}) \pm 0.81\,(\text{syst.})$ \\
\hline
\end{tabular}
\end{table}

\begin{figure}[b]
\includegraphics[width=0.47\textwidth]{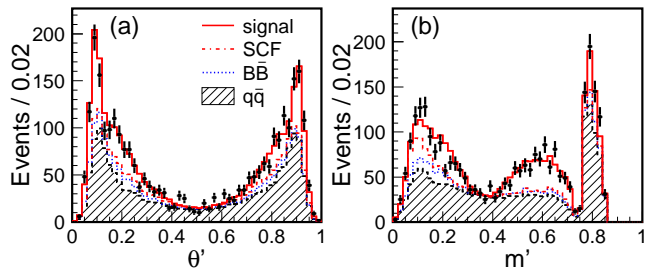}
\caption{Distributions of $\theta'$ (a) and $m'$ (b) with
fit results.
The histograms are cumulative. Solid, dot-dashed, dotted and dashed
hatched histograms correspond to correctly reconstructed signal,
SCF, $B\bbar$, and continuum PDFs, respectively.
}
\label{fig:dalitz_plot}
\end{figure}

\begin{figure*}[htb]
\includegraphics[width=0.288\textwidth]{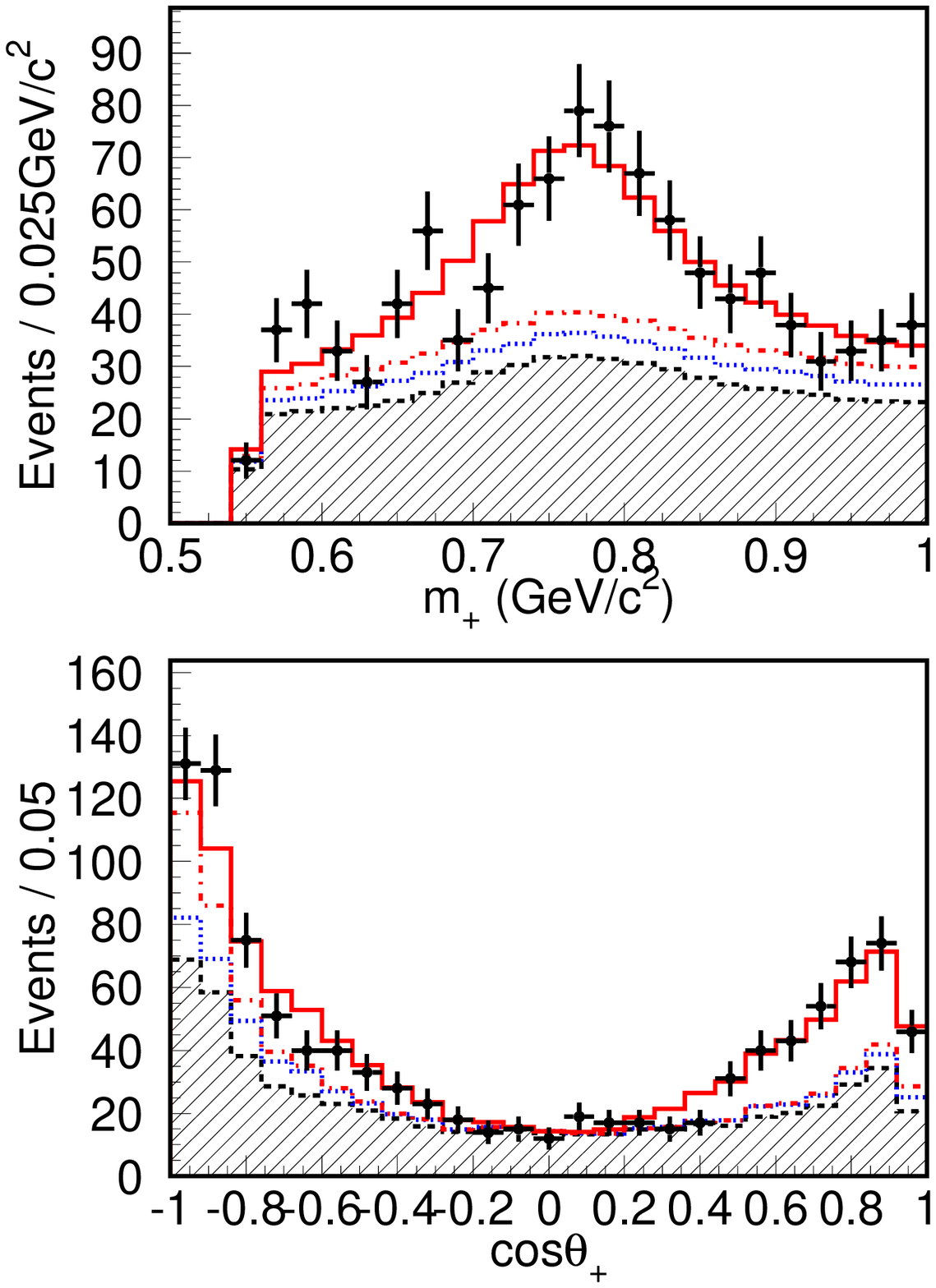}
\includegraphics[width=0.288\textwidth]{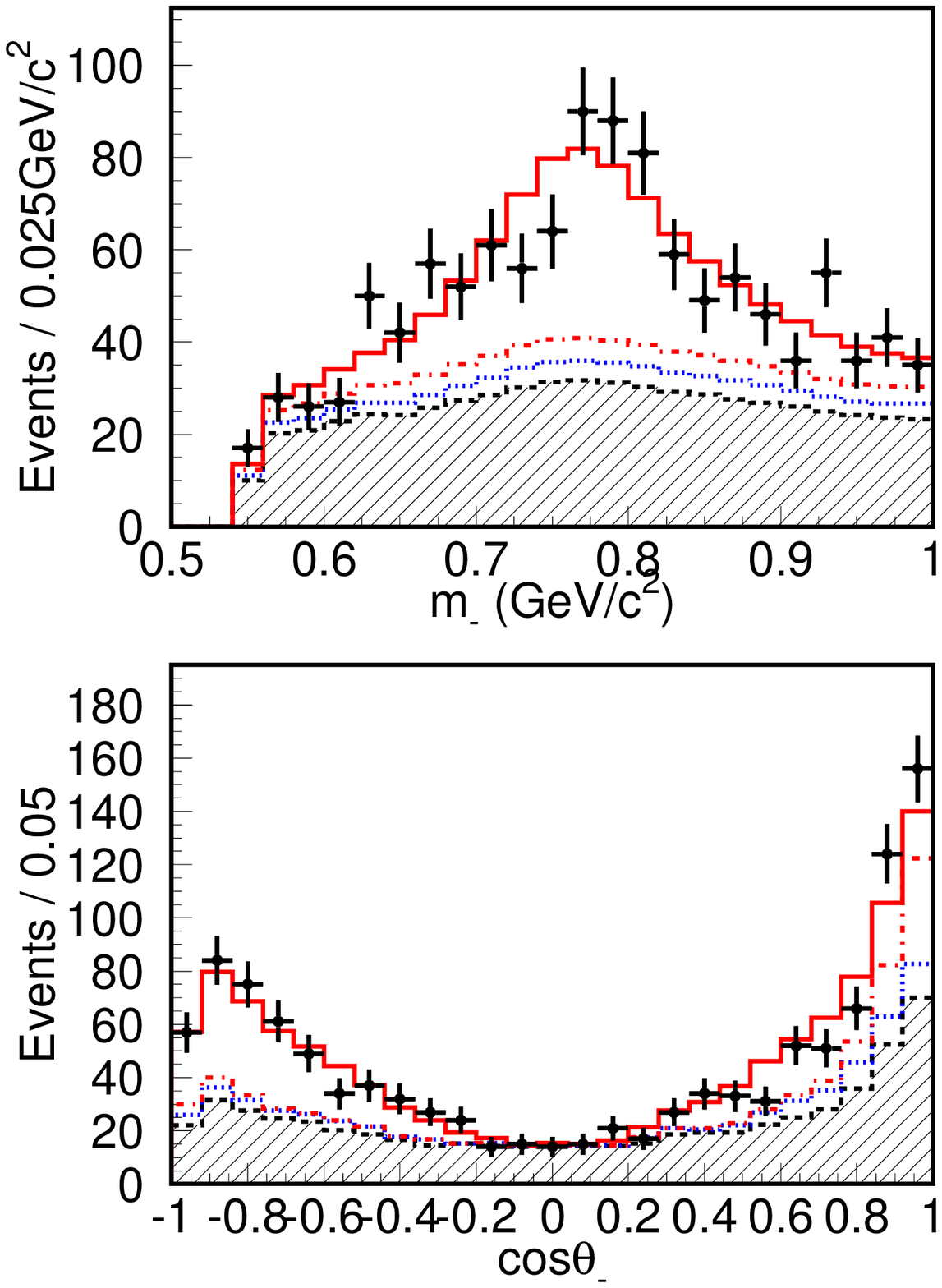}
\includegraphics[width=0.288\textwidth]{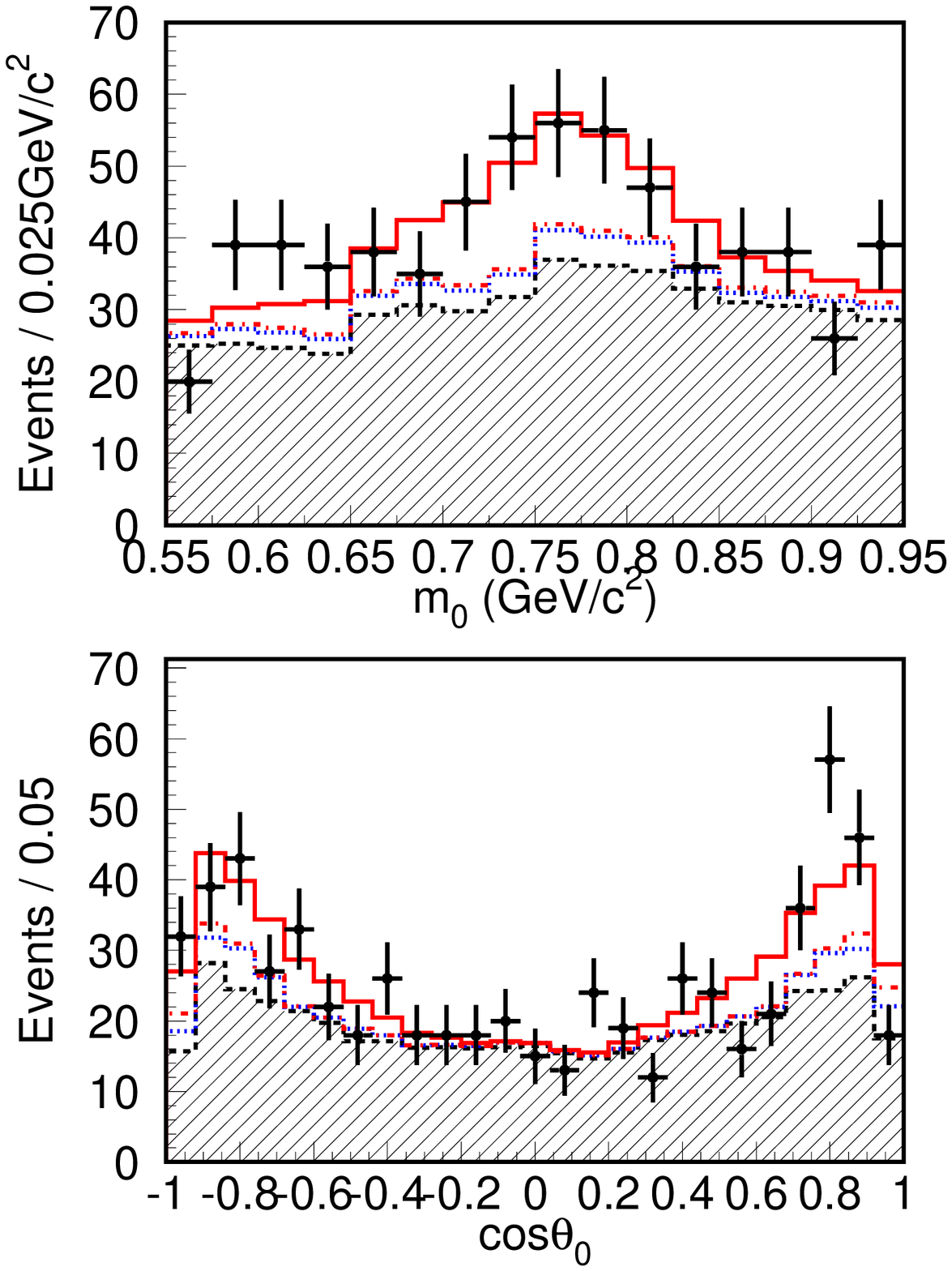}
\caption{
Mass (upper) and helicity (lower) distribution of
$\rho^+ \pi^-$ (left), $\rho^- \pi^+$ (middle),
and $\rho^0 \pi^0$ (right) enhanced regions.
The histograms are cumulative. Solid, dot-dashed, dotted and dashed
hatched histograms correspond to correctly reconstructed signal,
SCF, $B\bbar$, and continuum PDFs, respectively.
}
\label{fig:mass_helicity_plot}
\end{figure*}
\begin{figure*}[htbp]
\includegraphics[width=0.288\textwidth]{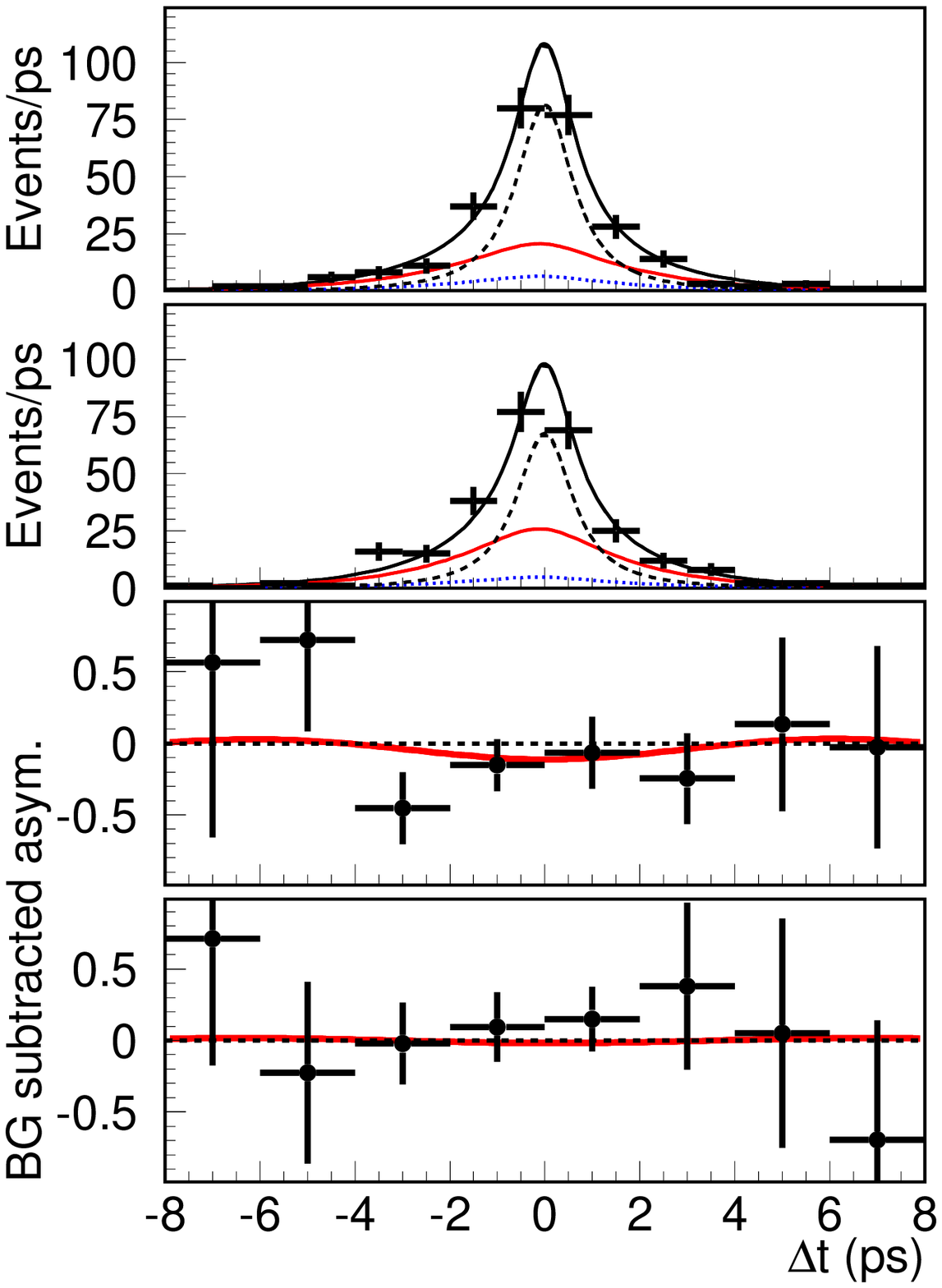}
\includegraphics[width=0.288\textwidth]{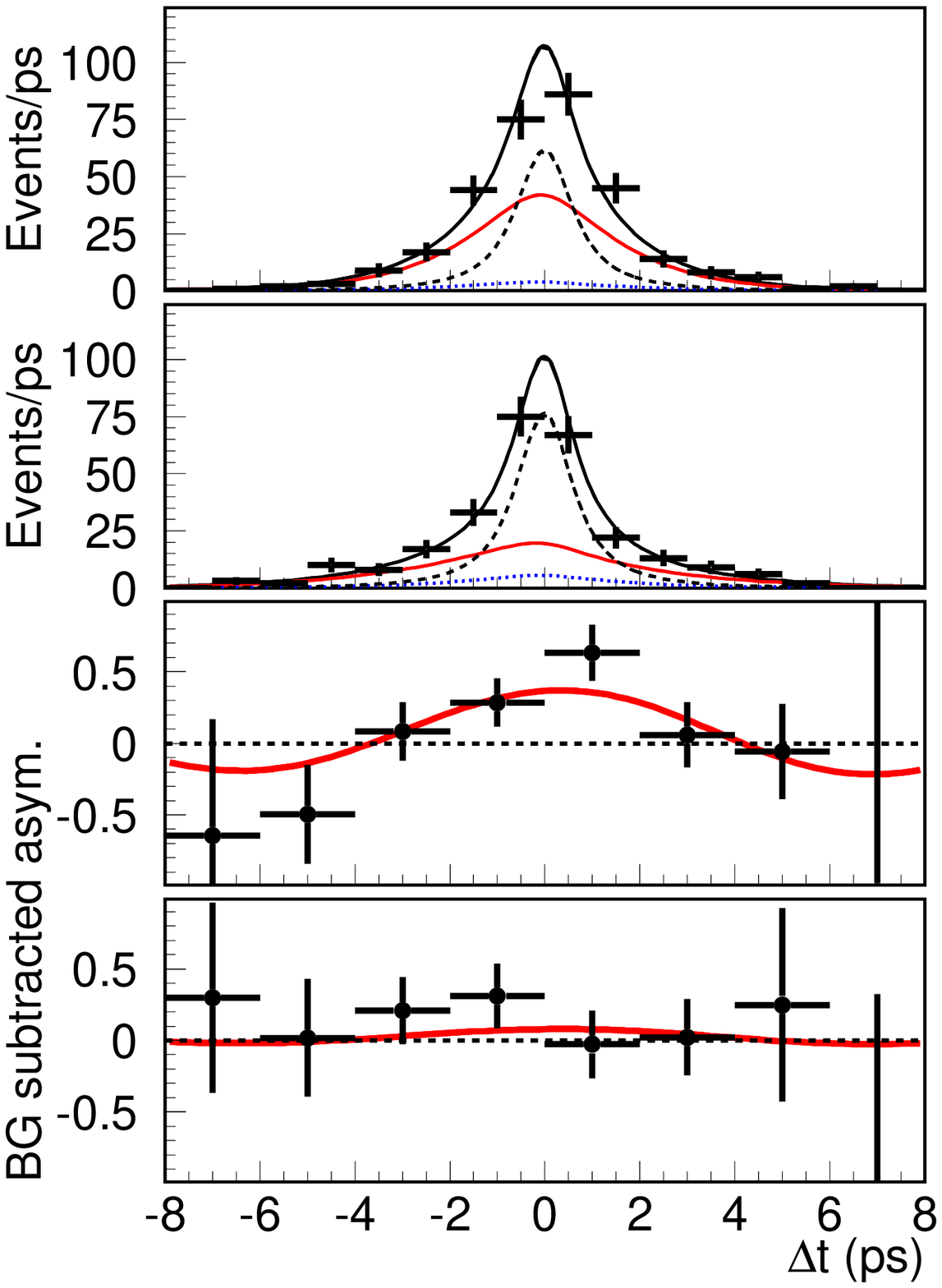}
\includegraphics[width=0.288\textwidth]{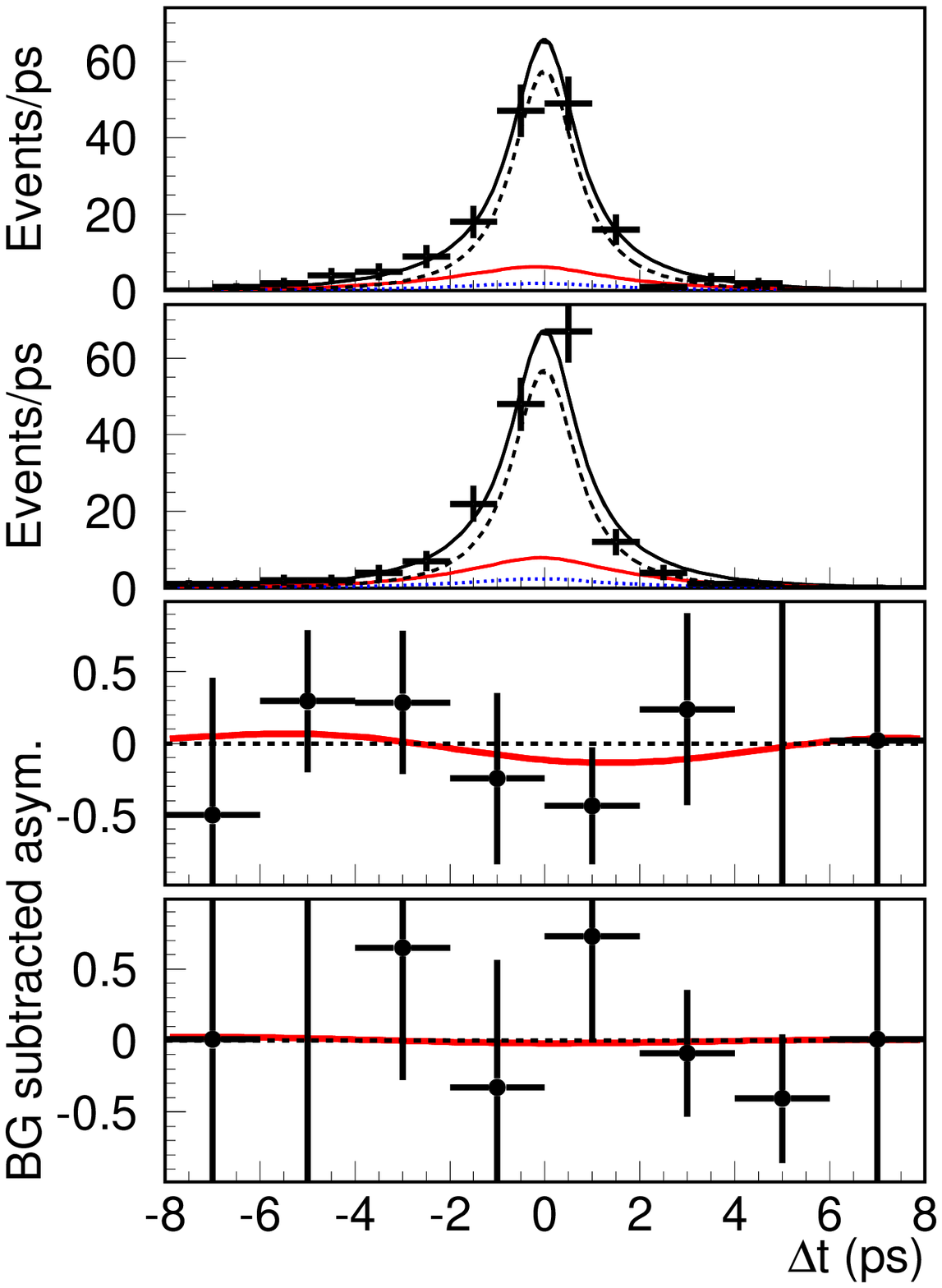}
\caption{
Proper time distributions of good tag ($r>0.5$) regions
for $f_\mathrm{tag} = \bz$ (upper) and
$f_\mathrm{tag} = \bzb$ (middle upper),
in $\rho^+\pi^-$ (left), $\rho^- \pi^+$ (middle),
$\rho^0\pi^0$ (right) enhanced regions,
where solid (red), dotted, and dashed curves
correspond to signal, continuum, and $B\bbar$ PDFs.
The middle lower and lower plots
show the background-subtracted asymmetries
in the good tag ($r>0.5$) and poor tag ($r<0.5$) regions, respectively.
The significant asymmetry in the $\rho^- \pi^+$ enhanced region
(middle)
corresponds to a non-zero value of $U^-_-$.
\label{fig:dt_plot_result}
}
\end{figure*}

\subsection{Treatment of statistical errors}
With a MC study, we check the pull distributions,
where the pull is defined
as the residual divided by the MINOS error.
Here, the MINOS error,
which corresponds to the deviation from the best fit parameter
when $-2\ln(\mathcal{L}/\mathcal{L}_\mathrm{max})$ increases by one,
is an estimate of the statistical error.
Although the pull is expected to follow
a Gaussian distribution with unit width,
we find that the width of the pull distribution
tends to be significantly larger than one for the interfering
parameters
due to small statistics.
We verify with MC that with high statistics the widths of the pull
distributions are unity.
To restore the pull width to unity,
we multiply the MINOS errors of the interfering parameters by a factor
of 1.17,
which is the average pull width for the interfering parameters
obtained above,
and quote the results as the statistical errors.
For the non-interfering terms,
we quote the MINOS errors without any correction factor.


\section{Systematic Uncertainties
\label{sec:systematic_errors}
}
Tables~\ref{tbl:systematics1}--\ref{tbl:systematics3}
list the systematic errors for the 26 time-dependent
Dalitz plot parameters.
The total systematic error is obtained by adding each
source of systematic uncertainty in quadrature.

\begin{table*}[htbp]
\caption{
Table of systematic errors (1).
The notation ``$<0.01$'' means that the value is small and less than
0.01, and thus not visible for the number of significant digits shown
here. We calculate the total systematic error including these small
contributions.
\label{tbl:systematics1}
}
\newcolumntype{R}{>{\raggedleft\arraybackslash}X}
\begin{tabularx}{150mm}{l|*{8}{R@{\hspace{4mm}}}}
\hline
\hline
& \multicolumn{1}{c}{$U^+_-$}  & \multicolumn{1}{c}{$U^+_0$}  & \multicolumn{1}{c}{$U^{+,\mathrm{Re}}_{+-}$}  & \multicolumn{1}{c}{$U^{+,\mathrm{Re}}_{+0}$}  & \multicolumn{1}{c}{$U^{+,\mathrm{Re}}_{-0}$}  & \multicolumn{1}{c}{$U^{+,\mathrm{Im}}_{+-}$}  & \multicolumn{1}{c}{$U^{+,\mathrm{Im}}_{+0}$}  & \multicolumn{1}{c}{$U^{+,\mathrm{Im}}_{-0}$}  \\
\hline
$\rho'$ and $\rho''$  & 0.01 & 0.01 & 0.31 & 0.19 & 0.19 & 0.21 & 0.39 & 0.30  \\
SCF  & 0.01 & 0.02 & 0.31 & 0.09 & 0.11 & 0.11 & 0.12 & 0.11  \\
Signal Dalitz  & 0.06 & 0.01 & 0.15 & 0.20 & 0.18 & 0.13 & 0.10 & 0.10  \\
BG Dalitz  & 0.02 & 0.01 & 0.17 & 0.11 & 0.11 & 0.14 & 0.10 & 0.19  \\
Other $\pi\pi\pi$  & 0.04 & 0.02 & 0.06 & 0.08 & 0.07 & 0.10 & 0.09 & 0.07  \\
BG fraction  & 0.02 & 0.01 & 0.08 & 0.04 & 0.07 & 0.06 & 0.03 & 0.11  \\
Physics  & 0.02 & $<0.01$ & 0.01 & 0.01 & 0.02 & 0.01 & 0.01 & 0.01  \\
BG $\Delta t$  & $<0.01$ & $<0.01$ & 0.03 & 0.01 & 0.01 & 0.02 & 0.01 & 0.01  \\
Vertexing  & 0.03 & 0.01 & 0.03 & 0.05 & 0.02 & 0.09 & 0.05 & 0.07  \\
Resolution  & $<0.01$ & $<0.01$ & 0.03 & 0.05 & 0.02 & 0.02 & 0.02 & 0.03  \\
Flavor tagging  & $<0.01$ & $<0.01$ & $<0.01$ & $<0.01$ & $<0.01$ & $<0.01$ & $<0.01$ & 0.01  \\
Fit bias  & 0.02 & 0.02 & 0.10 & 0.11 & 0.07 & 0.06 & 0.24 & 0.22  \\
TSI  & $<0.01$ & $<0.01$ & 0.01 & 0.02 & 0.02 & 0.02 & 0.01 & $<0.01$  \\
\hline
Total  & 0.09 & 0.04 & 0.52 & 0.35 & 0.33 & 0.34 & 0.51 & 0.47  \\
\hline
\end{tabularx}
\end{table*}
\begin{table*}[htbp]
\caption{
Table of systematic errors (2).
The notation ``$<0.01$'' means that the value is small and less than
0.01, and thus not visible for the number of significant digits shown
here. We calculate the total systematic error including these small
contributions.
\label{tbl:systematics2}
}
\newcolumntype{R}{>{\raggedleft\arraybackslash}X}
\begin{tabularx}{164mm}{l|*{9}{R@{\hspace{4mm}}}}
\hline
\hline
& \multicolumn{1}{c}{$U^-_+$}  & \multicolumn{1}{c}{$U^-_-$}  & \multicolumn{1}{c}{$U^-_0$}  & \multicolumn{1}{c}{$U^{-,\mathrm{Re}}_{+-}$}  & \multicolumn{1}{c}{$U^{-,\mathrm{Re}}_{+0}$}  & \multicolumn{1}{c}{$U^{-,\mathrm{Re}}_{-0}$}  & \multicolumn{1}{c}{$U^{-,\mathrm{Im}}_{+-}$}  & \multicolumn{1}{c}{$U^{-,\mathrm{Im}}_{+0}$}  & \multicolumn{1}{c}{$U^{-,\mathrm{Im}}_{-0}$}  \\
\hline
$\rho'$ and $\rho''$  & 0.01 & 0.02 & 0.04 & 0.53 & 0.29 & 0.42 & 0.70 & 0.31 & 0.59  \\
SCF  & 0.02 & 0.02 & 0.02 & 0.09 & 0.17 & 0.17 & 0.13 & 0.09 & 0.18  \\
Signal Dalitz  & 0.01 & 0.02 & 0.01 & 0.27 & 0.20 & 0.14 & 0.30 & 0.15 & 0.19  \\
BG Dalitz  & 0.04 & 0.03 & 0.02 & 0.28 & 0.32 & 0.22 & 0.30 & 0.20 & 0.30  \\
Other $\pi\pi\pi$ & 0.03 & 0.03 & 0.02 & 0.07 & 0.08 & 0.12 & 0.13 & 0.08 & 0.08  \\
BG fraction  & 0.02 & 0.04 & 0.01 & 0.18 & 0.17 & 0.14 & 0.22 & 0.13 & 0.11  \\
Physics  & 0.01 & 0.01 & $<\!0.01$ & 0.03 & 0.03 & 0.03 & 0.04 & 0.01 & 0.04  \\
BG $\Delta t$  & $<\!0.01$ & $<\!0.01$ & $<\!0.01$ & 0.02 & 0.03 & 0.02 & 0.03 & 0.02 & 0.03  \\
Vertexing  & 0.02 & 0.01 & 0.05 & 0.18 & 0.20 & 0.17 & 0.08 & 0.07 & 0.11  \\
Resolution  & 0.01 & 0.01 & $<\!0.01$ & 0.10 & 0.14 & 0.28 & 0.07 & 0.11 & 0.26  \\
Flavor tagging  & 0.01 & 0.01 & $<\!0.01$ & 0.03 & 0.03 & 0.03 & 0.05 & 0.03 & 0.02  \\
Fit bias  & $<\!0.01$ & 0.02 & $<\!0.01$ & 0.03 & 0.09 & 0.02 & 0.27 & 0.08 & 0.26  \\
TSI  & 0.03 & 0.03 & 0.01 & 0.06 & 0.03 & 0.01 & 0.05 & 0.04 & 0.02  \\
\hline
Total  & 0.07 & 0.08 & 0.08 & 0.72 & 0.60 & 0.65 & 0.91 & 0.47 & 0.83  \\
\hline
\end{tabularx}
\end{table*}
\begin{table*}[htbp]
\caption{
Table of systematic errors (3).
The notation ``$<0.01$'' means that the value is small and less than
0.01, and thus not visible for the number of significant digits shown
here. We calculate the total systematic error including these small
contributions.
\label{tbl:systematics3}
}
\newcolumntype{R}{>{\raggedleft\arraybackslash}X}
\begin{tabularx}{164mm}{l|*{9}{R@{\hspace{4mm}}}}
\hline
\hline
& \multicolumn{1}{c}{$I_+$}  & \multicolumn{1}{c}{$I_-$}  & \multicolumn{1}{c}{$I_0$}  & \multicolumn{1}{c}{$I^{\mathrm{Re}}_{+-}$}  & \multicolumn{1}{c}{$I^{\mathrm{Re}}_{+0}$}  & \multicolumn{1}{c}{$I^{\mathrm{Re}}_{-0}$}  & \multicolumn{1}{c}{$I^{\mathrm{Im}}_{+-}$}  & \multicolumn{1}{c}{$I^{\mathrm{Im}}_{+0}$}  & \multicolumn{1}{c}{$I^{\mathrm{Im}}_{-0}$}  \\
\hline
$\rho'$ and $\rho''$  & 0.02 & 0.02 & 0.03 & 0.82 & 0.64 & 0.55 & 0.46 & 0.56 & 0.48  \\
SCF  & 0.01 & 0.01 & 0.01 & 0.18 & 0.27 & 0.10 & 0.38 & 0.17 & 0.14  \\
Signal Dalitz  & 0.01 & 0.01 & 0.01 & 0.28 & 0.22 & 0.14 & 0.27 & 0.21 & 0.30  \\
BG Dalitz  & 0.01 & 0.01 & 0.01 & 0.29 & 0.35 & 0.26 & 0.28 & 0.26 & 0.34  \\
Other $\pi\pi\pi$  & 0.02 & 0.03 & 0.01 & 0.13 & 0.10 & 0.10 & 0.10 & 0.13 & 0.14  \\
BG fraction  & 0.01 & 0.01 & 0.01 & 0.13 & 0.24 & 0.19 & 0.16 & 0.15 & 0.25  \\
Physics  & 0.01 & 0.01 & $<\!0.01$ & 0.04 & 0.05 & 0.03 & 0.04 & 0.03 & 0.05  \\
BG $\Delta t$  & $<\!0.01$ & $<\!0.01$ & $<\!0.01$ & 0.05 & 0.04 & 0.03 & 0.05 & 0.04 & 0.09  \\
Vertexing  & 0.02 & 0.01 & 0.03 & 0.11 & 0.24 & 0.09 & 0.31 & 0.36 & 0.16  \\
Resolution  & 0.01 & 0.01 & 0.01 & 0.19 & 0.22 & 0.15 & 0.28 & 0.20 & 0.23  \\
Flavor tagging  & $<\!0.01$ & $<\!0.01$ & $<\!0.01$ & 0.04 & 0.07 & 0.04 & 0.04 & 0.07 & 0.03  \\
Fit bias  & $<\!0.01$ & 0.01 & $<\!0.01$ & 0.11 & 0.10 & 0.41 & 0.25 & 0.13 & 0.18  \\
TSI  & $<\!0.01$ & $<\!0.01$ & $<\!0.01$ & 0.09 & 0.04 & 0.06 & 0.05 & 0.18 & 0.05  \\
\hline
Total  & 0.04 & 0.04 & 0.05 & 0.98 & 0.92 & 0.80 & 0.89 & 0.85 & 0.81  \\
\hline
\end{tabularx}
\end{table*}

\subsection{\boldmath Radial excitations ($\rho'$ and $\rho''$)
\label{subsec:syst_rhop_rhopp}
}
The largest contribution for the interfering parameters
tends to come from radial excitations.
The systematic error related to the radial excitations
($\rho(1450)$ and $\rho(1700)$, or $\rho'$ and $\rho''$)
can be categorized into
three classes:
1) uncertainties from the lineshape variation,
i.e., the lineshape difference between
each decay amplitude,
2) uncertainties in
external parameters,
$m_{\rho(1450)}$, $\Gamma_{\rho(1450)}$,
$m_{\rho(1700)}$, $\Gamma_{\rho(1700)}$, and
3) uncertainties in the common lineshape parameters
$\beta$ and $\gamma$ used for the nominal fit.

In our nominal fit, we assume all six decay amplitudes
have the same contribution from $\rho(1450)$ and $\rho(1700)$,
i.e.,
we assume
Eq.~(\ref{equ:f_kappa_definition_with_unique_lineshape_assumption}).
This assumption, however, is not well grounded.
In general, the contributions from $\rho(1450)$ and $\rho(1700)$
can be different for each of the decay amplitudes and thus
the systematic uncertainty from this assumption
must be addressed.
Without the assumption about the higher resonances,
Eq.~(\ref{equ:f_kappa_definition_with_unique_lineshape_assumption})
becomes
\begin{equation}
\kakkoOverlinef {}_\kappa
= T^\kappa_1 \kakkoOverlineF {}^\kappa_\pi(s_\kappa) \;,
\end{equation}
where
\begin{equation}
\begin{split}
\kakkoOverlineF {}^\kappa_\pi(s)
\equiv & BW_{\rho(770)}(s) \\
& + (\beta + \Delta \kakkoOverlineBeta {}_\kappa)
BW_{\rho(1450)}(s) \\
& + (\gamma + \Delta \kakkoOverlineGamma {}_\kappa)
BW_{\rho(1700)}(s) \;. \\
\end{split}
\end{equation}
The variation of the contributions from radial excitations
is described by non-zero
$\Delta \kakkoOverlineBeta {}_\kappa$
and $\Delta \kakkoOverlineGamma {}_\kappa$,
which are 12 complex variables.
We generate various toy MC samples,
where the input $A^\kappa$ and $\overline{A}{}^\kappa$ are fixed
but the values of
$\Delta \kakkoOverlineBeta {}_\kappa$
and
$\Delta \kakkoOverlineGamma {}_\kappa$
are varied randomly
according to the constraints
on $\Delta \kakkoOverlineBeta {}_\kappa$
and
$\Delta \kakkoOverlineGamma {}_\kappa$;
these constraints are obtained from the
results in Sec.~\ref{sec:lineshape_determine}, which are
combined with the isospin relation~\cite{Lipkin:1991st,Gronau:1991dq}
to improve the constraints.
The statistics for each pseudo-experiment are set to be
large enough so that the statistical uncertainty is negligible.
We assign the variations and the biases
of the fit results
due to the
$\Delta \kakkoOverlineBeta {}_\kappa$
and
$\Delta \kakkoOverlineGamma {}_\kappa$
variation as systematic errors.

For the masses and widths
of the $\rho(1450)$ and $\rho(1700)$,
we use the values from
the PDG~\cite{Eidelman:2004wy}.
To estimate the systematic error originating from uncertainties
in their parameters,
we generate toy MC samples varying the input masses and widths.
We fit them using parameterizations with the masses and widths of the
nominal fit.
Here, we vary the masses by twice the PDG error
($\pm 50 \, \mathrm{MeV}/c^2$ for the $\rho(1450)$
and $\pm 40 \, \mathrm{MeV}/c^2$ for the $\rho(1700)$)
since the variations between
independent experiments are much larger than the $1\sigma$ PDG errors,
while we vary the widths by the $\pm 1\sigma$ PDG errors.
We quote the mean shift of the toy MC ensemble as the systematic error.
We also take into account the systematic errors
from the uncertainties in $\beta$ and $\gamma$
for the nominal fit (Eq.~(\ref{equ:nominal_beta_gamma}))
in the same way.

\subsection{SCF}
Systematic errors due to SCF
are dominated by the uncertainty
in the difference between data and MC;
these errors are determined from
$B \rightarrow D^{(*)}\rho$ control samples
that contain a single $\pi^0$ in the final state.
We vary the amount of SCF by its $1\sigma$ error, which is
$\pm 100\%$ for the CR SCF and ${}^{+30(60)}_{-30}\%$ for the NR SCF
in DS-I (DS-II),
where DS-I and DS-II denote the subsets of data taken with the
different detector configurations defined in
Sec.~1-1.
Here, NR represents the neutral-pion-replaced SCF and
CR represents the charged-pion-replaced SCF.
We quote the differences from the nominal
fit as the systematic error.
The event fraction for each $r$ region (for CR and NR),
the wrong tag fractions (for CR)
and lifetime used in the $\Delta t$ PDF (for CR), which are obtained from MC,
are also varied and the differences in the fit results are assigned as a
systematic error.

\subsection{Signal Dalitz PDF}
Systematic errors due to the Dalitz PDF for signal
is mainly from the Dalitz-plot-dependent efficiency.
We take account of
MC statistics in the efficiency
and uncertainty in the $\pi^0$ momentum
dependent efficiency correction, $\epsilon'(p_{\pi^0})$,
obtained from the control samples
of the decay modes
$\bzb\to\rho^-D^{(*)+}$, $\bzb\to\pi^-D^{(*)+}$, $B^-\to\rho^-D^{(*)0}$
and $B^-\to\pi^-D^{(*)0}$.
The Dalitz plot efficiency
obtained from MC is
found to have a small charge asymmetry ($\sim 3\%$ at most).
We use this asymmetric efficiency for our nominal fit.
To estimate the systematic error from the asymmetry,
we fit the data using a symmetric
efficiency and conservatively quote twice the difference
between symmetric and asymmetric efficiencies
as the systematic error.
The Dalitz plot efficiency is $r$-region dependent
and obtained as a product with the event fraction
in the corresponding region,
$\mathcal{F}^l_\mathrm{sig} \cdot \epsilon^l(m', \theta')$,
using MC.
The difference in the fraction
for data and MC is
estimated to be $\sim 10\%$
using the $B^0 \rightarrow D^{*-} \pi^+$
control sample.
The fractions are varied by $\pm 10\%$
to estimate the systematic error.

\subsection{Background Dalitz PDF}
The Dalitz plot for continuum background
has an uncertainty due to the limited statistics
of the sideband events,
which we use to model the PDF.
We estimate the uncertainty
by performing a toy MC study of sideband events.
With each MC pseudo-experiment,
we model the PDF in the same way as we do for real data.
Using the PDF, we fit the data in the signal region
and quote the variation of fit results as the systematic error.
The flavor-asymmetry parameters for the continuum background,
which are fitted from sideband events,
are varied by their uncertainties.
Systematic uncertainty from
the statistics of the $B\bbar$ MC,
which is used to model the $B\bbar$ Dalitz plot PDF,
is also taken into account.

\subsection{\boldmath
$B^0 \rightarrow \pi^+\pi^-\pi^0$ processes other than $B^0
\rightarrow (\rho\pi)^0$
\label{sec:syst_other_pipipi}
}
The primary contribution to the systematic errors
of the non-interfering parameters
tends to come from the $B^0 \rightarrow \pi^+\pi^-\pi^0$
decay processes that are not $B^0 \rightarrow (\rho\pi)^0$.
We take account of the contributions from
$B^0 \rightarrow f_0(980) \pi^0$,
$B^0 \rightarrow f_0(600) \pi^0$,
$B^0 \rightarrow \omega \pi^0$,
and
non-resonant $B^0 \rightarrow \pi^+ \pi^- \pi^0$.
Upper limits on their contributions
are determined from data,
except for $B^0 \rightarrow \omega \pi^0$,
for which we use world averages for
$\mathcal{B}(B^0 \rightarrow \omega \pi^0)$~\cite{unknown:2006bi}
and $\mathcal{B}(\omega \rightarrow \pi^+\pi^-)$~\cite{Eidelman:2004wy}.
For the mass and width parameters of the $f_0(600)$ resonance,
we use recent measurements by
BES~\cite{Ablikim:2004qn},
CLEO~\cite{Muramatsu:2002jp},
and E791~\cite{Aitala:2000xu}
and take the largest variation.
We find no significant signals for any of the above decay modes.
Using the $1\sigma$ upper limits as input,
we generate toy MC
for each mode
with the interference between
the $B^0 \rightarrow (\rho \pi)^0$
and the other
$B^0 \rightarrow \pi^+ \pi^- \pi^0$ mode taken into account.
We obtain the systematic error by
fitting the toy MC assuming
$B^0 \rightarrow (\rho \pi)^0$ only in the PDF.
Within the physically allowed regions,
we vary the $CP$ violation parameters
of the other $B^0 \rightarrow \pi^+ \pi^- \pi^0$ modes and
the relative phase difference
between $B^0 \rightarrow (\rho \pi)^0$
and the other $B^0 \rightarrow \pi^+ \pi^- \pi^0$ modes,
and use the largest deviation as the systematic error for each decay mode.

In the above procedure, we use relativistic Breit-Wigners
for the $f_0(980)$ and $f_0(600)$. To validate the estimated
systematic uncertainties, we investigate possible model dependence
as follows.
For the $f_0(980)$, we perform the same procedure using a
coupled-channel Breit-Wigner~\cite{Aitala:2000xt},
which takes account of the opening of the $K\overline{K}$ decay channel,
instead of a simple relativistic Breit-Wigner.
We observe no systematic increase in the uncertainties.
By changing the model for the $f_0(980)$,
the total systematic error for
the ``$\pipipi$ other than $\rho\pi$'' category increases
by at most 10\% in the non-interfering parameters,
which are the only parameters in which the systematic error
contribution from this category is significant.
Thus, we conclude the
model dependence of the $f_0(980)$ resonance parameterization is
negligibly small.
For the $f_0(600)$, the situation is more complicated because
there are not only possible variations of the resonance mass spectrum
but also uncertainties in the low mass $\pi^-\pi^+$ $S$-wave component;
this contribution may not be modeled by a
simple scalar resonance such as the $f_0(600)$ but by a more
sophisticated
description known as the $K$-matrix~\cite{Au:1986vs,Anisovich:2002ij}.
To address this issue in a model independent and conservative way, we
perform the
systematic error study assuming a hypothetical scalar resonance that has
exactly the same mass spectrum as the $\rho^0(770)$. Since this resonance,
as a $\pi^+\pi^-$ $S$-wave contribution,
is maximally similar to the $\rho^0\pi^0$ signal,
this procedure will lead to systematic
uncertainties larger than, or at least comparable to, any other model of
the $\pi^+\pi^-$ $S$-wave contribution.
We find that including the hypothetical resonance leads to no
significant increase
of the systematic uncertainties compared to those we assigned to
the $f_0(600)$ in our nominal systematic errors.
The increase compared to the case of the $f_0(600)$ is at most 30\%
in the
non-interfering parameters, which corresponds to only a 10\% increase
of the total systematic error for
the ``$\pipipi$ other than $\rho\pi$'' category.
This is because the discrimination power of the $\rho^0\pi^0$ signal
from these low mass $\pi^+\pi^-$ $S$-wave contribution in general
mainly comes from the helicity distribution and the result is not
sensitive to the details of the models of their mass distribution.
Thus, the systematic uncertainties that we have assigned for
a possible contribution from $\bz \to f_0(600)\pi^0$ are reasonable
estimates
of the systematic uncertainties from possible
$\bz \to $ (low mass $\pi^+\pi^-$ $S$-wave)$\pi^0$
contributions in general.

\subsection{Background fraction}
Systematic errors
due to the event-by-event $\dE$-$\mbc$ background fractions
are studied
by varying the PDF shape parameters;
the fraction of continuum background;
and a correction factor to the signal PDF shape,
which takes account of the data-MC difference,
by $\pm 1\sigma$.
We also vary the fractions of the $B\bbar$ background,
which are estimated with MC,
by $\pm 50\%$ ($\pm 20 \%$) for
$b\rightarrow c$ ($b\rightarrow u$) processes.

\subsection{Physics parameters}
We use world averages~\cite{Eidelman:2004wy,unknown:2006bi}
for the following physics parameters:
$\tau_{B^0}$ and $\dmd$ (used for signal and $B\bbar$ background $\Delta t$),
the CKM angles $\phi_1$ and $\phi_2$ (used in $B\bbar$ background),
and the branching fractions of $b\rightarrow u$ decay modes (used in
$B\bbar$ background).
The systematic error is assigned by varying these parameters by
$\pm 1\sigma$.
The charge asymmetry of $B^0 \rightarrow a_1^\pm \pi^\mp$,
for which we use zero in the nominal fit,
is varied over the physically allowed region,
i.e., $\pm 1$.

\subsection{\boldmath Background $\Delta t$ PDF}
Systematic errors
from uncertainties in the background $\Delta t$ shapes
for both continuum and $B\bbar$ backgrounds
are estimated by varying each parameter by $\pm 1\sigma$.

\subsection{Vertex reconstruction}
To determine the systematic error
that arises from uncertainties in the vertex reconstruction,
the track and vertex selection criteria are varied to search
for possible systematic biases.
In addition to the tracks, the IP constraint is also used in the vertex
reconstruction with the smearing due to $B$-flight distance taken into
account. The systematic error due to the IP constraint
is estimated by varying the smearing by
$\pm 10 \, \mu \mathrm{m}$.

\subsection{\boldmath Resolution function for the $\Delta t$ PDF}
Systematic errors due to uncertainties in the resolution function are
estimated by varying each resolution parameter obtained from data (MC)
by $\pm 1\sigma$ ($\pm 2\sigma$).
Systematic errors due to uncertainties in the wrong tag fractions
are also studied by varying the wrong tag fraction individually
for each $r$ region.

\subsection{Fit bias}
We observed fit biases due to small statistics for some of the fitted parameters.
Since these biases are much smaller than the statistical errors,
we do not correct for them but rather take them into account in the
systematic errors.
For each parameter, we estimate the size of the fit bias from a toy MC
study
and quote the bias in the systematic errors.
We also confirm that the bias
is consistent between toy MC and full detector MC simulation.

\subsection{Tag-side interference}
Finally, we investigate the effects of tag-side interference (TSI),
which is the interference between
CKM-favored and CKM-suppressed $B\rightarrow D$ transitions
in the $f_\mathrm{tag}$ final state~\cite{Long:2003wq}.
A small correction to the PDF for the signal distribution arises
from the interference.
We estimate the size of the correction using a
$\bz \rightarrow D^{*-}\ell^+ \nu$ control sample.
We then generate MC pseudo-experiments and make an
ensemble test to obtain the systematic biases.


\section{Quasi-two-body parameters
\label{sec:quasi_two_body}
}
One can deduce quasi-two-body $CP$-violation parameters from the fit
result of the time-dependent Dalitz plot analysis. In this section, we
obtain the quasi-two-body $CP$-violation parameters from
the results for the $U$ and $I$ parameters determined
in the previous section.

The time-dependent partial width for the quasi-two-body decay process of
$\bz \to \rho^\pm\pi^\mp$ is given by~\cite{Wang:2004va}
\begin{equation}
\begin{split}
\frac{d\Gamma}{d \Dt}
\propto & (1 \pm \mathcal{A}_{\rho\pi}^{CP}) \,  e^{-|\Dt|/\taubz}
\\
& \times \Bigl[
1
- q_\mathrm{tag} (\mathcal{C} \pm \Delta \mathcal{C}) \cos (\dmd \Dt)
\\
& \qquad
+ q_\mathrm{tag} (\mathcal{S} \pm \Delta \mathcal{S}) \sin (\dmd \Dt)
\Bigr] \;,
\end{split}
\end{equation}
where the upper (lower) sign is taken for
$\bz \to \rho^+\pi^-$ $(\rho^-\pi^+)$. The parameters
$\mathcal{A}_{\rho\pi}^{CP}$, $\mathcal{C}$, $\Delta \mathcal{C}$,
$\mathcal{S}$, and $\Delta \mathcal{S}$ characterize $CP$-violating and
charge-asymmetric properties of $\bz \to \rho^\pm\pi^\mp$;
$\mathcal{A}_{\rho\pi}^{CP}$ is a time- and flavor-integrated charge
asymmetry, $\mathcal{C}$ is a flavor-dependent direct $CP$-violation
parameter, $\mathcal{S}$ is a mixing-induced $CP$-violation parameter,
and $\Delta \mathcal{C}$ and $\Delta \mathcal{S}$ are $CP$-conserving
parameters (i.e., non-zero $\Delta \mathcal{C}$ or $\Delta \mathcal{S}$
does not imply $CP$ violation).
They are related to the parameters obtained in the time-dependent
Dalitz plot analysis as
\begin{equation}
\begin{split}
\mathcal{A}_{\rho \pi}^{CP} = & \frac{U^+_+ - U^+_-}{U^+_+ + U^+_-} \;,
\\
\mathcal{C} \equiv \frac{\mathcal{C}^+ + \mathcal{C}^-}{2} \;,
& \quad
\Delta \mathcal{C} \equiv \frac{\mathcal{C}^+ - \mathcal{C}^-}{2} \;,
\\
\mathcal{S} \equiv \frac{\mathcal{S}^+ + \mathcal{S}^-}{2} \;,
& \quad
\Delta \mathcal{S} \equiv \frac{\mathcal{S}^+ - \mathcal{S}^-}{2} \;,
\end{split}
\end{equation}
where
\begin{equation}
\mathcal{C}^+ = \frac{U^-_+}{U^+_+}\; , \quad \mathcal{C}^- = \frac{U^-_-}{U^+_-}\; , \quad
\mathcal{S}^+ = \frac{2 I_+}{U^+_+}\; , \quad \mathcal{S}^- = \frac{2 I_-}{U^+_-}\; .
\end{equation}
We obtain
\begin{eqnarray}
\mathcal{A}_{\rho \pi}^{CP}  & = & -0.12 \pm 0.05 \pm 0.04 \; ,\\
\mathcal{C}                  & = & -0.13 \pm 0.09 \pm 0.05 \; ,\\
\Delta \mathcal{C}           & = & +0.36 \pm 0.10 \pm 0.05 \; ,\\
\mathcal{S}                  & = & +0.06 \pm 0.13 \pm 0.05 \; ,\\
\Delta \mathcal{S}           & = & -0.08 \pm 0.13 \pm 0.05 \; ,
\end{eqnarray}
where first and second errors are statistical and systematic, respectively.
The correlation matrix is shown in Table \ref{tab:q2b_correlation_matrix}.
\begin{table}[b]
\caption{
Correlation matrix of the quasi-two-body parameters,
with statistical and systematic errors combined.
\label{tab:q2b_correlation_matrix}
}
\begin{tabular*}{8cm}{@{\extracolsep{\fill}}@{\hspace{3mm}}c@{\hspace{3mm}}|ccccc}
\hline
\hline
&
$\mathcal{A}_{\rho \pi}^{CP}$ &
$\mathcal{C}$                 &
$\Delta \mathcal{C}$          &
$\mathcal{S}$                 &
$\Delta \mathcal{S}$          \\
\hline
$\mathcal{A}_{\rho \pi}^{CP}$  & $+1.00$ \\
$\mathcal{C}$                  & $-0.17$ & $+1.00$ \\
$\Delta \mathcal{C}$           & $+0.09$ & $+0.16$ & $+1.00$ \\
$\mathcal{S}$                  & $+0.01$ & $-0.02$ & $-0.00$ & $+1.00$ \\
$\Delta \mathcal{S}$           & $-0.00$ & $-0.01$ & $-0.02$ & $+0.29$ & $+1.00$ \\
\hline
\end{tabular*}
\end{table}

The angle $\phi_2^\text{eff}$, which equals $\phi_2$ in the no-penguin
limit, can be defined as~\cite{Gronau:2004tm}
\begin{equation}
\phi_2^\text{eff} \equiv \frac{1}{2}
\left(
\phi_2^{\text{eff},+} + \phi_2^{\text{eff},-}
\right)
\end{equation}
with
\begin{equation}
2 \phi_2^{\text{eff},\pm} \pm \hat{\delta}
=
\arcsin  \left(
\frac{\mathcal{S} \pm \Delta \mathcal{S}}
{\sqrt{1 - (\mathcal{C} \pm \Delta \mathcal{C})^2}}
\right) \;,
\end{equation}
and
\begin{equation}
\hat{\delta} = \arg \left( A^{-*} A^+ \right) \;.
\end{equation}
Our measurement gives
\begin{equation}
\phi_2^\text{eff} = \left( \; 88.0 \pm 3.9 \pm 1.7 \; \right)^\circ \;,
\end{equation}
where $\phi_2^\text{eff} \sim 90^\circ$ would give values of
$\mathcal{S}$ and $\Delta \mathcal{S}$ consistent with zero.
There also exists a mirror solution
$\phi_2^\text{eff} = \left(2.0 \pm 3.9 \pm 1.7\right)^\circ$
due to the two-fold ambiguity in the arcsine.
In addition, other solutions
$\phi_2^\text{eff} \sim 45^\circ$ and $135^\circ$ are also allowed in
principle.
The additional solutions correspond to cases where
$2 \phi_2^{\text{eff},+} + \hat{\delta}$
and
$2 \phi_2^{\text{eff},-} - \hat{\delta}$
differ by $\sim 180^\circ$; they can be excluded by including a weak
theoretical assumption (flavor SU(3) or QCD factorization implies
a much smaller value)~\cite{Gronau:2004tm}.
The measured $\phi_2^\text{eff}$ can be used
to constrain $\phi_2$ in a model dependent way,
using a theoretical
assumption that puts a limit on the difference
$\phi_2 - \phi_2^\text{eff}$~\cite{Gronau:2004tm,Gronau:2004sj}.

\begin{figure}[tbp]
\begin{center}
\includegraphics[width=0.4\textwidth]{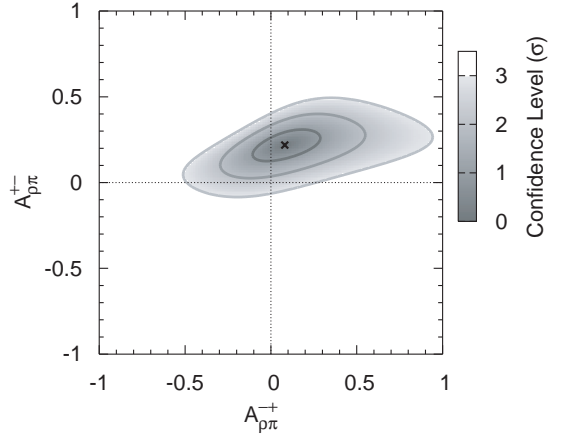}
\caption{
\label{fig:apm_amp_contour}
Contour plot of the confidence level for the direct $CP$ violation
parameters $\mathcal{A}_{\rho \pi}^{+-}$ vs.
$\mathcal{A}_{\rho \pi}^{-+}$.
}
\end{center}
\end{figure}

The direct $CP$-violation parameters for the process
$\bz \to \rho^\pm \pi^\mp$,  $\mathcal{A}_{\rho \pi}^{+-}$
and
$\mathcal{A}_{\rho \pi}^{-+}$,
are defined as
\begin{equation}
\mathcal{A}_{\rho\pi}^{+-}
=
\frac{
\Gamma (\bzb \to \rho^-\pi^+) - \Gamma (\bz \to \rho^+\pi^-)
}
{
\Gamma (\bzb \to \rho^-\pi^+) + \Gamma (\bz \to \rho^+\pi^-)
}
\;,
\end{equation}
and
\begin{equation}
\mathcal{A}_{\rho\pi}^{-+}
=
\frac{
\Gamma (\bzb \to \rho^+\pi^-) - \Gamma (\bz \to \rho^-\pi^+)
}
{
\Gamma (\bzb \to \rho^+\pi^-) + \Gamma (\bz \to \rho^-\pi^+)
} \;.
\end{equation}
One can transform the parameters $\mathcal{A}_{\rho \pi}^{CP}$,
$\mathcal{C}$, and $\Delta \mathcal{C}$ into
the direct $CP$ violation parameters as
\begin{equation}
\mathcal{A}_{\rho \pi}^{+-}
=
- \frac{\mathcal{A}_{\rho \pi}^{CP} + \mathcal{C} + \mathcal{A}_{\rho
\pi}^{CP} \Delta \mathcal{C}}
{1 + \Delta \mathcal{C} + \mathcal{A}_{\rho \pi}^{CP} \: \mathcal{C}} \;,
\end{equation}
\begin{equation}
\mathcal{A}_{\rho \pi}^{-+}
=
\frac{\mathcal{A}_{\rho \pi}^{CP} - \mathcal{C} - \mathcal{A}_{\rho
\pi}^{CP} \Delta \mathcal{C}}
{1 - \Delta \mathcal{C} - \mathcal{A}_{\rho \pi}^{CP} \: \mathcal{C}} \;.
\end{equation}
We obtain
\begin{eqnarray}
\mathcal{A}_{\rho \pi}^{+-} & = & +0.21 \pm 0.08 \pm 0.04 \;, \\
\mathcal{A}_{\rho \pi}^{-+} & = & +0.08 \pm 0.16 \pm 0.11 \;,
\end{eqnarray}
with a correlation coefficient of $+0.47$.
Our result differs from the case of no direct $CP$
asymmetry ($\mathcal{A}_{\rho \pi}^{+-} = 0$ and
$\mathcal{A}_{\rho \pi}^{-+} = 0$)
by $2.3$ standard deviations (Fig.~\ref{fig:apm_amp_contour}).

We also measure the $CP$ violating parameters
of the quasi-two-body $B^0 \rightarrow \rho^0 \pi^0$ decay process.
The time-dependent partial width for the process is given as
\begin{equation}
\begin{split}
\frac{d\Gamma}{d \Dt}
\propto e^{-|\Dt|/\taubz}
\Bigl[ &
1
+ q_\mathrm{tag} \mathcal{A}_{\rho^0\pi^0} \cos (\dmd \Dt)
\\
&
+ q_\mathrm{tag} \mathcal{S}_{\rho^0\pi^0} \sin (\dmd \Dt)
\Bigr] \;,
\end{split}
\end{equation}
where $\mathcal{A}_{\rho^0\pi^0}$ and $\mathcal{S}_{\rho^0\pi^0}$
are the parameters to be measured.
They are calculated from the parameters fitted in the time-dependent
Dalitz plot analysis as
\begin{equation}
\mathcal{A}_{\rho^0 \pi^0} = -\frac{U^-_0}{U^+_0} \; ,
\quad
\mathcal{S}_{\rho^0 \pi^0} = \frac{2 I_0}{U^+_0} \; .
\end{equation}
We obtain
\begin{eqnarray}
\mathcal{A}_{\rho^0 \pi^0} & = & -0.49 \pm 0.36 \pm 0.28 \;,\\
\mathcal{S}_{\rho^0 \pi^0} & = & +0.17 \pm 0.57 \pm 0.35 \;,
\end{eqnarray}
with a correlation coefficient of $-0.08$.
We observe a very small correlation between the quasi-two-body
$CP$-violation parameters of the processes
$\bz \to \rho^\pm\pi^\mp$ and $\bz \to \rho^0\pi^0$, whose absolute
values are less than about 0.02.
Our measurement of $\mathcal{A}_{\rho^0 \pi^0}$
is consistent with the previous measurement from Belle~\cite{Dragic:2006yv}.


\section{Branching Fraction Measurements
\label{sec:branch}
}

The number of $\bz \to \pipipi$ events, $N_\text{sig}$, obtained in
Sec.~\ref{sec:selection_and_reconstruction} is
\begin{equation}
N_\text{sig} = 971 \pm 42 \;.
\label{equ:br_signal_yield}
\end{equation}
The branching fraction for $\bz \to (\rho\pi)^0 \to \pipipi$ including
radial excitations is given as
\begin{equation}
\mathcal{B}(\bz \to \rho\pi^\mathrm{all})
= \frac{N_\text{sig}}{N_{\BB} \, \epsilon^\text{Det} \,
\epsilon^{\text{Veto}} \, \epsilon'{}^\text{KID}}
\;,
\label{equ:br_rhopi_all}
\end{equation}
where $\epsilon^\text{Det}$ and $\epsilon^{\text{Veto}}$ are the average
detection efficiency and the efficiency corresponding to the Dalitz veto
(i.e., the upper and lower bounds on $s_+$, $s_-$, and $s_0$),
respectively; $\epsilon'{}^{\text{KID}}$ is the efficiency correction
factor to take account of the KID difference between data and MC;
and $N_{\BB}$ is the number of $\BB$ pairs produced, where we assume
equal numbers of $\bz\bzb$ and $B^+ B^-$ pairs.
The branching fraction for the decay to a ground state
$\rho^\kappa(770) \pi^\sigma$ is given as
\begin{equation}
\begin{split}
\mathcal{B}(\bz \to \rho^\kappa(770) \pi^\sigma)
& =
f_{\rho_\text{all}^\kappa} f_{\rho^\kappa (770)}
\mathcal{B}(\bz \to \rho\pi^\mathrm{all})
\;,
\\
\Bigl( \quad [\kappa, \sigma] & = [\pm, \mp], [0, 0] \quad \Bigr)
\end{split}
\label{equ:br_rhopi_770}
\end{equation}
where $f_{\rho_\text{all}^\kappa}$ and $f_{\rho^\kappa(770)}$
are the fractions of
$\bz \to \rho^\kappa_\text{all} \pi^\sigma$
among all $\bz \to \rho\pi^\text{all} \to \pipipi$
and that of $\bz \to \rho^\kappa (770) \pi^\sigma$
among $\bz \to \rho_\text{all}^\kappa \pi^\sigma$, respectively.
Here, $\rho^\kappa_\text{all}$ symbolically represents the total
contribution from $\rho^\kappa(770)$, $\rho^\kappa(1450)$,
and $\rho^\kappa(1700)$.
The fraction $f_{\rho_\text{all}^\kappa}$ is calculated from the
parameters $U^+_\kappa$ and $U^{+, \RE(\IM)}_{\kappa\sigma}$
($\kappa, \sigma = +, -, 0$) in Table~\ref{tbl:dt_all_data},
while the coefficients representing contributions from radial
excitations $(\beta, \gamma)$ in Eq.~(\ref{equ:nominal_beta_gamma})
determine $f_{\rho^\kappa(770)}$.
More details of the formalism can be found in
Appendix~\ref{sec:br_formalism}.
The values for these coefficients are shown in
Table~\ref{tab:eff_factor_summary}.
Note that
$f_{\rho^\kappa (770)}$ is close to 1 and $|1-f_{\rho^\kappa (770)}|$ is
much smaller than the error in our result. This means that
it is reasonable to compare the central values of
our result with the preceding branching fraction measurements that do
not separate the contribution
from radial excitations.
However, the errors are not directly comparable.
\begin{table}[bp]
\begin{center}
\caption{
Summary of the coefficients used in the branching fraction
measurement.
\label{tab:eff_factor_summary}
}
\newcolumntype{C}{>{\centering\arraybackslash}X}
\begin{tabularx}{35mm}{@{\hspace{2mm}}cC}
\hline
\hline
$\epsilon^{\mathrm{Det}}$ & 0.10 \\
$\epsilon^{\mathrm{Veto}}$ & 0.84 \\
$\epsilon'{}^{\text{KID}}$ & 0.96 \\
\hline
$f_{\rho^\pm_\text{all}}$ & 0.89 \\
$f_{\rho^0_\text{all}}$ & 0.12 \\
\hline
$f_{\rho^\pm (770)}$ &0.99 \\
$f_{\rho^0 (770)}$   & 0.99 \\
\hline
\hline
\end{tabularx}
\end{center}
\end{table}

From (\ref{equ:br_signal_yield}) and the
coefficients in Table~\ref{tab:eff_factor_summary}, we obtain
\begin{equation}
\mathcal{B}(\bz \to \rho\pi^\text{all})
= (\BFall) \times 10^{-6} \;,
\end{equation}
and
\begin{equation}
\begin{split}
\mathcal{B}(\bz \to \rho^\pm (770)\pi^\mp) & = (\BFpm)  \times 10^{-6} \;,
\\
\mathcal{B}(\bz \to \rho^0 (770)\pi^0)   & = (\BFz)  \times 10^{-6}
\;.
\\
\end{split}
\end{equation}
Here, the first and second errors correspond to the statistical and
systematic errors, respectively, where statistical errors include
the contributions from the statistical errors on the number of events
($N_\text{sig}$) and uncertainties in the Dalitz parameters
($U^+_\kappa$ and $U^{+, \RE(\IM)}_{\kappa\sigma}$).
The correlation coefficient for the statistical errors between
$\mathcal{B}(\bz \to \rho^\pm (770)\pi^\mp)$ and
$\mathcal{B}(\bz \to \rho^0 (770)\pi^0)$ is $-0.09$.
The branching fractions obtained are consistent with our previous
measurements~\cite{Gordon:2002yt,Dragic:2006yv}.

\begin{table}[bp]
\def\Topspc{\rule{0pt}{11pt}}
\def\Btmspc{\rule[-6pt]{0pt}{0pt}}
\begin{center}
\caption{
\label{tbl:br_all}
Summary table of the systematic errors for the branching fraction
measurements.
A common factor of $\times 10^{-6}$ is omitted for simplicity.
}
\newcolumntype{C}{>{\centering\arraybackslash}X}
\begin{tabularx}{0.47\textwidth}{@{\hspace{2mm}}l@{\hspace{4mm}}*{3}{C@{\hspace{2mm}}}}
\hline
\hline
& $\rho\pi^\mathrm{all}$ & $\rho^{\pm}(770)\pi^\mp$ & $\rho^0(770)\pi^0$ \\
\hline
$\rho'$ and $\rho''$   & $\pm 2.8$  & $\pm 3.9$  & $\pm 0.5$ \\
Physics  Parameters & $\pm 0.1$  & $\pm 0.1$  & $\pm 0.0$ \\
Fit   & $\pm 1.7$  & $\pm 1.5$  & $\pm 0.2$ \\
Detection Efficiency   & $\pm 1.4$  & $\pm 1.3$  & $\pm 0.2$ \\
TDPA Systematic   & $\pm 0.2$  & $\pm 0.4$  & $\pm 0.4$ \\
Number of $\BB$   & $\pm 0.3$  & $\pm 0.3$  & $\pm 0.0$ \\
\hline
Total   & $\pm 3.6$  & $\pm 4.4$  & $\pm 0.7$ \\
\hline
\hline
\end{tabularx}
\end{center}
\end{table}

\subsection{Systematic uncertainties}
Table~\ref{tbl:br_all} summarizes the systematic errors for the
branching fraction measurement. We discuss each item in the table in the
following.

\subsubsection{Radial excitations ($\rho'$ and $\rho''$)}
As in the time-dependent analysis,
the systematic uncertainty related to the radial excitations come from
1) uncertainties due to lineshape variation, 2) uncertainties in masses
and widths of the $\rho(770)$ and its radial excitations, and 3) the common lineshape
parameters.
Their impacts are estimated in the same manner as done for the
time-dependent analysis. The impact from possible lineshape variation,
which is constrained by our data, is estimated by an MC
study. We adopt the same
uncertainties for the masses and widths of the $\rho(770)$ and radial
excitations as in the
time-dependent analysis and estimate their impact on the branching
fractions. The errors due to common lineshape parameters
$(\beta, \gamma)$ are also taken into account.

The uncertainty for this category is sizable since
the $\rho(770)$ and radial excitations are separated in the branching
fraction analysis.
This is an essential difference from the
time-dependent analysis, where we do not separate the $\rho(770)$ and
the radial excitations.

\subsubsection{Physics parameters}
This category includes
the systematic error from the uncertainties in the branching fractions
of the $\BB$ background components as well as those due to the possible
contribution from
the $\bz \to \pipipi$ processes other than $\bz \to (\rho\pi)^0$.

\subsubsection{Fit}
This category includes the uncertainties related to the extended
unbinned maximum likelihood fit, except for the items included in above
two categories. It consists of the uncertainties from the modeling of
PDF used in the fit, where the possible uncertainty in the SCF component
has a sizable
impact;
the fraction of the $\BB$ background component,
which is fixed in the nominal fit; and the fit bias.

\subsubsection{Detection efficiency}
The largest components of the detection efficiency systematic
uncertainty are the differences between data and MC.
We consider
differences in $\pi^0$ detection efficiency with $\pi^0$ momentum
dependence, the KID efficiency correction
($\epsilon'{}^\text{KID}$), the continuum event suppression cut, and
the vertex reconstruction efficiency.
All are estimated from control sample studies.
Small uncertainty due to the limited statistics of MC used in
calculation of the efficiency is also taken into account.
Table~\ref{tbl:branch_syst_eff} shows the breakdown of the systematic
uncertainty contributions from the above items.

\begin{table}[tbp]
\begin{center}
\caption{
Summary of the systematic uncertainties related to the detection
efficiency.
\label{tbl:branch_syst_eff}
}
\begin{tabular}{@{\hspace{4mm}}l@{\hspace{6mm}}c@{\hspace{4mm}}}
\hline
\hline
$\pi^0$ Detection & 4.7\% \\
Kaon Identification (KID) &   0.4\% \\
Continuum Suppression &    2.3\% \\
Vertex Reconstruction &  1.8\% \\
Dalitz Efficiency &  0.8\% \\
\hline
\hline
\end{tabular}
\end{center}
\end{table}

\subsubsection{TDPA systematic errors}
The systematic errors in the Dalitz plot parameters obtained in the
time-dependent Dalitz plot analysis (TDPA) listed in
Table~\ref{tbl:dt_all_data} propagate to the branching fractions.

\subsubsection{Number of $\BB$ pairs}
The number of accumulated $\BB$ pairs and its uncertainty are
$N_{\BB} = (449.3 \pm 5.7) \times 10^6$, assuming an equal
production rate for charged and neutral $B\overline{B}$ pairs from
the $\Upsilon(4S)$.
The uncertainty in the number of $B\overline{B}$ pairs propagates to the
branching fraction and
is taken into account as a systematic error.


\section{\boldmath Constraint on the CKM angle $\phi_2$
\label{sec:constraint_on_ph2}
}
\begin{figure}[b]
\begin{center}
\includegraphics[width=0.38\textwidth,height=0.232\textwidth]{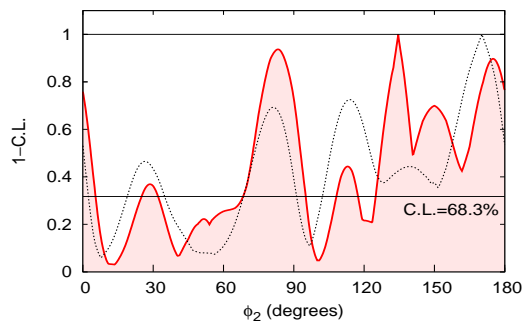}
\caption{
$1-\mathrm{C.L.}$ vs. $\phi_2$.
Dotted and solid curves
correspond to the result from
the time-dependent Dalitz plot analysis
only and that from
the Dalitz
and an isospin (pentagon) combined analysis,
respectively.
\label{fig:phi2_1_cl_plot}
}
\end{center}
\end{figure}

We constrain the CKM angle $\phi_2$ from our analysis
following the procedure described in
Ref.~\cite{Snyder:1993mx}.
With three $\bz \rightarrow (\rho\pi)^0$ decay modes,
we have 9 free parameters including $\phi_2$:
\begin{equation}
\begin{split}
9 = & \mathrm{(6\;complex\;amplitudes=12 \, d.o.f.) + \phi_2}
\\
& \mathrm{
- (1\;global\;phase)
- (1\;global\;normalization)}
\\
& \mathrm{- (1\;isospin\;relation=2 \, d.o.f.)}
\;,
\end{split}
\end{equation}
where we make use of an isospin relation
that relates neutral $B$
decay processes only~\cite{Lipkin:1991st,Gronau:1991dq}.
Parameterizing the 6 complex amplitudes with 9 free parameters,
we form a $\chi^2$ function using the 26 measurements
from our time-dependent Dalitz plot analysis as constraints.
We first optimize all the 9 parameters to obtain a minimum $\chi^2$,
$\chi^2_\mathrm{min}$; we then scan $\phi_2$
from $0^\circ$ to $180^\circ$ optimizing the other 8 parameters,
whose resultant minima are defined as $\chi^2(\phi_2)$;
the difference $\Delta \chi^2(\phi_2)$ is defined as
$\Delta \chi^2(\phi_2) \equiv \chi^2(\phi_2) - \chi^2_\mathrm{min}$.
Performing a toy MC study
following the procedure described in Ref.~\cite{Charles:2004jd},
we obtain the $\mathrm{1-C.L.}$ plot in Fig.~\ref{fig:phi2_1_cl_plot}
(dotted line) from the $\Delta \chi^2(\phi_2)$~\cite{Chi2ToyMC}.

In addition to the 26 observables obtained from our
time-dependent Dalitz plot analysis,
we use
the branching fraction
$\mathcal{B}(B^0 \rightarrow \rho\pi^\text{all})$
obtained in Sec.~\ref{sec:branch}
and the following
world average branching fractions and asymmetries:
$\mathcal{B}(B^+ \rightarrow \rho^+ \pi^0)$,
$\mathcal{A}(B^+ \rightarrow \rho^+ \pi^0)$,
$\mathcal{B}(B^+ \rightarrow \rho^0 \pi^+)$,
and
$\mathcal{A}(B^+ \rightarrow \rho^0 \pi^+)$~\cite{unknown:2006bi},
which are not correlated with our 26 observables.
With the 31 measurements above,
we perform a full combined
Dalitz and isospin (pentagon) analysis.
Having 5 related decay modes,
we have 12 free parameters including $\phi_2$:
\begin{equation}
\begin{split}
12 = &
\mathrm{(10\;complex\;amplitudes=20 \, d.o.f.) + \phi_2}
\\
& \mathrm{- (1\;global\;phase)}
\\
& \mathrm{- (4\;isospin\;relations=8 \, d.o.f.)} \;.
\end{split}
\end{equation}
The detail of the $\chi^2$ construction can be found
in appendix \ref{sec:phi2_chi2_parameterization}.
The $\chi^2_\mathrm{min}$ obtained is 10.2,
which is reasonable for
$\mathrm{31(measurements)}$ $ - $
$\mathrm{12(free\; parameters)} $ $= 19$ degrees of freedom.
Following the same procedure as above,
we obtain the
$\mathrm{1-C.L.}$ plot in Fig.~\ref{fig:phi2_1_cl_plot} (solid line).
We obtain $68^\circ < \phi_2 < 95^\circ$ as the 68.3\% confidence
interval consistent with the SM expectation.
Several SM-disfavored region
($0^\circ<\phi_2<5^\circ$, $25^\circ <\phi_2<32^\circ$,
and $108^\circ <\phi_2<180^\circ$) are also allowed.


\section{Conclusion}
Using $\lint \, \text{fb}^{-1}$ of data we have performed a
time-dependent Dalitz plot analysis
of the $B^0 \rightarrow \pi^+\pi^-\pi^0$ decay mode.
Combining our analysis and information
from charged $B$ decay modes,
a full Dalitz plot and isospin analysis is performed
to obtain a constraint on $\phi_2$ in a model-independent way.
We obtain $68^\circ < \phi_2 < 95^\circ$ at the 68.3\% confidence
interval for the solution consistent with the SM expectation.
However, a large CKM-disfavored region also remains.
In principle, with more data we may be able to remove all the
additional $\phi_2$ solutions.
From the result of the Dalitz plot analysis, we also obtain the
branching fractions for the decays $\bz \to \rho^\pm(770) \pi^\mp$
and $\bz \to \rho^0(770) \pi^0$.
These are the first branching fraction measurements of these processes
with the lowest resonance $\rho(770)$ explicitly separated from the
radial excitations.

\section*{Acknowledgments}

We thank the KEKB group for the excellent operation of the
accelerator, the KEK cryogenics group for the efficient
operation of the solenoid, and the KEK computer group and
the National Institute of Informatics for valuable computing
and Super-SINET network support. We acknowledge support from
the Ministry of Education, Culture, Sports, Science, and
Technology of Japan and the Japan Society for the Promotion
of Science; the Australian Research Council and the
Australian Department of Education, Science and Training;
the National Science Foundation of China and the Knowledge
Innovation Program of the Chinese Academy of Sciences under
contract No.~10575109 and IHEP-U-503; the Department of
Science and Technology of India;
the BK21 program of the Ministry of Education of Korea,
the CHEP SRC program and Basic Research program
(grant No.~R01-2005-000-10089-0) of the Korea Science and
Engineering Foundation, and the Pure Basic Research Group
program of the Korea Research Foundation;
the Polish State Committee for Scientific Research;
the Ministry of Education and Science of the Russian
Federation and the Russian Federal Agency for Atomic Energy;
the Slovenian Research Agency;  the Swiss
National Science Foundation; the National Science Council
and the Ministry of Education of Taiwan; and the U.S.\
Department of Energy.

\appendix

\section{PDFs for time-dependent Dalitz plot analysis
\label{sec:appendix_pdf_definitions}
}
In this section, we describe the details of the PDF for each component,
which appear in Eq.
(\ref{equ:dembc_dalitz_simultaneous-total_pdf}).

\subsection{Signal PDF
\label{sec:signal_pdf}
}
The PDF for signal events consists of
a PDF for the correctly reconstructed events
$\mathcal{P}_\mathrm{true}$ and
PDFs for SCF events
$\mathcal{P}_{i}$ ($i = \mathrm{NR,CR}$):
\begin{equation}
\label{equ:signal_pdf_definition}
\mathcal{P}_\mathrm{sig}(\vec{x}) =
\frac{\mathcal{P}_\mathrm{true}(\vec{x})
+\sum_{i=\mathrm{NR,CR}} \mathcal{P}_{i}(\vec{x})
}
{
\mathcal{N}_\mathrm{true}  + \sum_{i = \mathrm{NR,CR}} \mathcal{N}_i
}
\;,
\end{equation}
where NR and CR represent the
$\pi^0$ (neutral)
replaced and $\pi^\pm$ (charged) replaced SCF's, respectively,
and $\mathcal{N}$ are the integrals of the PDF's.

\subsubsection{PDF for correctly reconstructed events}
In terms of the event fractions
for the $l$-th flavor tagging region
($\mathcal{F}^l_\mathrm{true}$),
the Dalitz plot dependent efficiency
($\epsilon^l$), the $\pi^0$ momentum-dependent efficiency correction
taking account of the difference
between data and MC ($\epsilon'$),
wrong-tag fractions ($w_l$), and the differences in wrong-tag
fractions between $\bz$ and $\bzb$ ($\dwl$),
the PDF for correctly reconstructed events is given by
\begin{equation}
\begin{split}
\mathcal{P}_\mathrm{true}(\vec{x})
= & {\mathcal{F}}^l_\text{true}
\cdot \mathcal{P}_\mathrm{true}(\Delta E, \mbc; p_{\pi^0})
\\
& \cdot \epsilon_\mathrm{true}(m',\theta'; l) \, \epsilon'(p_{\pi^0})
\cdot |\mathrm{det} \boldsymbol{J}(m',\theta')|
\\
& \cdot
\mathcal{P}_\mathrm{true}( m',\theta';\Delta t, q_\mathrm{tag}; l) \;
,
\label{equ:app_pdf_true_pdf}
\end{split}
\end{equation}
where the Dalitz-$\Dt$ PDF
$\mathcal{P}_\mathrm{true} ( m', \theta';\Delta t, q_\mathrm{tag}; l)$
corresponds to the right-hand side of
Eq.~(\ref{equ:amplitude_time_dep_three_pi}) and is
\begin{equation}
\begin{split}
\mathcal{P}_\mathrm{true} ( & m', \theta';\Delta t, q_\mathrm{tag}; l)
\\
= & \frac{e^{-|\Dt|/\taubz}}{4\taubz}
\Biggl\{
(1-q_\mathrm{tag}\dwl)(|A_{3\pi}|^2+|\overline{A}_{3\pi}|^2)
\\
&
- q_\mathrm{tag} (1-2w_l) \left(|A_{3\pi}|^2-|\overline{A}_{3\pi}|^2\right)
\cos (\dmd \Dt)
\\
& + q_\mathrm{tag} (1-2w_l) 2
\mathrm{Im}\left(\frac{q}{p}\overline{A}_{3\pi}A^*_{3\pi}\right)\sin(\dmd \Dt)
\Biggr\} \;.
\end{split}
\label{equ:app_pdf_true_pdf_dalitz_dt}
\end{equation}
For the $\Dt$ PDF,
the above equation is convolved with the resolution
function~\cite{Tajima:2003bu}.
The terms $|A_{3\pi}|^2 \pm |\overline{A}_{3\pi}|^2$
and $\mathrm{Im}\left(\frac{q}{p}\overline{A}_{3\pi}A^*_{3\pi}\right)$
are expanded as in Eqs.~(\ref{equ:dalitz_parameterization_1}) and
(\ref{equ:dalitz_parameterization_2}).
The $\dE$-$\mbc$ PDF is normalized such that
\begin{equation}
\iint_\mathrm{signal \; region}
\hspace{-12mm} d\Delta E \; d\mbc \;
\mathcal{P}_\mathrm{true}(\Delta E, \mbc; p_{\pi^0}) = 1
\quad (\forall p_{\pi^0}) \;,
\end{equation}
since we define the Dalitz plot efficiency for events inside the signal region.
Since the PDF's in the $\Delta t$-$q_\mathrm{tag}$ direction
are also normalized to be unity,
the integral inside the signal region, $\mathcal{N}_\mathrm{true}$, is,
\begin{equation}
\mathcal{N}_\mathrm{true}
= \sum_l \mathcal{N}^l_\mathrm{true} \;,
\end{equation}
where
\begin{widetext}
\begin{equation}
\begin{split}
\label{equ:dembc_dalitz_simultaneous-normalization_dalitz}
\mathcal{N}^l_\mathrm{true}
& \equiv
\sum_{q_\mathrm{tag}}
\int \hspace{-1mm} d \Delta t
\iint_\mathrm{signal \; region}
\hspace{-15mm} d\Delta E \; d\mbc \;
\iint_\mathrm{SDP, \; Veto}
\hspace{-12mm}
dm' \; d\theta' \;
\mathcal{P}_\mathrm{true}(\vec{x})
\\
& =
\mathcal{F}^l_\mathrm{true}
\int \hspace{-2mm}
\int_\mathrm{SDP, \; Veto}
\hspace{-12mm}
dm' \; d\theta' \;
\epsilon_\mathrm{true}(m', \theta'; l) \, \epsilon'(p_{\pi^0})
\,
|\mathrm{det} \boldsymbol{J}| \,
\left( |A_{3\pi}|^2 + |\overline{A}_{3\pi}|^2 \right) \;,
\end{split}
\end{equation}
\end{widetext}
and the correlation between $p_{\pi^0}$ and $m'$ is properly
taken into account in the integration
on the last line.
The notation
$\int \int_\mathrm{SDP, \; Veto} dm' \; d\theta'$
means integration over the square Dalitz plot
with the vetoed region in the Dalitz plot taken into account.

The $\pi^0$ momentum dependent
$\dE$-$\mbc$ PDF,
$\mathcal{P}_\mathrm{true}(\dE, \mb; p_{\pi^0})$,
is modeled
using MC-simulated events in
a binned histogram interpolated in the $p_{\pi^0}$ direction,
to which a small correction obtained with
$\bzb\to\rho^-D^{(*)+}$
is applied
to account for the difference between MC and data.

The Dalitz plot distribution is smeared and distorted by
detection efficiencies and detector resolutions.
We obtain the signal Dalitz plot efficiency from MC
to take the former into account.
We introduce a dependence of the efficiency on the $r$ region,
$\epsilon^l_\mathrm{true}$,
since a significanct dependence is observed in MC.
Small corrections, $\epsilon'(p_{\pi^0})$,
are also applied to the MC-determined efficiency
to account for differences between MC and data.
We use $\bzb\to\rho^-D^{(*)+}$, $\bzb\to\pi^-D^{*-}$, $B^-\to\rho^-D^0$
and $B^-\to\pi^-D^0$ decays to obtain the correction factors.
The smearing in the Dalitz plot due to the finite detector resolutions
is small
compared to the widths of $\rho(770)$ resonances;
the smearing is confirmed by MC to be a negligibly small effect.

\subsubsection{PDF for SCF events}
Approximately 20\% of signal candidates are SCFs,
which are subdivided into  $\sim 4\%$ NR SCF
and $\sim 16\%$ CR SCF.
It is therefore important to develop a model that describes
the SCF component well.
The time-dependent PDF for SCF events
is defined as
\begin{equation}
\begin{split}
\mathcal{P}_{i}(\dE, & \mbc; m',\theta';\Delta t, q_\mathrm{tag}; l)
\\
= &
{\mathcal{F}}^l_{i}\cdot \mathcal{P}_{i}(\Delta E, \mbc; s_i)
\cdot
\mathcal{P}_i(m', \theta'; \Delta t, q_\mathrm{tag}) \;,
\\
& \qquad \qquad (i=\mathrm{NR, CR})
\end{split}
\label{equ:app_pdf_scf_pdf}
\end{equation}
with
\begin{equation}
\begin{split}
\mathcal{P}_i (m', \theta'; & \Delta t, q_\mathrm{tag} )
\\
= & \frac{e^{-|\Delta t|/\tau_i}}{4\tau_i}
\Bigl\{ (1-q_\mathrm{tag}\dwl^{i} ) \: \mathcal{P}^\mathrm{Life}_{i}(m', \theta')
\\
& - q_\mathrm{tag}(1-2 {w_l}^{i}) \: \mathcal{P}^{\mathrm{Cos}}_{i}(m', \theta')\cos(\dmd \Delta t)
\\
& + q_\mathrm{tag}(1-2 {w_l}^{i}) \: \mathcal{P}^{\mathrm{Sin}}_{i}(m', \theta')\sin(\dmd \Delta t) \Bigr\} \;.
\\
\end{split}
\label{equ:dt_pdf_of_SCF}
\end{equation}
where ${\mathcal{F}}^l_i$ is the event fraction in each tagging $r$-bin.

The $\dE$-$\mbc$ PDF is normalized inside the signal region as
\begin{equation}
\int \hspace{-2mm} \int_\mathrm{signal \; region}
\hspace{-12mm} d\Delta E \; d\mbc \;
\mathcal{P}_{i}(\Delta E, \mbc; s_i) = 1
\quad (\forall s_i) \;.
\end{equation}
As the PDF's in the $\Delta t$-$q_\mathrm{tag}$ direction
are normalized to unity
and $\sum_l {\mathcal{F}}^l_i = 1$,
the integral inside the signal region, $\mathcal{N}_i$, is
\begin{widetext}
\begin{equation} 
\mathcal{N}_i \equiv \sum_l \sum_{q_\mathrm{tag}}
\int \hspace{-1mm} d \Delta t
\int \hspace{-2mm} \int_\mathrm{signal \; region}
\hspace{-15mm} d\Delta E \; d\mbc \;
\iint_\mathrm{SDP, \; Veto}
\hspace{-12mm}
dm' \; d\theta' \;
\mathcal{P}_{i}^l(\dE, \mbc; m',\theta';\Delta t, q_\mathrm{tag})
=
\iint_\mathrm{SDP, \; Veto}
\hspace{-12mm}
dm' \; d\theta' \;
\mathcal{P}^\mathrm{Life}_{i}(m', \theta') \;.
\end{equation} 
\end{widetext}

We find that the $\dE$-$\mb$ distribution for SCF has a sizable correlation with
Dalitz plot variables,
but only in one of its two dimensions.
We thus introduce a model
with dependences on the Dalitz plot variable $s_i$.
The variable $s_\mathrm{CR} = s_\pm \equiv \max(s_+, s_-)$ is used,
because the CR SCF can be divided into a $\pi^+$ replaced SCF
and a $\pi^-$ replaced SCF, where $s_-$ ($s_+$) is used for
$\pi^+$ ($\pi^-$) replaced SCF.
Here, we exploit the fact that almost all of the $\pi^+$ ($\pi^-$)
replaced SCF distributes in the region of $s_+ < s_-$ ($s_+ > s_-$).
For the NR SCF, $s_\mathrm{NR} = s_0$.
This parameterization
models the correlation quite well,
with each of the parameters $s_i$
related to the kinematics of replaced tracks.

Since track ($\pi$) replacement
changes the measured kinematic variables,
the SCF events ``migrate'' in the Dalitz plot
from the correct (or generated) position
to the observed position.
Using MC,
we determine resolution functions
$R_i(m'_\mathrm{obs}, \theta'_\mathrm{obs}; m'_\mathrm{gen},
\theta'_\mathrm{gen})$
to describe this ``migration'' effect,
where $(m'_\mathrm{obs}, \theta'_\mathrm{obs})$
and $(m'_\mathrm{gen}, \theta'_\mathrm{gen})$
are the observed and the generated (correct)
positions in the Dalitz plot, respectively.
The resolution function satisfies the normalization condition of
\begin{equation}
\begin{split}
\iint_\text{SDP} \hspace{-5mm} dm'_\text{obs} d\theta'_\text{obs} \:
R_i(m'_\text{obs}, \theta'_\text{obs}; m'_\text{gen},
& \theta'_\text{gen}) = 1\;.
\\
& (\forall m'_\text{gen}, \theta'_\text{gen})
\end{split}
\label{equ:app_pdf_scf_resolution_normalization}
\end{equation}
Together with the efficiency function
$\epsilon_i(m'_\mathrm{gen}, \theta'_\mathrm{gen})$,
which is also obtained with MC,
the Dalitz plot PDF for SCF is described as
\begin{equation}
\begin{split}
\mathcal{P}^j_{i}(m', \theta')
= & \left[(R_i \cdot \epsilon_i) \otimes P^j_\mathrm{phys}\right](m',
\theta')
\\
\equiv &
\iint_\mathrm{SDP}
\hspace{-4mm}
dm'_\mathrm{gen} \; d\theta'_\mathrm{gen} \;
R_i(m', \theta'; m'_\mathrm{gen}, \theta'_\mathrm{gen})
\\
& \quad \cdot \epsilon_i (m'_\mathrm{gen}, \theta'_\mathrm{gen})
\cdot P^j_\mathrm{phys} (m'_\mathrm{gen}, \theta'_\mathrm{gen})
\;,
\\
(\; i= & \mathrm{CR, NR} \;, \quad j=\mathrm{Life, Cos, Sin}\;)
\label{equ:app_pdf_scf_dalitz_convl}
\end{split}
\end{equation}
\[
\]
where
\begin{equation}
\begin{split}
P^\mathrm{Life}_\mathrm{phys} (m'_\mathrm{gen}, \theta'_\mathrm{gen})
& =
|\mathrm{det}\boldsymbol{J}| (|A_{3\pi}|^2 + |\overline{A}_{3\pi}|^2) \;,
\\
P^\mathrm{Cos}_\mathrm{phys} (m'_\mathrm{gen}, \theta'_\mathrm{gen})
& =
|\mathrm{det}\boldsymbol{J}| (|A_{3\pi}|^2 - |\overline{A}_{3\pi}|^2) \;,
\\
P^\mathrm{Sin}_\mathrm{phys} (m'_\mathrm{gen}, \theta'_\mathrm{gen})
& =
|\mathrm{det}\boldsymbol{J}| 2\mathrm{Im}\left(\frac{q}{p}\overline{A}_{3\pi}A^*_{3\pi}\right) \;.
\end{split}
\end{equation}

For the NR SCF,
the shape of the $\Delta t$ PDF defined in
Eq.~(\ref{equ:dt_pdf_of_SCF})
is exactly the same as correctly reconstructed signal,
i.e., $\tau_\mathrm{NR} = \tau_\bz$,
${w_l}^\mathrm{NR} = w_l$, and $\dwl^\mathrm{NR} = \dwl$,
since the replaced $\pi^0$
is not used for either vertexing or flavor tagging.
On the other hand, for the CR SCF,
the $\Delta t$ PDF is different from correctly reconstructed
signal,
since the replaced $\pi^\pm$ is used for both
vertexing and flavor tagging.
Thus, we use MC-simulated CR SCF events to obtain
$\tau_\mathrm{CR}$,
${w_l}^\mathrm{CR}$, and $\dwl^\mathrm{CR}$,
which are different from those of correctly reconstructed signal events.
In particular,
$\dwl^\mathrm{CR}$
is opposite in sign
for the $\pi^+$ and $\pi^-$ replaced SCFs,
due to the fact that the replaced
$\pi^\pm$ tends to be directly used
for flavor tagging in the slow pion category.

\subsection{Continuum PDF}
The PDF for the continuum background is
\begin{equation}
\begin{split}
\mathcal{P}_{\qq} & (\Delta E, \mbc; m', \theta';\Delta t, q_\mathrm{tag}; l)
\\
=
& \mathcal{F}^l_{\qq}
\cdot \mathcal{P}_{\qq}^l(\dE, \mbc)
\cdot \mathcal{P}_{\qq}(m', \theta'; \Delta E, \mbc)
\\
& \cdot \left[ \frac{1+q_\mathrm{tag} A^l(m', \theta')}{2} \right]
\cdot \mathcal{P}_{\qq}(\Delta t) \;,
\end{split}
\end{equation}
where $\mathcal{F}^l_{\qq}$ is the event fraction
for each $r$ region obtained in the signal yield fit.
All the terms on the right hand side of the equation are normalized to be unity
so that
\begin{equation}
\begin{split}
\sum_l \sum_{q_\mathrm{tag}}
& \int \hspace{-1mm} d \Delta t
\iint_\mathrm{signal \; region}
\hspace{-15mm} d\Delta E \; d\mbc \;
\\
\times
& \iint_\mathrm{SDP, \; Veto}
\hspace{-12mm}
dm' \; d\theta' \;
\mathcal{P}_{\qq}(\Delta E, \mbc; m', \theta';\Delta t, q_\mathrm{tag}; l) = 1 .
\end{split}
\end{equation}

Since the allowed kinematic region is dependent on
$\dE$ and $\mbc$,
the Dalitz plot distribution is dependent on $\dE$ and $\mbc$.
We define a $\dE$-$\mbc$ independent PDF, $\mathcal{P}_{\qq}(m'_\mathrm{scale},
\theta')$,
where $m'_\mathrm{scale}$ is a re-defined SDP variable
with the kinematic effect taken into account as
\begin{equation}
m'_\mathrm{scale}
\equiv
\frac{1}{\pi}
\arccos
\left(
2 \frac{m_0 - m_0^{\mathrm{min}}}{m_0^{\mathrm{max}} -
m_0^{\mathrm{min}} + \dE + \Delta \mbc} - 1
\right),
\end{equation}
where
\begin{equation}
\Delta \mbc \equiv \mbc - m_{\bz} \;.
\end{equation}
Using the $\dE$-$\mbc$ independent PDF,
$\mathcal{P}_{\qq}(m', \theta'; \Delta E, \mbc)$
is described as
\begin{equation}
\begin{split}
\mathcal{P}_{\qq}(m', \theta'; \Delta E, \mbc)
=
& \frac{1}{\mathcal{N}_{\qq}(\dE + \Delta \mbc)}
\\
& \cdot \frac{\sin(\pi m')}{\sin (\pi
m'_\mathrm{scale})}
\cdot
\mathcal{P}_{\qq}(m'_\mathrm{scale},\theta')
\end{split}
\end{equation}
for the region,
$m_0^\mathrm{min} < m_0 < \min(m_0^\mathrm{max}, m_0^\mathrm{max}+\dE +
\Delta \mbc)$
($p_{\qq}=0$ otherwise),
where $\mathcal{N}_{\qq}(\dE + \Delta \mbc)$
and $\sin(\pi m')/\sin (\pi m'_\mathrm{scale})$
are a normalization factor and the Jacobian
for the parameter transformation $m'_\mathrm{scale} \rightarrow m'$,
respectively.
We obtain the $p_{\qq}(m'_\mathrm{scale},\theta')$
distribution from data in part of the sideband region,
$-0.1\,\mathrm{GeV} < \dE < 0.2\,\mathrm{GeV}$
and $5.2\,\mathrm{GeV}/c^2 < \mbc < 5.26\,\mathrm{GeV}/c^2$,
where the contribution from $B\bbar$ background is negligible.

Since we find significant flavor asymmetry
depending on the location in the Dalitz plot,
we introduce the following term to take account of it:
\begin{equation}
\frac{1+q_\mathrm{tag} A^l(m', \theta')}{2} \;,
\end{equation}
which is $r$ region dependent.
The asymmetry is anti-symmetric in the direction of $\theta'$,
i.e., $A^l(m', \theta') > 0$ ($A^l(m', \theta') < 0$)
in the region of $\theta'>0.5$ ($\theta < 0.5$),
and the size of the asymmetry is at most $\sim 20 \%$ in
the best $r$ region.
This effect is due to
the jet-like topology of continuum events;
when an event has a high momentum $\pi^-$ ($\pi^+$) on the $CP$ side,
the highest momentum $\pi$ on the tag side tends to have $+$ ($-$)
charge.
The highest momentum $\pi$ on the tag side with $+$ ($-$) charge
tags the flavor as $\bz$ ($\bzb$).
Since an event with a high momentum $\pi^-$ ($\pi^+$) resides
in the region $\theta'>0.5$ ($\theta' < 0.5$),
a continuum event in the region $\theta'>0.5$ ($\theta' < 0.5$)
tends to be tagged as $\bz$ ($\bzb$).
We again parameterize the asymmetry $A^l(m', \theta')$
in a $\dE$-$\mbc$ independent way as
\begin{equation}
A^l(m', \theta') = A^l(m', \theta'; \dE, \mbc)
= A^l(m'_\mathrm{scale}, \theta') \;,
\end{equation}
and model it with a two-dimensional polynomial,
whose coefficients are determined by a fit to data in the
$\dE$-$\mbc$ sideband region.

\subsection{\boldmath $B\bbar$ background PDF}
The treatment of $B\bbar$ background is
different for $CP$ eigenstate modes
and flavor-specific or charged modes.
The PDF for the $CP$ eigenstate modes is
\begin{equation}
\begin{split}
& \mathcal{P}^l_{\BB} (\Delta E, \mbc; m', \theta';\Delta t,
q_\mathrm{tag})
\\
& \, = \mathcal{F}^l_{\BB} \cdot \mathcal{P}_{\BB}(\Delta E, \mbc)
\cdot \mathcal{P}_{\BB}(m', \theta')
\cdot \mathcal{P}_{\BB}(\Delta t, q_\mathrm{tag}; l),
\end{split}
\end{equation}
where $\mathcal{P}_{\BB}(\Delta t, q_\mathrm{tag})$ is a time-dependent $CP$
violation PDF normalized as
\begin{equation}
\sum_{q_\mathrm{tag}} \int \hspace{-1mm} d\Delta t \;
\mathcal{P}_{\BB}(\Delta t, q_\mathrm{tag} ; l) = 1 \;.
\quad (\forall l)
\end{equation}
For the flavor specific or charged modes,
the PDF is
\begin{equation}
\begin{split}
\mathcal{P}_{\BB} & (\Delta E, \mbc; m', \theta';\Delta t,
q_\mathrm{tag}; l)
\\
= & \mathcal{F}^l_{\BB} \cdot \mathcal{P}_{\BB}(\Delta E, \mbc)
\\
&
\cdot \sum_{q_\mathrm{rec}}
\mathcal{P}_{\BB}(m', \theta'; q_\mathrm{rec})
\mathcal{P}_{\BB}(\Delta t, q_\mathrm{tag}, q_\mathrm{rec}; l) \;,
\end{split}
\end{equation}
where the Dalitz plot PDF $\mathcal{P}_{\BB}(m', \theta'; q_\mathrm{rec})$
is dependent on the true flavor of the $CP$
(fully reconstructed) side, $q_\mathrm{rec}$,
and the time dependent part is a mixing PDF (lifetime PDF with flavor asymmetry)
for flavor specific (charged) modes.
The $\Delta t$ PDF is normalized as
\begin{equation}
\sum_{q_\mathrm{tag}} \sum_{q_\mathrm{rec}}
\int \hspace{-1mm} d\Delta t \;
\mathcal{P}_{\BB}(\Delta t, q_\mathrm{tag}, q_\mathrm{rec}; l) = 1 \;.
\quad (\forall l)
\end{equation}

The $\dE$-$\mbc$ PDF and Dalitz plot PDF are obtained mode-by-mode
from MC.
The Dalitz plot PDF of the $CP$ eigenstate modes is assumed to have the following
symmetry
\begin{equation}
\mathcal{P}_{\BB}(m', \theta')
= \mathcal{P}_{\BB}(m', 1-\theta') \;,
\end{equation}
while
that of
flavor specific and charged modes is assumed to have the following symmetry
\begin{equation}
\mathcal{P}_{\BB}(m', \theta'; q_\mathrm{rec})
= \mathcal{P}_{\BB}(m', 1-\theta'; -q_\mathrm{rec}) \;.
\end{equation}
The total PDF of the $B\bbar$ background is a linear combination
of each mode with efficiencies and branching fractions
taken into account.


\section{Correlation matrix of the fit result
\label{sec:appendix_correlation}
}
Tables
\ref{tab:correl_mtx_26_1}-\ref{tab:correl_mtx_26_3}
show the correlation matrix for the 26 parameters
determined in the time-dependent Dalitz plot analysis,
corresponding to the
total error matrix with statistical and systematic
error matrices combined.
We assume no correlations for the systematic errors.
\begin{table*}[htbp] 
\caption{
Correlation matrix (1) of the 26 fitted parameters,
with statistical and systematic errors combined.
\label{tab:correl_mtx_26_1}
}
\begin{tabular*}{14cm}{@{\extracolsep{\fill}}@{\hspace{5mm}}l@{\hspace{5mm}}|cccccccc}
\hline
\hline
& $U^+_-$                  & $U^+_0$                  & $U^{+,\mathrm{Re}}_{+-}$ & $U^{+,\mathrm{Re}}_{+0}$ & $U^{+,\mathrm{Re}}_{-0}$ & $U^{+,\mathrm{Im}}_{+-}$ & $U^{+,\mathrm{Im}}_{+0}$ & $U^{+,\mathrm{Im}}_{-0}$ \\
\hline
$U^+_-$                  & $+1.00$ \\
$U^+_0$                  & $+0.22$ & $+1.00$ \\
$U^{+,\mathrm{Re}}_{+-}$ & $+0.06$ & $+0.04$ & $+1.00$ \\
$U^{+,\mathrm{Re}}_{+0}$ & $+0.10$ & $+0.02$ & $+0.02$ & $+1.00$ \\
$U^{+,\mathrm{Re}}_{-0}$ & $-0.04$ & $-0.11$ & $+0.01$ & $+0.01$ & $+1.00$ \\
$U^{+,\mathrm{Im}}_{+-}$ & $+0.08$ & $+0.03$ & $+0.12$ & $+0.02$ & $-0.00$ & $+1.00$ \\
$U^{+,\mathrm{Im}}_{+0}$ & $-0.03$ & $-0.08$ & $-0.00$ & $+0.13$ & $+0.02$ & $-0.00$ & $+1.00$ \\
$U^{+,\mathrm{Im}}_{-0}$ & $-0.14$ & $-0.08$ & $-0.02$ & $-0.02$ & $+0.10$ & $-0.01$ & $+0.01$ & $+1.00$ \\
\hline
$U^-_+$                  & $+0.05$ & $+0.02$ & $+0.00$ & $-0.02$ & $+0.00$ & $-0.02$ & $-0.01$ & $-0.01$ \\
$U^-_-$                  & $-0.23$ & $-0.08$ & $-0.03$ & $-0.04$ & $-0.02$ & $-0.03$ & $+0.01$ & $+0.03$ \\
$U^-_0$                  & $+0.05$ & $+0.10$ & $+0.01$ & $+0.00$ & $-0.04$ & $+0.01$ & $-0.06$ & $-0.08$ \\
$U^{-,\mathrm{Re}}_{+-}$ & $-0.03$ & $-0.01$ & $-0.03$ & $-0.00$ & $-0.00$ & $-0.04$ & $+0.00$ & $+0.01$ \\
$U^{-,\mathrm{Re}}_{+0}$ & $-0.04$ & $-0.01$ & $-0.01$ & $-0.12$ & $-0.00$ & $+0.00$ & $-0.01$ & $+0.02$ \\
$U^{-,\mathrm{Re}}_{-0}$ & $-0.02$ & $-0.04$ & $-0.00$ & $-0.00$ & $+0.06$ & $-0.00$ & $+0.01$ & $+0.08$ \\
$U^{-,\mathrm{Im}}_{+-}$ & $-0.04$ & $-0.02$ & $-0.05$ & $-0.01$ & $-0.00$ & $+0.00$ & $+0.00$ & $+0.01$ \\
$U^{-,\mathrm{Im}}_{+0}$ & $-0.03$ & $-0.09$ & $-0.01$ & $-0.01$ & $+0.02$ & $-0.00$ & $-0.04$ & $+0.01$ \\
$U^{-,\mathrm{Im}}_{-0}$ & $+0.01$ & $-0.02$ & $+0.00$ & $+0.00$ & $+0.00$ & $-0.00$ & $+0.00$ & $-0.25$ \\
\hline
$I_+$                    & $+0.00$ & $+0.00$ & $-0.02$ & $-0.01$ & $-0.00$ & $-0.05$ & $-0.01$ & $-0.00$ \\
$I_-$                    & $+0.06$ & $+0.03$ & $-0.01$ & $+0.01$ & $-0.02$ & $+0.05$ & $-0.00$ & $+0.04$ \\
$I_0$                    & $+0.01$ & $+0.01$ & $+0.00$ & $+0.02$ & $-0.00$ & $+0.00$ & $-0.02$ & $-0.02$ \\
$I^{\mathrm{Re}}_{+-}$   & $-0.04$ & $-0.01$ & $+0.01$ & $-0.00$ & $+0.00$ & $-0.16$ & $+0.00$ & $+0.00$ \\
$I^{\mathrm{Re}}_{+0}$   & $+0.00$ & $+0.02$ & $+0.00$ & $-0.13$ & $-0.01$ & $-0.00$ & $-0.00$ & $-0.00$ \\
$I^{\mathrm{Re}}_{-0}$   & $-0.06$ & $+0.01$ & $-0.01$ & $-0.01$ & $-0.12$ & $+0.00$ & $-0.01$ & $-0.29$ \\
$I^{\mathrm{Im}}_{+-}$   & $-0.02$ & $-0.01$ & $+0.13$ & $-0.00$ & $+0.00$ & $+0.00$ & $+0.00$ & $+0.00$ \\
$I^{\mathrm{Im}}_{+0}$   & $-0.01$ & $-0.03$ & $+0.00$ & $+0.00$ & $+0.01$ & $+0.01$ & $+0.04$ & $+0.01$ \\
$I^{\mathrm{Im}}_{-0}$   & $-0.06$ & $-0.04$ & $-0.01$ & $-0.02$ & $-0.09$ & $-0.02$ & $+0.01$ & $+0.08$ \\
\hline

\end{tabular*}
\end{table*} 
\begin{table*}[htbp] 
\caption{
Correlation matrix (2) of the 26 fitted parameters,
with statistical and systematic errors combined.
\label{tab:correl_mtx_26_2}
}
\begin{tabular*}{15cm}{@{\extracolsep{\fill}}@{\hspace{5mm}}l@{\hspace{5mm}}|ccccccccc}
\hline
\hline
& $U^-_+$                  & $U^-_-$                  & $U^-_0$                  & $U^{-,\mathrm{Re}}_{+-}$ & $U^{-,\mathrm{Re}}_{+0}$ & $U^{-,\mathrm{Re}}_{-0}$ & $U^{-,\mathrm{Im}}_{+-}$ & $U^{-,\mathrm{Im}}_{+0}$ & $U^{-,\mathrm{Im}}_{-0}$ \\
\hline
$U^-_+$                  & $+1.00$ \\
$U^-_-$                  & $-0.06$ & $+1.00$ \\
$U^-_0$                  & $+0.00$ & $-0.01$ & $+1.00$ \\
$U^{-,\mathrm{Re}}_{+-}$ & $-0.07$ & $+0.01$ & $-0.00$ & $+1.00$ \\
$U^{-,\mathrm{Re}}_{+0}$ & $-0.21$ & $+0.03$ & $-0.08$ & $+0.02$ & $+1.00$ \\
$U^{-,\mathrm{Re}}_{-0}$ & $+0.01$ & $-0.12$ & $-0.16$ & $-0.00$ & $+0.02$ & $+1.00$ \\
$U^{-,\mathrm{Im}}_{+-}$ & $+0.03$ & $+0.03$ & $-0.00$ & $+0.20$ & $-0.01$ & $-0.00$ & $+1.00$ \\
$U^{-,\mathrm{Im}}_{+0}$ & $-0.02$ & $+0.01$ & $-0.03$ & $+0.00$ & $+0.01$ & $+0.01$ & $+0.00$ & $+1.00$ \\
$U^{-,\mathrm{Im}}_{-0}$ & $+0.00$ & $-0.03$ & $+0.02$ & $-0.00$ & $-0.00$ & $+0.14$ & $-0.00$ & $+0.01$ & $+1.00$ \\
\hline
$I_+$                    & $-0.02$ & $-0.01$ & $+0.00$ & $+0.03$ & $+0.01$ & $+0.00$ & $-0.02$ & $+0.01$ & $+0.00$ \\
$I_-$                    & $-0.00$ & $-0.01$ & $+0.01$ & $+0.03$ & $-0.00$ & $-0.05$ & $+0.00$ & $-0.01$ & $-0.07$ \\
$I_0$                    & $+0.00$ & $-0.01$ & $+0.07$ & $-0.00$ & $-0.02$ & $+0.02$ & $-0.00$ & $-0.05$ & $-0.06$ \\
$I^{\mathrm{Re}}_{+-}$   & $+0.02$ & $+0.01$ & $-0.00$ & $-0.02$ & $-0.00$ & $+0.00$ & $-0.15$ & $+0.00$ & $+0.00$ \\
$I^{\mathrm{Re}}_{+0}$   & $-0.01$ & $+0.00$ & $+0.02$ & $+0.00$ & $+0.09$ & $-0.01$ & $-0.00$ & $+0.16$ & $+0.01$ \\
$I^{\mathrm{Re}}_{-0}$   & $-0.01$ & $+0.08$ & $+0.08$ & $+0.01$ & $-0.00$ & $-0.12$ & $+0.01$ & $-0.00$ & $+0.21$ \\
$I^{\mathrm{Im}}_{+-}$   & $+0.02$ & $+0.04$ & $-0.00$ & $+0.04$ & $-0.00$ & $-0.00$ & $+0.04$ & $+0.00$ & $-0.00$ \\
$I^{\mathrm{Im}}_{+0}$   & $+0.03$ & $+0.00$ & $-0.03$ & $-0.01$ & $-0.28$ & $+0.00$ & $+0.01$ & $-0.03$ & $+0.01$ \\
$I^{\mathrm{Im}}_{-0}$   & $-0.00$ & $+0.01$ & $-0.01$ & $-0.00$ & $+0.01$ & $+0.18$ & $+0.00$ & $+0.02$ & $+0.11$ \\
\hline

\end{tabular*}
\end{table*} 
\begin{table*}[htbp] 
\caption{
Correlation matrix (3) of the 26 fitted parameters,
with statistical and systematic errors combined.
\label{tab:correl_mtx_26_3}
}
\begin{tabular*}{15cm}{@{\extracolsep{\fill}}@{\hspace{5mm}}l@{\hspace{7mm}}|ccccccccc}
\hline
\hline
& $I_+$                    & $I_-$                    & $I_0$                    & $I^{\mathrm{Re}}_{+-}$   & $I^{\mathrm{Re}}_{+0}$   & $I^{\mathrm{Re}}_{-0}$   & $I^{\mathrm{Im}}_{+-}$   & $I^{\mathrm{Im}}_{+0}$   & $I^{\mathrm{Im}}_{-0}$   \\
\hline
$I_+$                    & $+1.00$ \\
$I_-$                    & $-0.06$ & $+1.00$ \\
$I_0$                    & $+0.00$ & $+0.01$ & $+1.00$ \\
$I^{\mathrm{Re}}_{+-}$   & $-0.04$ & $-0.06$ & $-0.00$ & $+1.00$ \\
$I^{\mathrm{Re}}_{+0}$   & $+0.04$ & $-0.00$ & $-0.14$ & $-0.00$ & $+1.00$ \\
$I^{\mathrm{Re}}_{-0}$   & $-0.02$ & $+0.21$ & $+0.01$ & $-0.01$ & $+0.00$ & $+1.00$ \\
$I^{\mathrm{Im}}_{+-}$   & $-0.07$ & $-0.01$ & $-0.00$ & $-0.35$ & $-0.00$ & $+0.00$ & $+1.00$ \\
$I^{\mathrm{Im}}_{+0}$   & $-0.15$ & $+0.01$ & $-0.09$ & $+0.01$ & $-0.23$ & $-0.00$ & $+0.01$ & $+1.00$ \\
$I^{\mathrm{Im}}_{-0}$   & $+0.01$ & $-0.14$ & $-0.23$ & $+0.01$ & $+0.04$ & $-0.06$ & $+0.00$ & $+0.03$ & $+1.00$ \\
\hline

\end{tabular*}
\end{table*} 


\section{Formalism for branching fraction measurement
\label{sec:br_formalism}
}
In this section, we describe the formalism for the factors used in the
branching fraction calculation of Eqs.~(\ref{equ:br_rhopi_all}) and
(\ref{equ:br_rhopi_770}).

\subsection{Detection Efficiency}
Since the detection efficiency is Dalitz plot dependent, we use the
detection efficiency averaged over Dalitz plot, $\epsilon^{\text{Det}}$,
for the branching fraction measurement.
From Eqs.~(\ref{equ:signal_pdf_definition}),
(\ref{equ:app_pdf_true_pdf}), (\ref{equ:app_pdf_true_pdf_dalitz_dt}),
(\ref{equ:app_pdf_scf_pdf}), (\ref{equ:dt_pdf_of_SCF}), and
(\ref{equ:app_pdf_scf_dalitz_convl}), the PDF integrated over
$\dE$-$\mbc$ (in the signal region) and $\Dt$, and summed over $q_\mathrm{tag}$
and $l$ is
\begin{equation}
\mathcal{P}_\text{sig}(m', \theta')
=
\frac{\mathcal{P}_\text{true}(m', \theta')
+ \sum_{i = \text{CR, NR}} \mathcal{P}_i(m', \theta')}
{\mathcal{N}_\text{true} + \sum_{i = \text{CR, NR}}
\mathcal{N}_i}
\;,
\end{equation}
with
\begin{equation}
\begin{split}
\mathcal{P}_\text{true} & (m', \theta')
\\
& =
\sum_l {\mathcal{F}}^l_\text{true}
\epsilon_\text{true}(m', \theta'; l)
\epsilon'(p_{\piz})
P_\text{phys}(m', \theta') \;,
\end{split}
\end{equation}
\begin{equation}
\mathcal{P}_i(m', \theta')
=
\sum_l {\mathcal{F}}^l_i
\left[
(\epsilon_i \cdot R_i) \otimes
P_\text{phys}
\right](m', \theta')
\;,
\end{equation}
and
\begin{equation}
\begin{split}
P_\text{phys} & (m', \theta')
\\
= &
|\boldsymbol{J}(m', \theta')|
(|A_{3\pi}|^2+|\overline{A}_{3\pi}|^2)
\\
= &
|\boldsymbol{J}(m', \theta')|
\biggl\{
\sum_\kappa U^+_\kappa |f_\kappa|^2
\\
& \qquad
+ \sum_{\kappa<\sigma}
\left( U^{+, \RE}_{\kappa\sigma} \RE [f{}_\kappa f{}^*_\sigma]
-
U^{+, \IM}_{\kappa\sigma} \IM [f{}_\kappa f{}^*_\sigma]
\right)
\biggr\} \;.
\end{split}
\label{equ:branch_formalizm_Pphys}
\end{equation}
Consequently,
$\mathcal{P}_\text{sig}(m', \theta')$ can be symbolically rewritten
as
\begin{equation}
\mathcal{P}_\text{sig}(m', \theta')
=
\frac{[\epsilon^\text{sig} \otimes P_\text{phys}](m', \theta')}
{\mathcal{N}_\text{true} + \sum_{i = \text{CR, NR}}
\mathcal{N}_i}
\;,
\end{equation}
where
\begin{equation}
\begin{split}
[\epsilon^\text{sig} & \otimes P_\text{phys}](m', \theta')
\\
\equiv &  \sum_l \mathcal{F}_\text{true}^l
\epsilon_\text{true}(m', \theta'; l)
\epsilon'(p_{\piz})
P_\text{phys}(m', \theta')
\\
& +
\sum_{i=\text{CR, NR}} \sum_l
\mathcal{F}_i^l
\left[
(\epsilon_i \cdot R_i) \otimes
P^\text{phys}
\right](m', \theta') \;.
\end{split}
\end{equation}
Note that $R_i$ is normalized as
Eq.~(\ref{equ:app_pdf_scf_resolution_normalization}).

Consequently, the detection efficiency averaged over the Dalitz plot
$\epsilon^\text{Det}$ is
\begin{eqnarray}
\epsilon^\text{Det}
& = &
\frac{
\displaystyle
\iint_\text{SDP, Veto} \hspace{-10mm} dm' d\theta' \:
[\epsilon^\text{sig} \otimes P^\text{phys}](m', \theta')
}
{
\displaystyle
\iint_\text{SDP, Veto} \hspace{-10mm} dm' d\theta' \:
P^\text{phys}(m', \theta')
}
\\
& = &
\frac{
\mathcal{N}_\text{true} + \sum_{i = \text{CR, NR}}
\mathcal{N}_i
}
{
\displaystyle
\iint_\text{SDP, Veto} \hspace{-10mm} dm' d\theta' \:
P^\text{phys}(m', \theta')
} \;.
\end{eqnarray}

\subsection{Dalitz Veto Efficiency}
The efficiency corresponding to the Dalitz veto,
$\epsilon^\text{Veto}$, is simply calculated as
\begin{equation}
\epsilon^\text{Veto}
=
\frac{
\displaystyle
\iint_\text{SDP, Veto} \hspace{-10mm} dm' d\theta' \:
P_\text{phys}(m', \theta')
}
{
\displaystyle
\iint_\text{SDP, Whole} \hspace{-12mm} dm' d\theta' \:
P_\text{phys}(m', \theta')
}
\; .
\end{equation}

\subsection{\boldmath Fraction of $\rho^\kappa_{\text{all}} \pi^\sigma$}
The fraction of
$\bz \to \rho^\kappa_\text{all} \pi^{\sigma}$
normalized to $\bz \to \pipipi$
is
\begin{equation}
f_{\rho_\text{all}^\kappa}
=
\frac{
\displaystyle
\iint_\text{SDP, Whole} \hspace{-12mm} dm' d\theta' \:
P^\text{phys}_{\text{all}, \kappa}(m', \theta')
}
{
\displaystyle
\iint_\text{SDP, Whole} \hspace{-12mm} dm' d\theta' \:
P_\text{phys}(m', \theta')
} \;,
\end{equation}
where
\begin{equation}
P^\text{phys}_{\text{all}, \kappa}(m', \theta')
=
|\boldsymbol{J}(m', \theta')| \:
U^+_\kappa |f_\kappa|^2 \; .
\end{equation}

\subsection{\boldmath Fraction of $\rho^\kappa (770) \pi^\sigma$}
The fraction of $\bz \to \rho^\kappa (770) \pi^{\sigma}$
normalized to $\bz \to \rho_\text{all}^\kappa \pi^{\sigma}$
is
\begin{equation}
f_{\rho^\kappa(770)}
=
\frac{
\displaystyle
\iint_\text{SDP, Whole} \hspace{-12mm} dm' d\theta' \:
P^\text{phys}_{(770), \kappa}(m', \theta')
}
{
\displaystyle
\iint_\text{SDP, Whole} \hspace{-12mm} dm' d\theta' \:
P^\text{phys}_{\text{all}, \kappa}(m', \theta')
} \;,
\end{equation}
where
\begin{equation}
P^\text{phys}_{(770), \kappa}(m', \theta')
=
|\boldsymbol{J}(m', \theta')| \:
U^+_\kappa |f^{\rho(770)}_\kappa|^2 \; .
\end{equation}
The function $f^{\rho(770)}_\kappa$ is defined as
\begin{equation}
f^{\rho(770)}_\kappa
= T^\kappa_{J = 1} BW_{\rho(770)}(s_\kappa) \;,
\end{equation}
corresponding to the $\rho(770)$ part of $f_\kappa$ defined in
Eqs.~(\ref{equ:f_kappa_definition_with_unique_lineshape_assumption})
and (\ref{equ:fpi_lineshape_beta_gamma}).

\subsection{Summary}
By using the expressions described above, Eqs.~(\ref{equ:br_rhopi_all})
and (\ref{equ:br_rhopi_770}) are rewritten as
\begin{equation}
\mathcal{B}(\rho\pi^\text{all})
=
\frac{N_\text{sig}}{N_{\BB}}
\frac{
\displaystyle
\iint_\text{SDP, Whole} \hspace{-12mm} dm' d\theta' \:
P_\text{phys}(m', \theta')
}{
\mathcal{N}_\text{true} + \sum_{i = \text{CR, NR}}
\mathcal{N}_i
}
\;,
\end{equation}
and
\begin{equation}
\mathcal{B}(\rho^\kappa (770) \pi^\sigma)
=
\frac{N_\text{sig}}{N_{\BB}}
\frac{
\displaystyle
\iint_\text{SDP, Whole} \hspace{-12mm} dm' d\theta' \:
P^\text{phys}_{(770), \kappa}(m', \theta')
}
{
\mathcal{N}_\text{true} + \sum_{i = \text{CR, NR}}
\mathcal{N}_i
} \;.
\end{equation}


\section{\boldmath Method of $\phi_2$ constraint
\label{sec:phi2_chi2_parameterization}
}

\subsection{Formalism}
We define amplitudes as
\begin{eqnarray}
A^+    & \equiv & A(B^0\rightarrow \rho^+ \pi^-) \;, \\
A^-    & \equiv & A(B^0\rightarrow \rho^- \pi^+) \;, \\
A^0    & \equiv & A(B^0\rightarrow \rho^0 \pi^0) \;, \\
A^{+0} & \equiv & A(B^+\rightarrow \rho^+ \pi^0) \;, \\
A^{0+} & \equiv & A(B^+\rightarrow \rho^0 \pi^+) \;,
\end{eqnarray}
and
\begin{eqnarray}
\overline{A}{}^+ & \equiv & \frac{q}{p} A(\overline{B}{}^0\rightarrow \rho^+ \pi^-) \;, \\
\overline{A}{}^- & \equiv & \frac{q}{p} A(\overline{B}{}^0\rightarrow \rho^- \pi^+) \;, \\
\overline{A}{}^0 & \equiv & \frac{q}{p} A(\overline{B}{}^0\rightarrow \rho^0 \pi^0) \;, \\
A^{-0} & \equiv & \frac{q}{p} A(B^-\rightarrow \rho^- \pi^0) \;, \\
A^{0-} & \equiv & \frac{q}{p} A(B^-\rightarrow \rho^0 \pi^-) \;.
\end{eqnarray}
These amplitudes are obtained from
1) 26 measurements determined in the time-dependent Dalitz plot
analysis as well as
2) branching fractions and asymmetry measurements,
and give a constraint on $\phi_2$.

Equations~(\ref{equ:fit_params_first})--(\ref{equ:fit_params_last})
define the relations between the amplitudes for the neutral modes
and the parameters determined in the time-dependent Dalitz plot analysis.
The relations between the branching fractions and asymmetries,
and the amplitudes are
\begin{equation}
\mathcal{B}(\rho \pi^\text{all}) =
c
\cdot
\sum_{\kappa = +, -, 0}
\left( |A^\kappa|^2 + |\overline{A}{}^\kappa|^2 \right)
\cdot
\tau_{B^0}
\label{equ:phi2_constraint_br_pm_definition}
\;,
\end{equation}
\begin{equation}
\mathcal{B}(\rho^+ \pi^0) =
c
\cdot
\left(
|A^{+0}|^2 + |A^{-0}|^2
\right)
\cdot
\tau_{B^+} \;,
\end{equation}
\begin{equation}
\mathcal{B}(\rho^0 \pi^+) =
c
\cdot
\left(
|A^{0+}|^2 + |A^{0-}|^2
\right)
\cdot
\tau_{B^+} \;,
\end{equation}
\begin{equation}
\mathcal{A}(\rho^+ \pi^0)
= \frac{|A^{-0}|^2 - |A^{+0}|^2}{|A^{-0}|^2 + |A^{+0}|^2} \;,
\end{equation}
\begin{equation}
\mathcal{A}(\rho^0 \pi^+)
= \frac{|A^{0-}|^2 - |A^{0+}|^2}{|A^{0-}|^2 + |A^{0+}|^2} \;,
\end{equation}
where $c$ is a constant
and the lifetimes $\tau_{B^0}$ and $\tau_{B^+}$ are
introduced to take account of the total width difference
between $B^0$ and $B^+$.
Note that we do not use quasi-two-body parameters
related to neutral modes except for $\mathcal{B}(\rho^\pm \pi^\mp)$,
since they are included in the Dalitz plot parameters.

The amplitudes are expected to follow $SU(2)$ isospin symmetry
to a good approximation \cite{Lipkin:1991st,Gronau:1991dq}
\begin{equation}
\begin{array}{l}
A^+ + A^- + 2A^0 = \widetilde{A}{}^+ + \widetilde{A}{}^- + 2\tilde{A}{}^0 \\
\qquad = \sqrt{2} (A^{+0} + A^{0+})
= \sqrt{2} (\widetilde{A}{}^{-0} + \widetilde{A}{}^{0-}) \;,
\end{array}
\label{equ:phi2_constraint_isospin_relation_1}
\end{equation}
\begin{equation}
A^{+0} - A^{0+} - \sqrt{2} (A^+ - A^-)
=
\widetilde{A}^{-0} - \widetilde{A}^{0-} - \sqrt{2} (\widetilde{A}^- - \widetilde{A}^+) \;,
\label{equ:phi2_constraint_isospin_relation_2}
\end{equation}
where
\begin{equation}
\begin{split}
\widetilde{A}{}^\kappa \equiv e^{-2i\phi_2} \overline{A}{}^\kappa \;,
\quad
& \widetilde{A}{}^{-0} \equiv e^{-2i\phi_2} A^{-0} \;,
\\
& \mathrm{and}
\quad
\widetilde{A}{}^{0-} \equiv e^{-2i\phi_2} A^{0-} \;.
\label{equ:phi2_constraint_tilde_A_def}
\end{split}
\end{equation}
Note that there is an inconsistency
in Eq.~(\ref{equ:phi2_constraint_isospin_relation_2})
between Ref.~\cite{Lipkin:1991st} and Ref.~\cite{Gronau:1991dq};
we follow the treatment of Ref.~\cite{Lipkin:1991st}.

\subsection{Parameterization}
Here we give two examples of the parameterization
of the amplitudes.
The first example may be more intuitive,
while the second example is well behaved in the fit.
The results are independent of the parameterizations
with respect to the constraint on $\phi_2$.

In the following, we set the constant $c$ to be unity and
discard the normalization condition
$|A^+|^2 + |\overline{A}{}^+|^2 = 1$
instead.
This is equivalent to letting $c$ be a free parameter in the
$\phi_2$ constraint fit, keeping the normalization
$|A^+|^2 + |\overline{A}{}^+|^2 = 1$.
We adopt the former for simplicity. Note that
Eqs.~(\ref{equ:fit_params_first})--(\ref{equ:fit_params_last})
become
\begin{eqnarray*}
U^\pm_{\kappa}
& = & \left(|A^\kappa|^2 \pm |\overline{A}{}^\kappa|^2\right) / N \;,
\\
I_\kappa & = & \mathrm{Im}
\left[
\overline{A}{}^{\kappa} A^{\kappa *}
\right]  / N \;,
\\
U^{\pm,\mathrm{Re}(\mathrm{Im})}_{\kappa\sigma}
& = & \mathrm{Re}(\mathrm{Im})
\left[ A^\kappa A^{\sigma *} \pm \overline{A}{}^\kappa
\overline{A}{}^{\sigma *}
\right]  / N \;,
\\
I^\mathrm{Re(Im)}_{\kappa \sigma} & = & \mathrm{Re(Im)}
\left[
\overline{A}{}^\kappa A^{\sigma *}
\! - \! (+) \, \overline{A}{}^\sigma A^{\kappa *}
\right]  / N \;.
\end{eqnarray*}
\[
\left( \; N \equiv |A^+|^2 + |\overline{A}{}^+|^2 \; \right)
\]
in this case.

\subsubsection{Amplitude parameterization}
We can parameterize the amplitudes as follows~\cite{Lipkin:1991st}
\begin{eqnarray}
A^+ & = & e^{-i\phi_2} T^+ + P^+ \;, \label{equ:phi2_constraint_ampl_param_begin} \\
A^- & = & e^{-i\phi_2} T^- + P^- \;, \\
A^0 & = & e^{-i\phi_2} T^0 - \frac{1}{2}(P^+ + P^-) \;, \\
\sqrt{2} A^{+0} & = & e^{-i\phi_2} T^{+0} + P^+ - P^- \;,
\end{eqnarray}
\begin{equation}
\sqrt{2} A^{0+}  =  e^{-i\phi_2} (T^+ + T^- + 2T^0 - T^{+0})
- P^+ + P^- \;,
\end{equation}
and
\begin{eqnarray}
\overline{A}{}^+ & = & e^{+i\phi_2} T^- + P^- \;, \\
\overline{A}{}^- & = & e^{+i\phi_2} T^+ + P^+ \;, \\
\overline{A}{}^0 & = & e^{+i\phi_2} T^0 - \frac{1}{2}(P^+ + P^-) \;, \\
\sqrt{2} A^{-0} & = & e^{+i\phi_2} T^{+0} + P^+ - P^- \;,
\end{eqnarray}
\begin{equation}
\sqrt{2} A^{0-}  =  e^{+i\phi_2} (T^+ + T^- + 2T^0 - T^{+0})
- P^+ + P^- \;,
\label{equ:phi2_constraint_ampl_param_end}
\end{equation}
where the overall phase is fixed with the convention $\mathrm{Im}T^+ = 0$.
Thus, there are 6 complex amplitudes, $T^+, T^-, T^0, P^+, P^-$, and
$T^{+0}$,
corresponding to 11 degrees of freedom;
and $\phi_2$, corresponding to 12 degrees of freedom in total.
This parameterization automatically satisfies the isospin relations
without loss of generality,
i.e., the isospin relations are the only assumption here.

\subsubsection{Geometric parameterization}
\begin{figure}[tbp]
\begin{center}
\includegraphics[width=0.35\textwidth]{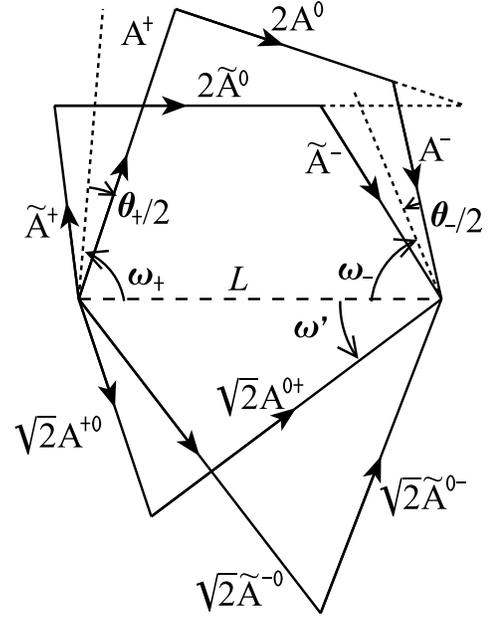}
\caption{Complex pentagons formed from the $B \rightarrow \rho\pi$ decay
amplitudes.
\label{fig:phi2_constraint_geometry}
}
\end{center}
\end{figure}

We can parameterize the amplitudes using the
geometric arrangement of Fig.~\ref{fig:phi2_constraint_geometry}
that satisfies the isospin relation of
Eq.~(\ref{equ:phi2_constraint_isospin_relation_1}).
This figure is equivalent to Fig.~3 of
Ref.~\cite{Gronau:1991dq},
except that
the sides corresponding to $B^0 \rightarrow \rho^-\pi^+$
and $B^0 \rightarrow \rho^0\pi^0$
are swapped.
This difference is not physically significant.
We apply this modification only to obtain a better behaved
parameterization;
the parameterization here uses
the angles $\omega_-$ and $\theta_-$
related to the process $\bz(\bzb) \rightarrow \rho^-\pi^+$,
which are better behaved than those related to
$\bz(\bzb) \rightarrow \rho^0\pi^0$.

To parameterize the amplitudes, we use $\phi_2$ and
the following 11 geometric parameters:
\begin{equation}
\omega_+, \omega_-, \omega', \theta_+, \theta_-, b_+, b_-, b', a_+,
a_-, L,
\end{equation}
where $b$ and $a$ {\it imply} branching fraction and asymmetry, respectively.
In terms of these parameters,
the amplitudes can be described as follows
\begin{eqnarray}
A^+ & = & e^{i(\omega_+ + \theta_+/2)} \sqrt{b_+  (1 - a_+) / 2} \;, \\
\widetilde{A}{}^+ & = & e^{i(\omega_+ - \theta_+/2)} \sqrt{b_+  (1 + a_+) / 2} \;, \\
A^- & = & e^{i(\omega_- + \theta_-/2)} \sqrt{b_-  (1 - a_-) / 2} \;, \\
\widetilde{A}{}^- & = & e^{i(\omega_- - \theta_-/2)} \sqrt{b_-  (1 + a_-) / 2} \;, \\
A^0 & = & (L - A^+ - A^-) / 2 \;, \\
\widetilde{A}^0 & = & (L - \widetilde{A}{}^+ - \widetilde{A}{}^-)/2 \;, \\
A^{0+} & = & e^{i\omega'} \sqrt{b'/2} \;, \\
A^{+0} & = & \frac{L}{\sqrt{2}} - A^{0+} \;, \\
\widetilde{A}^{-0} & = & \frac{L}{\sqrt{2}} - \widetilde{A}{}^{0-} \;,
\end{eqnarray}
and
\begin{equation}
\begin{split}
\widetilde{A}^{0-} = \frac{L}{2\sqrt{2}}
- \Bigl[ &
A^{+0} - A^{0+}
- \sqrt{2} (A^+ - A^-)
\\
& \quad + \sqrt{2} (\widetilde{A}{}^- - \widetilde{A}{}^+)
\Bigr] / 2 \;.
\end{split}
\label{equ:phi2_constraint_geom_util_unused_isospin}
\end{equation}
Equation~(\ref{equ:phi2_constraint_geom_util_unused_isospin})
exploits the isospin relation of
Eq.~(\ref{equ:phi2_constraint_isospin_relation_2}),
which Fig.~\ref{fig:phi2_constraint_geometry} does not incorporate geometrically.
The phase $\phi_2$ enters when the $\widetilde{A}$'s are converted
into $\overline{A}$'s with Eq.~(\ref{equ:phi2_constraint_tilde_A_def}).
When we perform the analysis only with the time-dependent Dalitz plot
observables and without the information from charged decay modes,
we remove the parameters $\omega'$ and $b'$ from the fit
and fix $L$ to be a constant.

This geometric parameterization has
a substantial advantage
in terms of required computational resources,
compared to the parameterization
based on the $T$ and $P$ amplitudes described in the
previous section.
In the procedure to constrain $\phi_2$,
the minimum $\chi^2$ has to be calculated for each
value of $\phi_2$.
To avoid local minima, initial values of the parameters
in the minimization have to be scanned.
This inflates the computing time,
which increases exponentially with the number of parameters.
However, the number of parameters to be scanned
decreases in the geometric parameterization.
Among the 11 parameters except for $\phi_2$, five of them,
$b_+, b_-, b', a_+$, and $a_-$,
are related to the branching fractions and asymmetries.
Since in most cases they do not have multiple solutions,
we do not have to scan their initial values.
In addition,
the optimum initial value for $L$
can also be determined using other parameters
and $b_0$,
the nominal branching fraction of $B^0 \rightarrow
\rho^0\pi^0$,
from the following relation
\begin{equation}
b_0 = \left|L - e^{i\omega_+} \sqrt{b_+/2}
- e^{i\omega_-} \sqrt{b_-/2}\right|^2 \;,
\end{equation}
up to a two-fold ambiguity.
Here $b_0$ is calculated using the input parameters as
\begin{equation}
b_0 =
\frac{U^+_0}{U^+_+ + U^+_-} \cdot
\frac{\mathcal{B}(\rho^\pm \pi^\mp)}{c \cdot \tau_{B^0}} \;,
\end{equation}
based on Eqs.~(\ref{equ:fit_params_first})
and (\ref{equ:phi2_constraint_br_pm_definition}).
The explicit solution for the optimal initial value of $L$ is
\begin{equation}
\begin{split}
L & = \mathrm{Re}\gamma \pm \sqrt{b_0 - \left(\mathrm{Im} \gamma
\right)^2} \;.
\\
\Bigl(
\quad \text{where} \;\;
& \gamma \equiv
e^{i\omega_+} \sqrt{b_+/2}
+ e^{i\omega_-} \sqrt{b_-/2}
\quad
\Bigr)
\end{split}
\end{equation}
When $b_0 - \left(\mathrm{Im} \gamma \right)^2 < 0$,
there is no real-valued solution and
$L = \mathrm{Re}\gamma$ is the optimum initial value.
With the optimum values calculated above,
the initial value of $L$ does not have to be scanned,
except for the two-fold ambiguity.
Consequently, the number of parameters
to be scanned in this parameterization is only five,
corresponding to
$\omega_+, \omega_-, \omega', \theta_+$, and $\theta_-$,
while 10 of the 11 parameters
have to be scanned in the $T$ and $P$ amplitude parameterization.
This leads to a substantial reduction
of the computational resources required.


\bibliographystyle{apsrev.bst}
\bibliography{cites}

\begin{thebibliography}{43}
\bibitem[{Cha()}]{ChargeConjugate}
\bibinfo{note}{Throughout this paper, the inclusion of the charge conjugate
  decay mode is implied unless otherwise stated.}
\bibitem[{Fpi()}]{Fpi_strictly_speakiing}
\bibinfo{note}{Strictly speaking, the difference between the $\pi^\pm$ and
  $\pi^0$ masses is taken into acccount and thus the form factors
  $F_{\pi}(s_\pm)$ and $F_{\pi}(s_0)$ are slightly different. The essential
  point here is, however, that one unique set of $(\beta, \gamma)$ is used for
  all six $\kakkoOverlineF_\kappa$.}
\bibitem[{Chi()}]{Chi2ToyMC}
\bibinfo{note}{$\Delta \chi^2(\phi_2)$ is usually expected to follow a $\chi^2$
  distribution with one degree of freedom and thus the cumulative $\chi^2$
  distribution for one degree of freedom is usually used to convert $\Delta
  \chi^2(\phi_2)$ into a $\mathrm{1-C.L.}$ plot. A toy MC study shows, however,
  that this is not the case for $B\rightarrow \rho \pi$, and an analysis with
  this assumption yields confidence intervals with undercoverage. Thus, we
  perform a dedicated Toy MC study to obtain the confidence interval.}
\end{thebibliography}

\end{document}